\documentclass[a4paper,11pt]{article}

\usepackage[margin=1in]{geometry} 

\usepackage{amsmath}
\usepackage{amsthm}
\usepackage{amssymb}
\usepackage{lineno}
\usepackage{subcaption}  

\usepackage{dutchcal}

\usepackage{placeins}

\usepackage{etoolbox}

\usepackage{graphicx}
\usepackage{bm}
\usepackage{lineno}
\usepackage{caption}

\usepackage{booktabs}

\usepackage[breaklinks]{hyperref}

\usepackage{multirow}

\usepackage{comment}
\usepackage{xargs}                      

\hypersetup{
	unicode,
	colorlinks=true,
	linkcolor=blue,
	filecolor=magenta,      
	citecolor=blue,
	urlcolor=cyan,
	pdfauthor={ S. A. Ahizi},
	pdftitle={Real-time identification of parametric sloshing-induced heat and mass transfer in a horizontally oriented cylindrical tank},
	pdfsubject={Real-time identification of parametric sloshing-induced heat and mass transfer in a horizontally oriented cylindrical tank},
	pdfkeywords={article, template, simple},
	pdfproducer={LaTeX},
	pdfcreator={pdflatex}
	colorlinks=true,
	citecolor=blue,
	linkcolor=blue,
	filecolor=magenta,      
	urlcolor=blue,
}

\usepackage[most]{tcolorbox}
\usepackage{booktabs}
\usepackage{amsmath}
\usepackage{wrapfig}

\usepackage{mathtools}
\tcbset{enhanced,colback=cyan!2!white,colframe=cyan!90!white,coltitle=blue!20!black,fonttitle=\bfseries}

\usepackage{listings}
\usepackage{hyperref}
\hypersetup{
	colorlinks=true,
	linkcolor=blue,
	filecolor=magenta,      
	citecolor=blue,
	urlcolor=cyan,
}

\usepackage[sort&compress,authoryear,round]{natbib}
\usepackage{algorithm, algpseudocode} 
\usepackage{mathrsfs} 

\usepackage{siunitx}
\newcolumntype{Z}[1]{S[scientific-notation=fixed, fixed-exponent=#1, table-format=1.2]}

\title{\LARGE \textbf{Real-time identification of parametric sloshing-induced heat and mass transfer in a horizontally oriented cylindrical tank}}

\author{
	Samuel Akatchi Ahizi$^{1,2}$\thanks{Corresponding author: samuel.ahizi@vki.ac.be} \and
	Francisco Monteiro$^{1,2}$ \and
	Ramòn Abarca$^{3}$ \and
	Miguel Alfonso Mendez$^{1,2,4}$
}

\date{%
	\begingroup
	\small
	\textit{%
		$^1$ Environmental and Applied Fluid Dynamics, von Karman Institute for Fluid Dynamics, Belgium \\
		$^2$ Experimental Aerodynamics and Propulsion Lab, Universidad Carlos III de Madrid, Spain \\
		$^3$ Airbus Operations S.L., Spain \\
		$^4$ Aero-Thermo-Mechanics Laboratory, École Polytechnique de Bruxelles, Université Libre de Bruxelles, Belgium}
	\endgroup
}

\begin{document}
	\maketitle
	
	\begin{abstract}

	Vertical forcing of partially filled tanks can induce parametric sloshing. Under non-isothermal conditions, the resulting mixing can disrupt the thermal stratification between liquid and vapor, leading to enhanced heat and mass transfer and large pressure fluctuations. This work presents an experimental investigation of sloshing-induced heat and mass transfer in a horizontally oriented cylindrical tank under vertical harmonic excitation. This configuration is particularly relevant for cryogenic fuel storage in aircraft and ground transportation, yet its thermodynamic response under parametric sloshing remains largely uncharacterized. {The present study provides the first experimental characterization of the sloshing-induced pressure drop and associated heat and mass transfer in this geometry.} Decoupled isothermal and non-isothermal experimental campaigns are carried out across multiple fill levels and forcing amplitudes, near resonance of the first longitudinal symmetric mode $(2,0)$, using a hydrofluoroether fluid (3M\textsuperscript{\texttrademark} Novec\textsuperscript{\texttrademark} HFE-7000). To quantify heat and mass transfer, a lumped thermodynamic model is combined with an Augmented-state Extended Kalman Filter (AEKF), enabling real-time, time-resolved inference of Nusselt numbers. A critical forcing threshold is identified: below it, the fluid remains quiescent and thermally stratified; above it, parametric resonance drives strong sloshing, complete thermal destratification, and a rapid pressure drop. At $50\%$ fill, the dominant $(2,0)$ response intermittently alternates with a planar $(1,0)$ mode, indicating subharmonic mode interaction. The inferred Nusselt numbers increase by several orders of magnitude after destratification, and pressure-rate analysis confirms that condensation governs the pressure evolution.
		
		\vspace{7mm}
		\noindent\textbf{Keywords:} Cryogenic, sloshing, modeling, experiments, scaling, Kalman filter
	\end{abstract}
	
	\section{Introduction}
	\label{Sec:Intro}
	Cryogenic fuels such as liquid hydrogen (LH$_2$), liquid methane (LCH$_4$), and liquefied natural gas (LNG) are increasingly considered for next-generation aviation, maritime, trucking, and space applications \citep{brewer1991hydrogen, saurav, AHLUWALIA202313308, simonini2024cryogenic}. In these contexts, partially filled propellant tanks are subjected to vehicle motions and external accelerations that trigger sloshing. Besides potentially compromising vehicle stability \citep{bauer, KOLAEI2014263, TOUMI20091026}, sloshing also has important thermodynamic consequences: it disrupts the natural thermal stratification, thus intensifying heat and mass exchange between liquid, vapor, and walls, and leading to rapid pressure fluctuations \citep{LUDWIG2013, Das2009-vertical, moran1994}.
	
	Most sloshing studies have focused on upright cylindrical and spherical tanks under lateral excitation, primarily motivated by space-propulsion challenges during the early space race. In these geometries, isothermal sloshing has been extensively investigated, resulting in detailed regime maps that relate the free-surface response to geometry and excitation parameters \citep{miles1984internally, miles1984resonantly}. 
	Beyond hydrodynamic regime mapping, interest in the thermodynamic consequences of sloshing was sparked by the pioneering work of \cite{moran1994}, who showed that resonant wave motion in a large spherical LH$_2$ tank can trigger a rapid collapse of ullage pressure. A similar behavior was later documented by \cite{lacapere2009experimental} in tanks partially filled with LN$_2$ and LOx, where sloshing destroyed the thermal stratification and enhanced interfacial heat and mass transfer. The link between de-stratification and pressure drop was further analyzed by  \cite{arndt} for upright cylindrical tanks with LN$_2$, who identified enhanced vapor condensation-triggered by the destruction of thermal stratification-as the driving mechanism of the pressure decay, and showed that the effect is mitigated by the presence of an inert gas in the ullage volume. A critical wave-amplitude threshold for the de-stratification (and hence the sloshing induced pressure drop) was identified by \cite{LUDWIG2013}, while more recent works have provided detailed flow visualizations in both harmonic \citep{himeno2011} and impulsive excitations \citep{himeno2018}, extensive experimental campaigns and simplified thermodynamic models \citep{vanForeest2014}, high fidelity numerical simulations \citep{konopka2016analysis} and inverse methods to identify the heat and mass transfer coefficients from experimental data \citep{MARQUES2023} to name but a few.
	
	By comparison, the literature on horizontal cylindrical tanks is strikingly sparse, especially in the case of vertical acceleration, despite this configuration is particularly relevant for aeronautical \citep{catherines_vibrations_1975} and ground transportation applications \citep{rissi}. Horizontally oriented cylindrical vessels are particularly attractive for fuselage-integrated LH$_2$ storage in future aircraft \citep{Colozza}, offering structural integration benefits and efficient usage of internal volume while minimizing aerodynamic penalties. Vertical sloshing dynamics in these geometries presents a much stronger dependence of the natural frequency to the fill ratio and aspect ratio. Although a few works investigating isothermal sloshing response in such a geometry under longitudinal \citep{Luo_2023} and angular \citep{GROTLE2018512} acceleration, on under realistic forcing during road transportation \citep{takeda2012synchronous}, the case of vertical acceleration has so far only focused on 2D tank \citep{Saltari2024,colville_faraday_2025}.
	
	In vertical sloshing, also known as parametric sloshing, the excitation works indirectly by altering the effective gravity, which normally acts as the restoring force in other sloshing conditions. Instead of pushing the liquid directly, vertical vibrations periodically weaken and strengthen this restoring mechanism, so the system’s ability to resist motion itself becomes time-dependent. This makes vertical sloshing qualitatively different from lateral sloshing: resonance does not occur when the forcing frequency simply matches a natural mode, but rather when the modulation of gravity interacts with the free-surface dynamics in a way that destabilizes them. This type of forcing is closely linked to the classical Faraday instability \citep{faraday1831forms, Douady_1990}, where a vertically oscillated liquid develops standing surface waves at half the excitation frequency. In confined tanks, exponential growth of sloshing modes occurs when the forcing amplitude exceeds a critical threshold, and the excitation frequency is close to twice the natural frequency of the leading mode \citep{benjamin, FRANDSEN200453}. This parametric resonance can trigger complex behaviors such as mode competition if multiple eigen-frequencies are excited simultaneously \citep{Ciliberto, Ibrahim_2005} or period-tripling states arising from non-linear interactions \citep{JIANG_PERLIN_SCHULTZ_1998}. These phenomena have only recently been been observed in a two-dimensional horizontal geometry by \cite{colville_faraday_2025}.
	
	To the author’s knowledge, the thermodynamic impact of vertical excitation-and in particular the relationship between vertical excitation and the associated pressure drop in cryogenic tanks-remains largely unexplored, having been first investigated by \cite{Das2009-vertical} for upright cylindrical tanks and more recently studied by \cite{Saltari2024} for a two-dimensional slice mimicking a horizontal cylinder. 
	
	No literature could be found on the thermodynamic consequences of vertical sloshing in full three-dimensional horizontal tanks. The present work addresses this gap by providing an experimental analysis of the thermodynamic response of a horizontal cylinder under parametric sloshing. In addition, we infer the effective Nusselt number associated with the sloshing-induced pressure drop by applying an Extended Kalman Filter (EKF, see \cite{Pei2017} for a gentle introduction) to assimilate experimental data into a reduced thermodynamic model. 
	
	More specifically, differently from the approach in \cite{Saltari2024}, the Nusselt number inferred with the EKF method is not algebraically and instantaneously linked to pressure and temperature data, which means that deficiencies in the thermodynamic model are not directly and rigidly imprinted onto the coefficient. Instead, the EKF treats the Nusselt number as a hidden, time-varying state that can adjust dynamically as measurements are assimilated. On the other hand, differently from the approach in \cite{MARQUES2023}, the proposed method does not require a full-horizon optimization over an entire sloshing episode, but rather provides online, adaptive estimates with quantified uncertainty, making it more flexible for transient regimes and real-time applications. While the EKF has been used in inverse heat transfer problems to estimate time-varying parameters \citep{emery_2004, wen_2020,bonini_2024}, to the author's knowledge this has never been used for heat and mass transfer analysis in non-isothermal sloshing.
	
	The experiments were carried out using HFE-7000, a liquid often employed as a surrogate for cryogenic fluids due to its low surface tension and contact angle \citep{Fiorini2022,Fiorini2023}, as well as its strong sensitivity of saturation pressure to temperature, which enhances thermo–fluid coupling within a laboratory-accessible temperature range \citep{Monteiro2025,Peveroni2020}.
	
	The experimental campaign employed two identical test sections. The first, built in transparent material, enabled direct visualization of the free-surface motion and characterization of the sloshing regime across different fill ratios, forcing amplitudes, and dimensionless forcing frequencies. The second, constructed in steel, was used for the non-isothermal measurements under identical excitation conditions, enabling precise thermal control and pressure monitoring.
	Splitting the analysis into two tanks implicitly assumes a one-way coupling: the sloshing dynamics influence the thermodynamics through vapor generation and pressure fluctuations, while non-isothermal effects are not considered to significantly alter the sloshing flow field. In other words, thermal and phase-change processes are treated as passive responses to the hydrodynamics rather than active drivers of the liquid motion. This assumption has been adopted by \cite{GROTLE2018512} and shown to remain valid even under severe sloshing conditions, such as those encountered in microgravity, by \cite{Monteiro2025}.
	
	The rest of the article is organized as follows. Section \ref{sec2} states the problem. Section \ref{sec3} presents the lumped-order thermodynamic model used for the inference. Section \ref{sec4} places the operating conditions on parametric-stability maps and identifies the key dimensionless groups that allow the present results to be related to a broader range of conditions.  Section \ref{sec5} details the methodology, including the experimental setup and operating procedure while Section \ref{sec6} covers data processing and heat transfer coefficient inference. Section \ref{sec7} reports the results while  Section \ref{sec.conclusions} closes with conclusions and avenues for future research.
	
	\section{Problem statement and definitions}\label{sec2}
	
\begin{figure}[ht]
		\centering
		\includegraphics[width=0.55\linewidth]{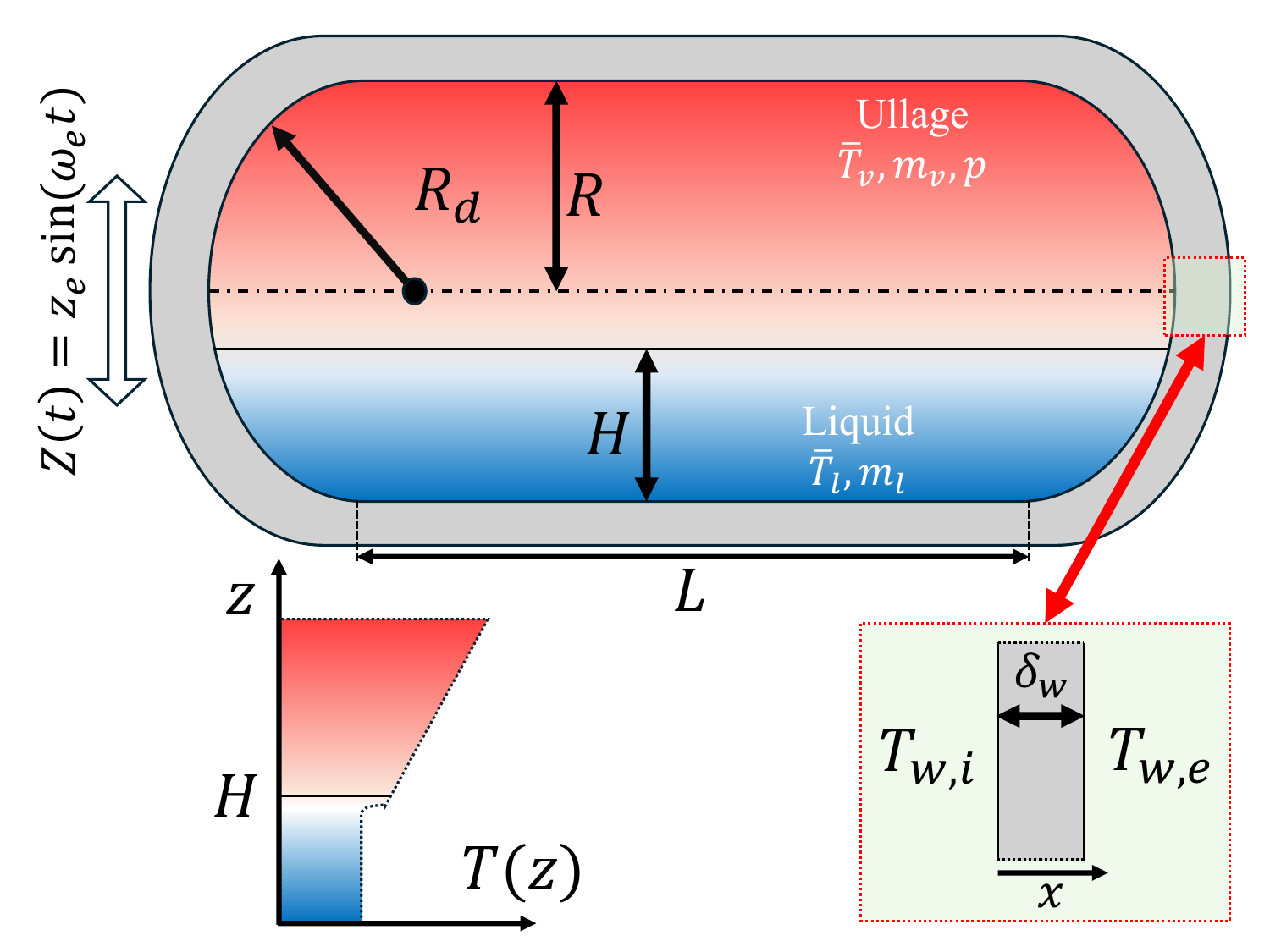}
		\caption{Problem schematic: cylindrical tank of radius \(R\), length \(L\), dome radius \(R_d\), wall thickness $\delta_w$ filled to height \(H\), subjected to vertical harmonic excitation $Z(t)=z_e \cos(\omega_et)$ with qualitative representation of the initial fluid vertical thermal stratification. The tank wall has a thickness $\delta_w$ with inner and external surface averaged temperatures $T_{w,i}$ and $T_{w,e}$, respectively}
		\label{fig:tank_schematic}
	\end{figure}

	\autoref{fig:tank_schematic} sketches the configuration under study: a horizontal cylindrical tank of radius \(R\) and length \(L\), closed at each end by domes of radius \(R_d\).   Throughout this paper, the subscripts $l$, $v$, and
	$w$ refer to the liquid, vapor, and tank wall, respectively, while the subscript $i$ is used for quantities at the gas-vapor interface.
	
	The vessel contains a liquid of density \(\rho_l\) and dynamic viscosity \(\mu_l\), filled to a height \(H\) corresponding to a volume $V_l$. Above the liquid, the ullage volume $V_v$ is occupied by its single-species vapor, with pressure \(p\), density \(\rho_v\), and viscosity \(\mu_v\). 
	At equilibrium, both the liquid and the vapor are at the saturation temperature $T_{\text{sat}}(p)$. Usually, however, the saturation temperature is preserved only at the interface, that is $T_i=T_\mathrm{sat}(p)$. The bulk liquid is typically sub-cooled ($T_l<T_i$), while the vapor is superheated ($T_v>T_i$), and both phases are thermally stratified. Such a configuration naturally develops in long-term storage due to self-pressurization or external pressurization. A qualitative temperature profile is shown in \autoref{fig:tank_schematic}. In the following, we focus on the mass-averaged temperatures on the vapor $\overline{T}_v$ and the liquid $\overline{T}_l$.
	
	At time $t=0$, an external excitation triggers sloshing. We here focus on a vertical harmonic displacement $Z(t)=z_e\cos (\omega_et)$. Sloshing enhances the heat transfer both from the vapor to the interface and from the interface to the liquid bulk. As the latter contribution is stronger than the former, a net condensation is produced. This, together with the cooling of the vapor, results in a pressure drop in the tank. Literature shows (see \cite{MARQUES2023} and \cite{ECOS_ahizi} and references cited therein) that this drop is mostly driven by condensation and develops on time scales much shorter than those of conduction through the tank wall thickness. One can thus expect that the wall retains much of the imprint of the pre-sloshing stratification during the sloshing and thus acts as a thermal buffer: as the liquid intermittently wets wall regions formerly in contact with vapor, and the vapor sweeps regions formerly wetted by liquid, the solid warms the liquid and cools the vapor. After sloshing, mixing reduces the temperature differences \(\Delta T_v = \overline{T}_v - T_i\) and \(\Delta T_l = T_i - \overline{T}_l\); in the limit of perfect mixing, both vanish as the phases homogenize around the saturated interface temperature.
	
	On the fluid side, other parameters of interest are thus the specific heat at constant volume for the liquid and the vapor $c_{v,l}$ and $c_{v,v}$ and the latent heat of vaporization $\mathcal{L}_v$. On the solid side, the relevant properties are the tank's mass \(m_w\) and thickness $\delta_w$, as well as the solid's conductivity $\kappa_w$, specific heat \(c_{w}\), and thermal diffusivity \(\alpha_w\). We assume that the tank is subject to an external power $Q_{ext}$ throughout the whole experiment.
	
	The goal of this work is to analyze how different sloshing conditions can lead to different degrees of mixing and, consequently, to different levels of heat-transfer enhancement, and to understand how these translate into pressure drops, with particular attention to the relative contributions of condensation and wall cooling. Furthermore, we seek to quantify these enhancements in terms of relevant Nusselt numbers, inferred from limited and noisy sensor data. This inference partially relies on a simplified thermodynamic model, which is briefly introduced in the following section.

	\section{Lumped thermodynamic modeling}\label{sec3}
	A lumped-parameter thermodynamic model is adopted to describe the system under non-equilibrium conditions. The domain is partitioned into three control volumes-vapor, liquid, and wall-while the liquid–vapor interface is idealized as an infinitesimally thin, massless layer maintained at saturation. Each control volume is characterized by a mass-averaged temperature \(\overline{T}_k(t)\), with \(k \in \{w,l,v\}\) denoting wall, liquid, and vapor, respectively. Interfacial heat and mass transfer are represented through fluxes, and the temporal evolution of \(\overline{T}_k\) is obtained from energy balances. These balances are written explicitly for the vapor and liquid phases, whereas at the interface the temperature is constrained by the saturation relation, \(T_i = T_{\mathrm{sat}}(p)\). For the wall, the temperature field is computed from a one-dimensional conduction model, as detailed in the following section.
	
	Since the tank is rigid, its internal volume remains constant, hence one has the sum of volume variations in the liquid and the vapor give $\mathrm d V_l/\mathrm dt= - \mathrm d V_v/\mathrm dt$. However, considering that the volume of condensed mass during the sloshing event is negligible compared to the volumes of liquid and vapor, we here neglect volume variations.  
	Concerning the masses in the two volumes, because the tank is closed during the sloshing event, any change in fluid mass arises solely from phase change and mass conservation yields
	\begin{equation}\label{eq:massVariation}
		\frac{\mathrm d m_l}{\mathrm dt} = -\,\frac{\mathrm d m_v}{\mathrm dt} = \dot m_{\mathrm c} \, ,
	\end{equation}
	where \(\dot m_{\mathrm c}\) denotes the net mass exchange rate across the interface, taken as positive for condensation (\(\dot m_{\mathrm c} > 0\)) and negative for evaporation (\(\dot m_{\mathrm c} < 0\)).
	
	The conservation of internal energy for the liquid and vapor control volumes, neglecting volume variations, reads  
	\begin{subequations}\label{eq:energy_balances} 
		\begin{align} 
			\frac{\mathrm d}{\mathrm dt}\!\left(m_l c_{v,l}\,\overline T_l\right) 
			&= Q_{w,l} - Q_{l,i} +\mathcal{h}^{\mathrm{sat}}_l \dot m_{\mathrm c} \;, 
			\label{eq:liquid_balance}\\[6pt] 
			\frac{\mathrm d}{\mathrm dt}\!\left(m_v c_{v,v}\,\overline T_v\right) 
			&= Q_{w,v} - Q_{v,i} \; -\mathcal{h}^{\mathrm{sat}}_v \dot m_{\mathrm c}, 
			\label{eq:gas_balance}
		\end{align} 
	\end{subequations} where $\mathcal{h}^{\mathrm{sat}}_v$ and $\mathcal{h}^{\mathrm{sat}}_l$ are the specific enthalpies at the saturation conditions for the vapor and liquid respectively.
	The convective heat fluxes $Q_{w,l}, Q_{l,i},Q_{w,v},Q_{v,i}$ are computed using Newton's law. Specifically, the ones between vapor and interface ($Q_{v,i}$) and between liquid and interface 
	($Q_{l,i}$) are
	
	\begin{equation}
		\label{heat_volumes}
		Q_{l,i}= h_{l,i} A_i \bigl( \overline{T}_l-T_i \bigr)  \,\, \mbox{and} \,\, Q_{v,i}= h_{v,i} A_i \bigl(\overline{T}_{v}- T_i \bigr)\,,
	\end{equation} where $A_i$ is the interface area, measured in quiescent conditions and $ h_{l,i}$ and $ h_{v,i}$ are the unknown heat transfer coefficients. By convention, these fluxes are taken as positive when directed toward the interface. Therefore, the net heat exchange with the interface can be related to the mass exchange rate by 
	
	\begin{equation}
		\dot m_{\mathrm{c}} = -\frac{Q_{v,i}+Q_{l,i}}{\mathcal{L}_{v}} \, ,
		\label{eq:mcond_def}
	\end{equation}
	with $\mathcal{L}_v=\mathcal{h}^{\mathrm{sat}}_v-\mathcal{h}^{\mathrm{sat}}_l$ the latent heat of vaporization. On the other hand, the heat exchanges with the wall require the definition of a wall temperature. Although this can vary significantly between the portion in contact with the liquid and the one in contact with the vapor, we here interpret it as a inner surface averaged temperature $T_{w,i}$, so that 
	
	\begin{equation}
		\label{heat_WALL}
		Q_{w,v}= h_{w,v} A_{w,v} \bigl( T_{w,i}-\overline{T}_v \bigr)  \, \mbox{and} \, Q_{w,l}= h_{w,l} A_{w,l} \bigl(T_{w,i}- \overline{T}_l \bigr)\,.
	\end{equation} where the unknown heat transfer coefficients and corresponding areas are defined as before. The evolution of the wall temperature is described using a one-dimensional conduction model across the wall thickness, with the main goal of allowing the model to reproduce the filtering effects of the wall between the internal heat exchange and the temperature on the outer surface of the wall. While not essential for the thermodynamic modeling point of view, this formulation was important for the inference based on measurements sensors on the external walls. 
	
	Defining as $T_w(x,t)$ the surface average wall temperature averaged across the thickness axis $x$, with $x=0$ the inner surface of the tank such that $T_{w,i}(t)=T_{w}(0,t)$ in \eqref{heat_WALL} and $T_{w,e}(t)=T_{w}(\delta_w,t)$ the external one (see Figure \ref{fig:tank_schematic}), the heat equation reads  
	
	\begin{subequations}
		\begin{align}
			\partial_t T_w(x,t) &= \alpha_w\partial_{xx} T_w(x,t),\\[6pt] 
			\partial_xT_w(0,t) &= -\frac{h_{wl}A_{wl}(T_{w,i}-\overline T_l) + h_{wv}A_{wv}(T_{w,i}-\overline T_v)}{\kappa_wA_w} ,\\[6pt] 
			\partial_xT_w(\delta_w,t) &=  \frac{Q_{ext}}{\kappa_wA_w}\,.
		\end{align}\label{eq:PDE}
	\end{subequations}

	Finally, the tank pressure is related to the vapor state through the equation of state
	\begin{equation}
		\label{eq:p_state}
		p = f\biggl(\rho_v = \frac{m_v}{V_v},\overline T_v\biggr).
	\end{equation}
	
	The combination of eqs \eqref{eq:massVariation}, \eqref{eq:energy_balances} with the heat closures \eqref{heat_volumes} and \eqref{heat_WALL}, and the energy balance closure at the interface 
	\eqref{eq:mcond_def}, the 1D problem \eqref{eq:PDE} for the wall and the equation of state \eqref{eq:p_state} results in a closed system of equations for the variables $m_l,m_v,p,\overline{T}_l,\overline{T}_v, T_w$. 
	
	\section{Scaling considerations}\label{sec4}
	
	A scaling analysis is performed to cast the present results in dimensionless form, enabling comparison with existing sloshing studies and providing a framework for future work, as no comparable data exist for vertically forced sloshing in horizontal cylindrical tanks. We first address the scaling of the sloshing dynamics (Section~\ref{sec4p1}), then the non-isothermal aspects (Section~\ref{sec4p2}), and finally summarize the parameters in Section~\ref{sec4p3}. We first focus on the scaling of the sloshing dynamics Section~\ref{sec4p1} then move to the non-isothermal aspects in Section~\ref{sec4p2}. A summary of the scaling parameters is provided in Section~\ref{sec4p3}.
	
	\subsection{Sloshing dynamics}\label{sec4p1}
	
	Disregarding capillary and density difference effects, the scaling of the interface dynamics $\eta(\bm{r},t)$, with $\bm{r}$ denoting the in-plane (horizontal) coordinate vector, depends on three geometrical parameters, one kinematic parameter and two dimensionless numbers (see also \cite{leo,ahizi}):  
	
	\begin{equation}
		\label{scaling_eta}
		\frac{\eta(\bm{r}, t)}{R} = f\left(\frac{L}{R},\frac{H}{2R}, \frac{z_e}{R}, \frac{\omega_e}{\omega_{1,0}},\mathrm{Re}, \mathrm{Fr}\right)
	\end{equation} where \(L/R\) is the cylinder aspect ratio, \(H/R\) the fill level, \(z_e/R\) the dimensionless forcing amplitude, 
	and \(\omega_e/\omega_{1,0}\) the ratio between the forcing frequency and the frequency of the lowest sloshing mode, with the first index referring to the longitudinal direction and the second to the transverse direction in the cylindrical geometry. The remaining dimensionless numbers are the Reynolds number $\mathrm{Re}=\rho_lz_e \omega_e R/\mu_l$ and the Froude number $\mathrm{Fr}=z_e \omega_e /\sqrt{g R}$, computed by taking the tank radius as the reference length and $U=z_e \omega_e$ as the reference velocity.
	
	Particularly delicate is the dimensionless pair ($\omega_e/\omega_{1,0}, \mathrm{Fr}$), which jointly characterizes the type of response: the frequency ratio controls the proximity to parametric resonance, while the Froude number sets the effective forcing strength. Contrary to laterally forced sloshing, where resonance occurs when the forcing frequency matches one of the natural modes, vertically forced sloshing exhibits a parametric character. In vertical sloshing, instability is not restricted to isolated resonance points but arises over continuous ranges of forcing frequencies. This gives rise to bands in the parameter space ($\omega_e/\omega_{1,0},\mathrm{Fr}$) where the interface response grows exponentially. Viscous dissipation sets a threshold in forcing amplitude for the instability to develop, shifting the boundaries of the instability regions.
	
	In a linearized, inviscid context, expanding the interface dynamics in terms of eigenfunction $\xi_{n,m}(\bm{r})$
	and their temporal evolution $b_{n,m}(t)$:
	
	\begin{equation}
		\label{expansion}
		\eta(\bm{r},t)=\sum^{\infty}_{n,m} \xi_{n,m}(\bm{r})b_{n,m}(t)\,,
	\end{equation} it follows from the orthogonality of the modes that each amplitude evolves independently.

	For mode \((n,m)\) one obtains the second-order ODE (see Appendix \ref{appA} for the full derivation)
	\begin{equation}
		\ddot b_{n,m}(t) + \omega_{n,m}^2 \,\frac{g_{\mathrm{eff}}(t)}{g}\, b_{n,m}(t) = 0,
		\label{eq:modal_ODE}
	\end{equation}
	where \(\omega_{n,m}\) is the natural frequency of the mode in quiescent conditions,  and \(g_{\mathrm{eff}}(t) = g - z_e\omega^2_e\cos(\omega_e t)\) is the effective gravity induced by the vertical excitation. Rescaling the time such that $\tau=\omega_e t/2$, equation \eqref{eq:modal_ODE} can be turned into the canonical Mathieu equation \citep{benjamin}:
	\begin{equation}
		\label{mathieu_2}
		\frac{d^2b_{n,m}}{d\tau^2}  + (a-2 q \cos(2\tau))b_{n,m}=0\,,
	\end{equation}with $a= (2\omega_{n,m}/\omega_e)^2$ and $q=2\omega^2_{n,m} z_e/g$. Floquet analysis provides the boundaries between the stable and unstable solutions of the Mathieu equation in the plane ($a$, $q$). The instability regions, also called Mathieu tongues, emanate from the points $a\simeq n^2$, $n=1,2,\dots$. Each tongue corresponds to a parametric resonance, with the first tongue ($n=1$) being the widest and most relevant for sloshing. In parametric sloshing, the stability diagram is often expressed in terms of the frequency ratio $\omega_e/\omega_{n,m}$ and the dimensionless vertical acceleration $\hat{a}_v = z_e \omega_e^2/g = \mathrm{Fr}^2 R/z_e$. 
	
	In this representation, the instability regions are centered around $\omega_e \simeq 2\omega_{n,m}/n$, and their extent in $\hat{a}_v$ determines the threshold amplitude required for modal growth.
	
\begin{figure}[htbp]
		\centering
		\includegraphics[width=0.6\linewidth]{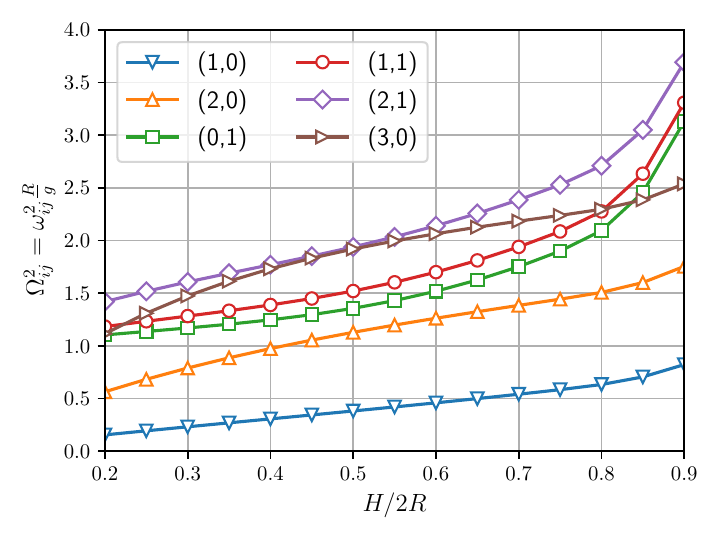}
		\caption{Dimensionless natural frequencies \(\Omega_{ij}^2\) of a flat-ended horizontal cylinder of aspect ratio $L/R=5$ against fill height. }
		\label{fig:tank_frequencies}
	\end{figure}

	The challenge in computing these maps and thus identifying the operating conditions lies in the difficulties in computing the natural frequencies $\omega_{n,m}$. For the case of horizontal cylinders with flat ends, the complete eigenfunction derivation was provided in \cite{HASHEMINEJAD2017338,han_semi-analytical_2021} and the natural frequencies can be computed numerically. This was done for a tank with (\(L/R=5\)) as in the present investigation, and the results are reported in \autoref{fig:tank_frequencies} in terms of dimensionless frequencies $\Omega^2_{n,m}={\omega^2_{n,m}R/g}$.

	For the curved-end horizontal cylinder considered here, however, analytical eigenfunctions are not available and the natural frequencies were therefore determined experimentally, as described in Section~\ref{sec6p1}. The retrieved values were then used to construct the stability maps.

\begin{figure*}[t]                
		\centering
		\begin{subfigure}[b]{0.49\textwidth} 
			\includegraphics[width=\textwidth]
			{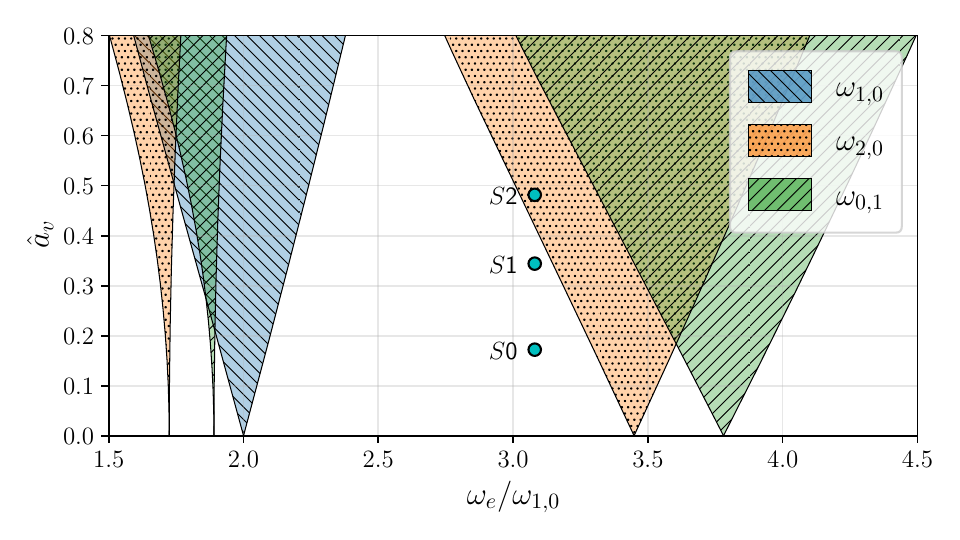}
			\caption[H/(2R)=0.5]{\(\displaystyle \frac{H}{2R}=0.5\)}
			\label{fig:stab_bound_theo_50_percent}
		\end{subfigure}  
		\begin{subfigure}[b]{0.49\textwidth} 
			\includegraphics[width=\textwidth]
			{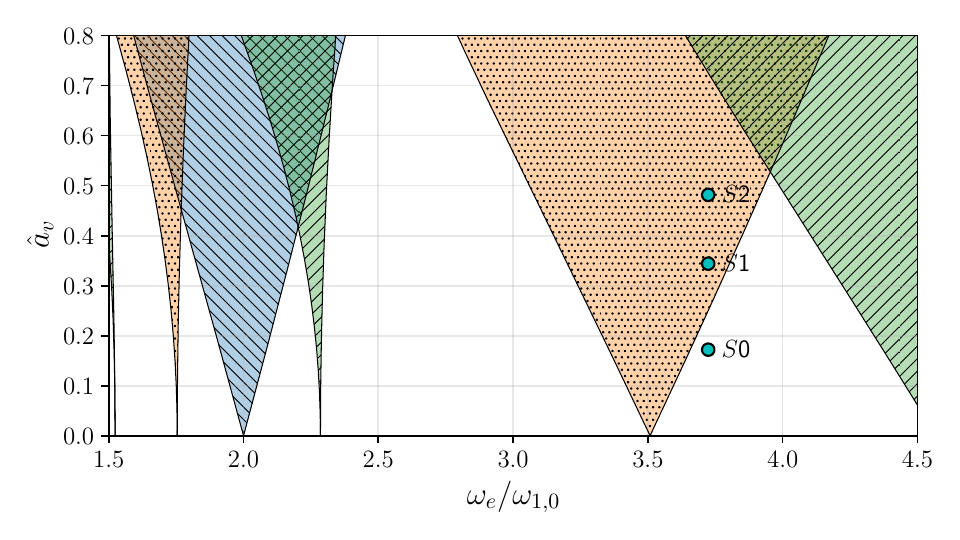}
			\caption[H/(2R)=0.5]{\(\displaystyle \frac{H}{2R}=0.5\)}
			\label{fig:stab_bound_exp_50_percent}
		\end{subfigure}
		\begin{subfigure}[b]{0.49\textwidth}
			\includegraphics[width=\textwidth]
			{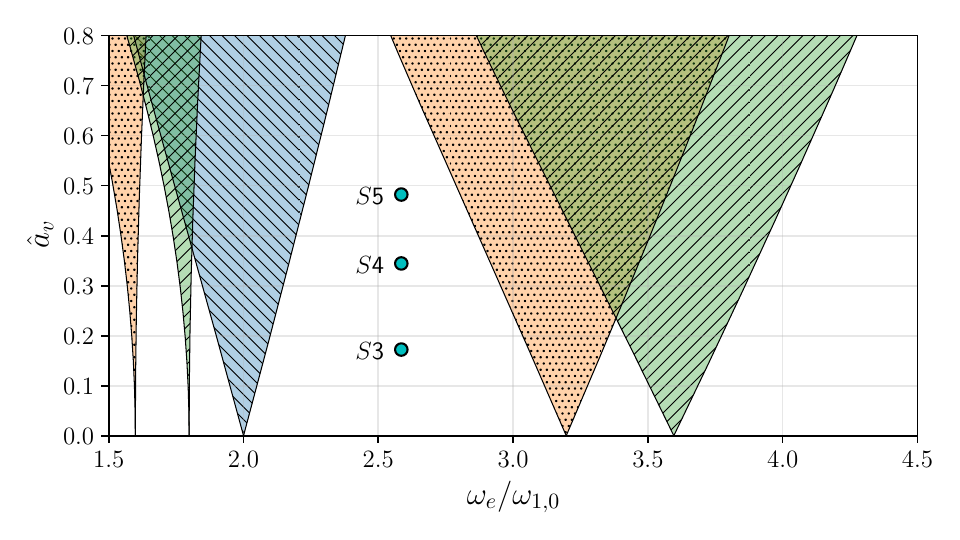}
			\caption[H/(2R)=0.7]{\(\displaystyle \frac{H}{2R}=0.7\)}
			\label{fig:stab_bound_theo_70_percent}
		\end{subfigure}
		\begin{subfigure}[b]{0.49\textwidth}
			\includegraphics[width=\textwidth]
			{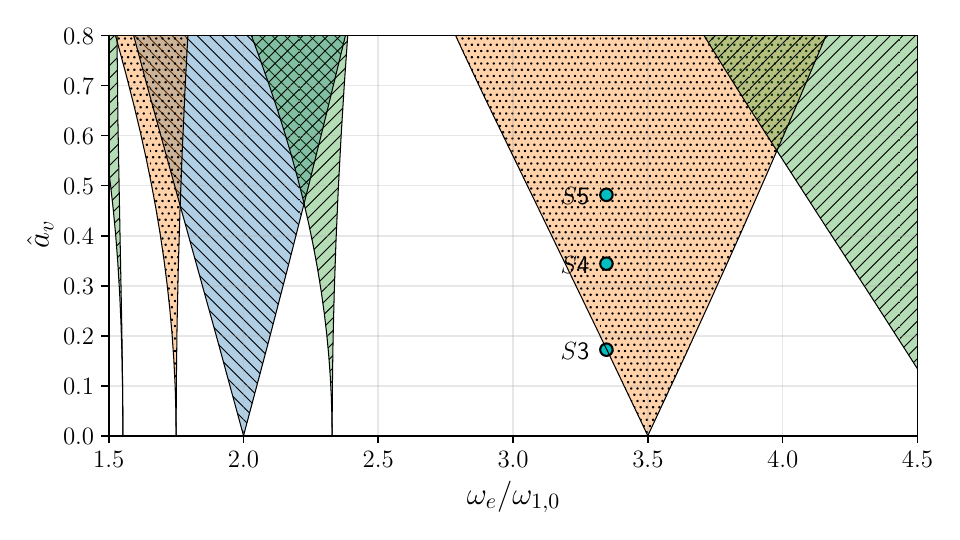}
			\caption[H/(2R)=0.7]{\(\displaystyle \frac{H}{2R}=0.7\)}
			\label{fig:stab_bound_exp_70_percent}
		\end{subfigure}
		\caption{Parametric‐instability regions for a harmonically excited horizontal cylindrical tank, shown in the non‐dimensional plane of acceleration $\hat{a}_v=\textrm{Fr}^2\frac{R}{z_e}=\frac{z_e\omega^2_e}{g}$ versus excitation frequency ratio $\omega_e/\omega_{1,0}$. Filled areas mark unstable solutions of Mathieu’s equation for three sloshing modes obtained from POD decomposition:  $\omega_{(1,0)}$ (blue, \textbackslash\textbackslash\textbackslash  hatching), $\omega_{(2,0)}$ (orange, dot hatching), and $\omega_{(0,1)}$ (green, /// hatching).  Experimental testing conditions are identified by blue dots and further detailed in Table~\ref{tab:ini_cond}. Panels:
			\subref{fig:stab_bound_theo_50_percent} and \subref{fig:stab_bound_theo_70_percent} theoretical frequencies based on geometry with flat ends;
			\subref{fig:stab_bound_exp_50_percent} and \subref{fig:stab_bound_exp_70_percent} experimentally measured frequencies.}
		\label{fig:stab_bound}
	\end{figure*}

	Figure~\ref{fig:stab_bound} compares the stability maps for a cylinder with flat ends (panels a,c), obtained from the eigenfunction expansion of \cite{HASHEMINEJAD2017338}, with those for the present curved-end geometry (panels b,d), built using the natural frequencies experimentally retrieved. Two fill levels are shown, \(H/2R=0.5\) and \(H/2R=0.7\) and markers indicate the investigated points. The maps highlight the sensitivity of the relative positions of the instability regions to the fill level. Across all scenarios considered, the primary instability tongues of the modes \((2,0)\) and \((0,1)\) overlap at relatively low forcing acceleration ($\hat a_v<0.6$). This overlap implies that both modes are susceptible to arise in parametric resonance. This phenomenon, referred to as mode competition, is known to promote complex, mode-coupling dynamics \citep{Ciliberto}. While this typically occurs for higher-order modes in upright geometries, horizontal cylinders are particularly prone to this phenomenon at low order, as their natural frequencies differ by only a few percent (see \autoref{fig:tank_frequencies}). For example, for straight ends at \(H/2R=0.5\), the \((2,0)\) and \((0,1)\) boundaries already overlap for \(\hat{a}_v \gtrsim 0.2\) in the undamped linear prediction (see \autoref{fig:stab_bound_theo_50_percent}).

	\subsection{Thermodynamic aspects}\label{sec4p2}
	
	The scaling of the thermodynamic model in Section \ref{sec3} allows to identify important dimensionless quantities governing the thermodynamic impact of sloshing. To quantify the maximum possible pressure drop, it is worth to first identify the idealized equilibrium conditions, that is the thermodynamic state one would obtain at the end of a complete thermal destratification. 
	
	Combining the internal energy balance for both liquid and vapor volumes \autoref{eq:energy_balances} with a lumped description of the wall one gets 
	
	\begin{equation}
		\frac{\mathrm dU_t}{\mathrm dt}=\frac{\mathrm d(m_lc_{v,l}\overline T_l+m_vc_{v,v}\overline T_v+m_wc_w\overline{T}_{w}])}{\mathrm dt}   =Q_{ext}\approx 0\,,
	\end{equation} under the assumption that the external heat ingress is negligible compared to the energy exchanged between the control volumes during the sloshing event or that this exchange impacts the volumes at time scales that are much faster than the impact of the heat ingress. Hence the equilibrium internal energy equals the initial one, and the energy balance sets
	\begin{equation}
		\label{U_final_ideal}
		\begin{split}
			U_t(0)&= m_l(0)c_{v,l}\overline T_l(0)+m_v(0)c_{v,v}\overline T_v(0)+m_wc_w\overline T_w(0)\\
			&=U_t^{eq} = \left[m_l(T_{eq})c_{v,l}+ m_v(T_{eq})c_{v,v} + m_w(T_{eq})c_w\right]T_{eq}\,. \\
		\end{split}
	\end{equation}
	
	Eq \eqref{U_final_ideal} can be solved for the equilibrium temperature $T_{eq}$ such that $\overline T_l(0)\leq T_{eq} \leq \overline T_v(0)$. This equilibrium point can be used to set a reference temperature for the scaling together with the associated saturation pressure $p_{eq}=p_\mathrm{sat}(T_{eq})$. The temperature variations $\Delta T= \overline{T}_v(0)-\overline{T}_l(0)$ and pressure variations $\Delta p=p(0)-p_{eq}$ are thus the largest possible. 
	
	Accordingly, appropriate dimensionless thermodynamic state variables are:
	
	\begin{equation}
		\hat{p} \;=\; \frac{p-p_{eq}}{\Delta p}, \quad
		\Theta   \;=\; \frac{T-T_{eq}}{\Delta T}\,.
	\end{equation}
	
	As for the time scale, it appears natural to use the period of the forcing frequency, hence the dimensionless time reads $ \hat{t}={\omega_e} t/({2\pi})$. Introducing these quantities in \eqref{eq:energy_balances}, scaling the masses with respect to their initial values ($\hat m_k(0)={m_k}/{m_k(0)}$) and the condensation rate by the initial vapor mass ($\hat{\dot m}_c={2\pi\dot m_c}/({\omega_em_v(0)})$), and treating the specific heats as constants for simplicity, gives

	\begin{subequations}\label{eq:energy_balances_dimensionless} 
		\begin{align} 
			\frac{\mathrm d}{\mathrm d \hat{t}}\!\left(\hat{m}_l \,\overline{\Theta}_l\right) 
			&= \Pi^h_{w,l} (\Theta_{w,i}-\overline{\Theta}_{l}) - \Pi^h_{l,i} (\overline{\Theta}_{l}-\Theta_{i}) - \Pi^\mathrm{sat}_l\,\hat{\dot m}_{\mathrm c}, 
			\label{eq:liquid_balance_dimensionless}\\[6pt] 
			\frac{\mathrm d}{\mathrm d \hat{t}}\!\left(\hat{m}_v \,\overline{\Theta}_v\right) 
			&= \Pi^h_{w,v} (\Theta_{w,i}-\overline{\Theta}_{v}) - \Pi^h_{v,i} (\overline{\Theta}_{v}-\Theta_{i}) + \Pi^\mathrm{sat}_v\,\hat{\dot m}_{\mathrm c},
			\label{eq:gas_balance_dimensionless}
		\end{align} 
	\end{subequations} where 
	
	\begin{align}
		\Pi^h_{w,l} &= \frac{2\pi h_{w,l} A_{w,l} }{\omega_e \,c_{v,l} \,m_l(0)} 
		=\text{Nu}_{w,l} \biggl( \frac{2\pi \kappa_l A_{w,l} }{\omega_e \,c_{v,l} \,m_l(0) R}\biggr)
		=\text{Nu}_{w,l} \Pi^{Nu}_{w,l}, \nonumber \\ 
		\Pi^h_{l,i} &= \frac{2\pi h_{l,i} A_{i} }{\omega_e \,c_{v,l} \,m_l(0)}
		=\text{Nu}_{l,i} \biggl( \frac{2\pi \kappa_l A_{i} }{\omega_e \,c_{v,l} \,m_l(0) R}\biggr)
		=\text{Nu}_{l,i} \Pi^{Nu}_{l,i}, \nonumber \\
		\Pi^h_{w,v} &= \frac{2\pi h_{w,v} A_{w,v} }{\omega_e \,c_{v,v}\, m_v(0)}
		=\text{Nu}_{w,v} \biggl( \frac{2\pi \kappa_v A_{w,v} }{\omega_e \,c_{v,v} \,m_v(0) R}\biggr)
		=\text{Nu}_{w,v} \Pi^{Nu}_{w,v}, \nonumber \\
		\Pi^h_{v,i} &= \frac{2\pi h_{v,i} A_{i} }{\omega_e \,c_{v,v}\, m_v(0)}
		=\text{Nu}_{v,i} \biggl( \frac{2\pi \kappa_v A_{i} }{\omega_e \,c_{v,v} \,m_v(0) R}\biggr)
		=\text{Nu}_{v,i} \Pi^{Nu}_{v,i}
	\end{align}
	are dimensionless representations of the heat transfer coefficients, here formulated in a way that introduces the associated Nusselt numbers, and 
	
	\begin{equation}
		\begin{aligned}
			\Pi^\mathrm{sat}_{l} &= \frac{ \mathcal{h}^{\mathrm{sat}}_l  \,m_v(0)}{ \,c_{v,l} \,\Delta T\,m_l(0)}, \\ 
			\Pi^\mathrm{sat}_{v} &= \frac{\mathcal{h}^{\mathrm{sat}}_v }{ \,c_{v,v}\, \Delta T }
		\end{aligned}
	\end{equation} are the dimensionless groups controlling the impact of the latent heat exchange on the temperature variation of the two volumes. Note that the term $m_v(0)/m_l(0)\ll1$ makes the first contribution negligible. Similarly, the scaling of \eqref{eq:mcond_def} gives

	\begin{equation}
		\hat{\dot m}_{\mathrm c}=- \Pi^{c}_{v}  (\overline{\Theta}_{v}-{\Theta}_{i})+\Pi^{c}_{l} ({\Theta}_{i}-\overline{\Theta}_{l}),
	\end{equation} with 
	
	\begin{equation}
		\begin{aligned}
			\Pi^{c}_{v}  &= \frac{2\pi h_{v,i} A_{i} \Delta T}{\omega_e \, m_v(0)\,\mathcal{L}_v}=\biggl( \frac{2\pi \kappa_v A_{i} }{\omega_e \,c_{v,v} \,m_v(0) R}\biggr)\biggl(\frac{c_{p,v}\Delta T}{\mathcal{L}_v}\biggr)\biggl(\frac{c_{v,v}}{c_{p,v}}\biggr)\\
			&=\Pi^h_{v,i}\, \mathrm{Ja}\,\frac{1}{\gamma},
		\end{aligned}
	\end{equation}
	
	and

	\begin{equation}
		\begin{aligned}
			\Pi^{c}_{l}  &= \frac{2\pi h_{l,i} A_{i} \Delta T}{ \omega_e m_v(0) \mathcal{L}_v}=\biggl( \frac{2\pi \kappa_l A_{i} }{\omega_e \,c_{v,l} \,m_l(0) R}\biggr)\biggl( \frac{c_{v,l}\Delta T m_l(0)}{\mathcal{L_v}m_v(0)} \biggr)\\
			&=\Pi^h_{l,i}\, \mathrm{Ja}\,\frac{1}{\gamma}\frac{ m_l(0)c_{v,l}}{m_v(0)c_{v,v}}.
		\end{aligned}
	\end{equation}
	
	The scaling introduces six dimensionless groups involving products of Nusselt numbers with quantities that can be evaluated a priori. Although the Nusselt numbers themselves cannot generally be predicted a priori, they are expected to depend on the sloshing dynamics (via the dimensionless interface dynamics $\hat{\eta}=\eta/R$, scaling as in \eqref{scaling_eta}) and on the thermal properties of both fluids. By analogy with forced convection, one may therefore write
	
	\begin{equation}
		\text{Nu}_{w,l},\text{Nu}_{l,i},\text{Nu}_{w,v},\text{Nu}_{v,i}= g\biggl(\hat{\eta}(\hat{\bm{r}},\hat{t}), \text{Pr}_l,\text{Pr}_v,\text{Ri}_l \biggr)\,,
	\end{equation} where $\text{Pr}_l=\nu_l/\alpha_l$ and $\text{Pr}_v=\nu_v/\alpha_v$ are the Prandtl number of liquid and vapor and $\text{Ri}_l=g \beta \Delta T R/(z_e \omega_e)^2$ the Richardson number on the liquid side. The observation that heat-transfer contributions enter only through products of the form  \(\Pi^h_{\bullet,\bullet}=\text{Nu}_{\bullet,\bullet}\,\Pi^{Nu}_{\bullet,\bullet}\) implies that scaling only requires matching the \emph{effective} groups \(\Pi^h\), with no need to reproduce \(\text{Nu}\) \emph{and} \(\Pi^{Nu}\):  any rescaling
	\(\text{Nu}\to a\,\text{Nu}\) accompanied by \(\Pi^{Nu}\to \Pi^{Nu}/a\)
	leaves the equations invariant.
	
	Finally, on the fluid side, assuming that the initial conditions $\hat{p}(0)$ and $\hat{\Theta}_{v,l,w}$ are matched, the scaling of the pressure evolution requires introducing a dimensionless equation of state. The relevant dimensionless groups can be exposed by considering a linearization of the relation $p=f(\rho_v,\overline T_v)$ around the equilibrium state $(\rho_{v,eq},T_{eq},p_{eq})$, namely
	\[
	p - p_{eq} \;\approx\;
	\Bigl(\tfrac{\partial p}{\partial \rho}\Bigr)_{eq} (\rho_v-\rho_{v,eq})
	\;+\;
	\Bigl(\tfrac{\partial p}{\partial T}\Bigr)_{eq} (\overline T_v-T_{eq})\,.
	\]
	
	The assumption of linearity does not entail a loss of generality, since only the resulting dimensionless groups are retained.  
	Introducing the dimensionless density
	$\hat\rho=(\rho-\rho_{eq})/\Delta\rho$
	leads to the relation
	\begin{equation}
		\hat p
		= \Pi_{p\rho}\,\hat\rho + \Pi_{pT}\,\overline \Theta_v,
	\end{equation}
	with 
	\begin{equation}
		\qquad
		\Pi_{p\rho}=\frac{(\partial p/\partial \rho)_{eq}\,\Delta\rho}{\Delta p},\quad
		\Pi_{pT}=\frac{(\partial p/\partial T)_{eq}\,\Delta T}{\Delta p}.
	\end{equation}

	The last set of dimensionless numbers arises from the scaling of the one-dimensional heat equation within the solid wall along its thickness \eqref{eq:PDE}. Taking the tank shell thickness $\delta_w$ as the reference length such as $\zeta=x/\delta_w$, assuming $Q_{ext}\approx0$ and both fluid phases to be at equilibrium, i.e. $\overline\Theta_v= \overline\Theta_l=0$, gives 
	
	\begin{equation}
		\begin{cases}
			\partial_{\hat t} \Theta_w(\zeta,\hat t)&=\mathrm{Fo}\; \partial_{\zeta\zeta}\Theta_w(\zeta, \hat t)\\
			\partial_\zeta \Theta_w|_{\zeta=0}&=-\mathrm{Bi}\;\Theta_{w,i} \text{ and } \partial_\zeta \Theta_w|_{\zeta=1}=0\\
			\Theta_w(\zeta,0)&=\Theta_w^0
		\end{cases}
	\end{equation}\label{eq:heat_eq_nd} with $\mathrm{Fo}={2\pi\alpha_w}/({\omega_e\delta_w^2) }$ the Fourier number and $\mathrm{Bi}=\delta_w(h_{wl}A_{wl}+h_{wv}A_{wv})/({(A_{wl}+A_{wv})\kappa_w})$ the Biot number.

	Finally, we note that the lowest pressure reachable by the system is the saturation pressure at the initial bulk liquid temperature, $\Delta p^{\text{max}}=p(0)-p^{\text{sat}}\left(\overline{T}_l(0)\right)$. In practice, this limit cannot be attained because of heat exchange with the walls, so that $\Delta p = p(0)-p_{eq}<\Delta p^{\text{max}}$. The magnitude of this departure can be estimated by comparing the sensible energy stored in the initial wall stratification with the characteristic energy scales of the fluid phases, namely the latent heat of condensation of the vapor and the sensible heat associated with liquid subcooling:
	\begin{equation}
		\begin{cases}
			E_w^{\mathrm{sens}} =   m_wc_{w}\Delta T,\\
			E_{v}^{\mathrm{lat}}=  m_v(0)\,\mathcal L_v,\\
			E_{l}^{\mathrm{sens}}= m_l(0)c_{p,l}\bigl[T_i(0)-\overline{T}_l(0)\bigr].
		\end{cases}
	\end{equation}
	
	This leads to the wall-to-fluid energy ratio
	\begin{equation}
		\varepsilon=\frac{E_w^{\text{sens}}}{E_v^{\text{lat}}+E_l^{\text{sens}}},
	\end{equation}
	which quantifies the wall influence on the thermodynamic exchanges. If $\varepsilon\ll1$, the effect of the wall is negligible and $\Delta p \approx \Delta p^{\text{max}}$ in case of ideal de-stratification.

	\subsection{Summary of Governing Dimensionless Groups}\label{sec4p3}
	
	The kinematic response is controlled by geometry/forcing (\(L/R,H/R,z_e/R,\omega_e/\omega_{1,0}\)), in particular the instability is governed by the pair \((\omega_e/\omega_{1,0},\hat a_v)\) setting onset of the instability. Thermal effects enter via four families:
	(i) effective heat–exchange groups \(\Pi^h\) (one‐cycle exchange vs capacity), factorable as \(\mathrm{Nu}\cdot\Pi^{Nu}\);
	(ii) enthalpy ratios \(\Pi^{\mathrm{sat}}\) (phase enthalpy advected vs sensible capacity);
	(iii) phase–change couplings \(\Pi^{c}\) (condensation rate vs cycle, latent–based);
	(iv) EOS sensitivities \(\Pi_{p\rho},\Pi_{pT}\). Wall response is set by \(\mathrm{Fo}\) and \(\mathrm{Bi}\), and the bounds of the thermodynamic evolution is dictated by \(\varepsilon\). The governing numbers are their definitions are summarized in \autoref{tab:scaling_summary_grouped}.

	\begin{table}[htbp]
		\centering
		\small
		\caption{Grouped dimensionless controls used in \S\ref{sec4}. Subscripts $\alpha\!\in\!\{l,v\}$ denote the fluid phase (liquid/vapor); $\beta\!\in\!\{w,i\}$ denotes wall/interface.}
		\resizebox{\textwidth}{!}{%
			\begin{tabular}{ccl}
				\toprule
				\textbf{Symbol} & \textbf{Definition} & \textbf{Physical meaning} \\
				\midrule
				\multicolumn{3}{c}{\emph{Geometry \& Kinematic response}}\\
				\midrule
				\addlinespace[2pt]
				& $L/R$ & Tank aspect ratio \\
				& $H/R$ & Liquid fill level \\
				& $z_e/R$ & Forcing amplitude–to–radius ratio \\
				& $\omega_e/\omega_{1,0}$ & Forcing–to–fundamental frequency ratio \\
				$\hat a_v$ & $z_e\omega_e^2/g$ & Acceleration amplitude  \\
				$\mathrm{Re}$ & $\rho_lz_e\omega_e R/\mu_l$ & Inertia–viscosity ratio  \\
				\midrule
				\multicolumn{3}{c}{\emph{Variables scaling}}\\
				\midrule
				\addlinespace[2pt]
				$\hat t$ & $\omega_e t/(2\pi)$ & Time measured in forcing periods\\
				$\Theta$ & $\dfrac{T-T_{eq}}{\overline T_v(0) - \overline T_l(0)}$ & Temperature scaled by initial stratification \\
				$\hat p$ & $\dfrac{p-p_{eq}}{p(0)-p_{eq}}$ & Pressure scaled by reachable drop \\
				\midrule
				\multicolumn{3}{c}{\emph{Fluid heat transfer processes}}\\
				\midrule
				\addlinespace[2pt]
				$\Pi^h_{j,k}$ & $\displaystyle \mathrm{Nu}_{j,k}\;\Pi^{Nu}_{j,k}$ &
				Forcing-to-convective time scale ratio with $j,k \in [w,l,v,i]$. \\
				$\mathrm{Nu}_{j,k}$ & $\displaystyle \frac{h_{j,k} R}{\kappa_j}$ & Convection–conduction ratio across path $j\to k$ \\
				$\Pi^{Nu}_{j,k}$ & $\displaystyle \frac{2\pi \kappa_j A_{j,k}}{\omega_e\,c_{v,j}\,m_\alpha(0)\,R}$ & Forcing-to-conduction time scale ratio \\
				$\Pi^{\mathrm{sat}}_v$ & $\displaystyle \frac{\mathcal h^{\mathrm{sat}}_v}{c_{v,v}\,\Delta T}$ & Heat transfer due to phase change (saturated enthalpy to sensible heat ratio) \\
				$\Pi^{\mathrm{sat}}_l$ & $\displaystyle \frac{\mathcal h^{\mathrm{sat}}_l}{c_{v,l}\,\Delta T}\,\frac{m_v(0)}{m_l(0)}$ & Liquid saturated enthalpy to sensible heat ratio weighted by mass ratio \\
				$\mathrm{Ri}$ & $g \beta \Delta T R/(z_e \omega_e)^2$ & Natural-to-forced convection ratio  \\
				$\mathrm{Pr}$ & $\rho_lz_e\omega_e R/\mu_l$ & Momentum-to-thermal diffusivity  \\
				\midrule
				\multicolumn{3}{c}{\emph{Phase–change processes}}\\
				\midrule
				\addlinespace[2pt]
				$\Pi^{c}_{v,i}$ & $\displaystyle \Pi^h_{v,i}\;\mathrm{Ja}\;\frac{1}{\gamma}$ &
				Evaporation sensitivity, linking ullage superheat to evaporation rate \\
				$\Pi^{c}_{l,i}$ & $\displaystyle \Pi^h_{l,i}\;\mathrm{Ja}\;\frac{1}{\gamma}\;\frac{m_l(0)c_{v,l}}{m_v(0)c_{v,v}}$ &
				Condensation sensitivity, linking liquid subcooling to condensation rate\\
				$\mathrm{Ja}$ & $\displaystyle \frac{c_{p,v}\,\Delta T}{\mathcal L_v}$ & Vapor sensible–to–latent heat ratio  \\
				$\gamma$ & $\displaystyle \frac{c_{p,v}}{c_{v,v}}$ & Vapor adiabatic index \\
				\midrule
				\multicolumn{3}{c}{\emph{Wall conduction}}\\
				\midrule
				\addlinespace[2pt]
				$\mathrm{Fo}$ & $\displaystyle \frac{2\pi \alpha_w}{\omega_e \delta_w^2}$ & Forcing–to–conduction through wall thickness time scales ratio  \\
				$\mathrm{Bi}$ & $\displaystyle \frac{h_{w,l}A_{w,l}+h_{w,v}A_{w,v}}{(A_{w,l}+A_{w,v})\kappa_w}\,\delta_w$ & Surface convection–to–internal conduction ratio  \\
				\midrule
				\multicolumn{3}{c}{\emph{Pressure sensitivities}}\\
				\midrule
				\addlinespace[2pt]
				$\Pi_{p\rho}$ & $\displaystyle \frac{(\partial p/\partial \rho)_{eq}\,\Delta\rho}{\Delta p}$ & Compressibility contribution: $p$ sensitivity to phase change \\
				$\Pi_{pT}$ & $\displaystyle \frac{(\partial p/\partial T)_{eq}\,\Delta T}{\Delta p}$ & Thermal contribution: $p$ sensitivity to ullage temperature varations \\
				\midrule
				\multicolumn{3}{c}{\emph{Pressure drop amplitude}}\\
				\midrule
				\addlinespace[2pt]
				$\varepsilon$ & $\displaystyle \frac{E_w^{\mathrm{sens}}}{E_v^{\mathrm{lat}}+E_l^{\mathrm{sens}}}$ &
				Wall–to–fluid energy ratio (if $\varepsilon\ll1$, $p_{eq}\approx p^{sat}(T_l(0))$) \\
				\bottomrule
			\end{tabular}
		}
		\label{tab:scaling_summary_grouped}
	\end{table}

	\section{Experimental Methodology}\label{sec5}

	The experimental methodology is structured to decouple the investigation of free-surface dynamics from the thermodynamic sloshing response. This strategy follows the assumption that, for the conditions studied, free-surface kinematics are governed primarily by inertia and gravity, with negligible feedback from the enhanced heat and mass processes. This assumption is commonly used in ground experiments (e.g., \cite{GROTLE2018512}) and is directly supported by recent microgravity evidence \citep{Monteiro2025}. Accordingly, the campaign is split into two complementary parts: first, an isothermal study in a transparent PMMA section to assess the free-surface response under vertical excitation, then a non-isothermal study in a fully instrumented metallic tank to quantify heat and mass transfer and the resulting thermodynamic evolution. We first describe the shaking apparatus (\S\ref{sec5p1}), then the isothermal test section (\S\ref{sec5p2}), the non-isothermal test section (\S\ref{sec5p3}), the experimental procedure (\S\ref{sec5p4}), and the test matrix (\S\ref{sec5p5}).

	\subsection{Shaking apparatus and working fluid}\label{sec5p1}
	
	The experimental campaign was conducted at the SHAKESPEARE (Shaking Apparatus for Kinetic Experiments of Sloshing Projects with Earthquake Reproduction) facility of the von Karman Institute. The table can replicate precise three-dimensional excitations with controlled amplitudes and frequencies. In this work, we impose a single-axis vertical harmonic displacement with $\omega_e=23.2$ rad/s with varying amplitudes in the range $z_e=3-9$mm. In both the isothermal and non-isothermal experiments, the working fluid is a hydrofluoroether, 3M\textsuperscript{\texttrademark} Novec\textsuperscript{\texttrademark} HFE-7000, known for its well-characterized thermophysical properties and widely employed to reproduce cryogenic-like thermophysical behavior under safe laboratory conditions.

	\subsection{Isothermal test section}\label{sec5p2}
	
	The test section, referred to as sloshing cell in the following, is made of a cylindrical body of length $L=312$mm, with an internal radius $R=67.25$ mm, closed with stainless steel domes of radius $R_d=100$ mm. The isothermal test section is built from a 15mm thick acrylic glass to ensure optical access. The free-surface evolution is recorded with a high-speed monochrome camera (JAI SP-12000M-CXP4-F) fitted with AF 35\,mm f/2D lenses, operated in back-lighting conditions using an LED screen of 290 W white LEDs. The videos have a resolution of $4096\times 3072$ pixels and a scaling factor of 8 pixels/mm are taken at a frequency of $100$ Hz and exposure of $1$ ms.

	\subsection{Non-isothermal test section}\label{sec5p3}
	
\begin{figure}[htbp]
		\centering
		\includegraphics[width=0.7\linewidth]{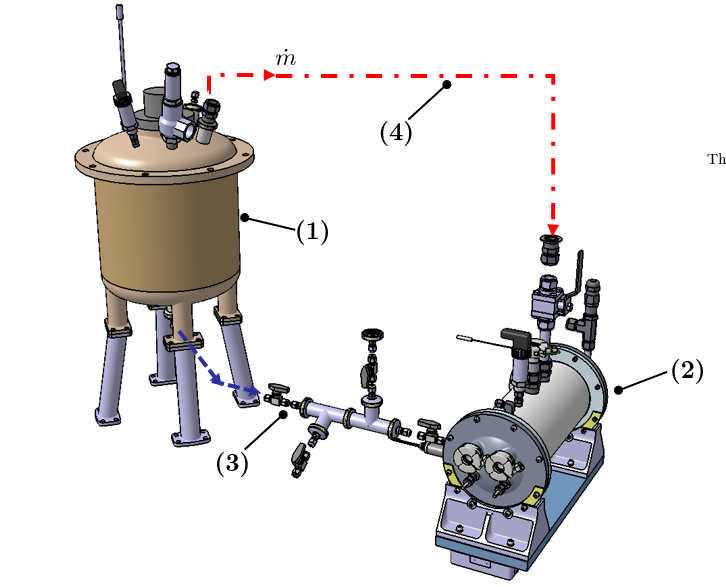}
		\caption{Non-isothermal experimental set-up composed of the \textit{Pressuring tank} (1), the sloshing cell (2), filling line (3), and pressuring line (4).}
		\label{fig:3d-nonsiso}
	\end{figure}

	The 3D view of the non-isothermal set ups is shown in Figure~\ref{fig:3d-nonsiso}. This consists of: (1) a large upright cylinder acting as pressuring tank, (2) the sloshing cell, (3) a filling line, and (4) a pressuring line. The pressurizing tank serves to establish the desired single-species environment in the sloshing cell, supplying both liquid HFE700 at room temperature or vapor at controlled superheated conditions.

	The connection lines are flexible, designed to handle mechanical stresses, and equipped with a set of valves to control the mass flows or isolate both tanks during the different phases of the experiment. The pressurizing line is insulated and fitted with an inline flexible heater (\textit{Vulcanic 26175.12}) to reduce condensation of the superheated vapor along the line.

	\begin{figure}[htbp]
		\centering
		\begin{subfigure}[t]{0.5\linewidth}
			\centering
			\includegraphics[width=0.9\linewidth]{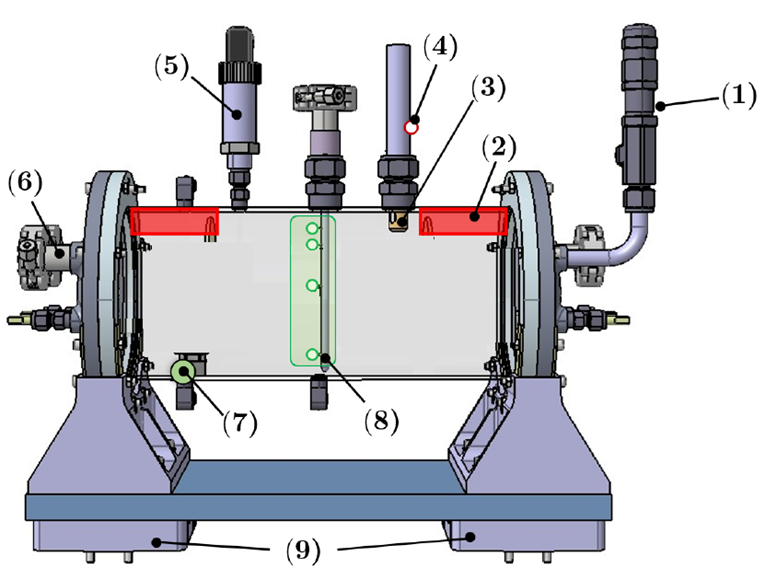}
			\caption{}
			\label{fig:sloshing-cell}
		\end{subfigure}\hfill
		\begin{subfigure}[t]{0.49\linewidth}
			\centering
			\includegraphics[width=0.85\linewidth]{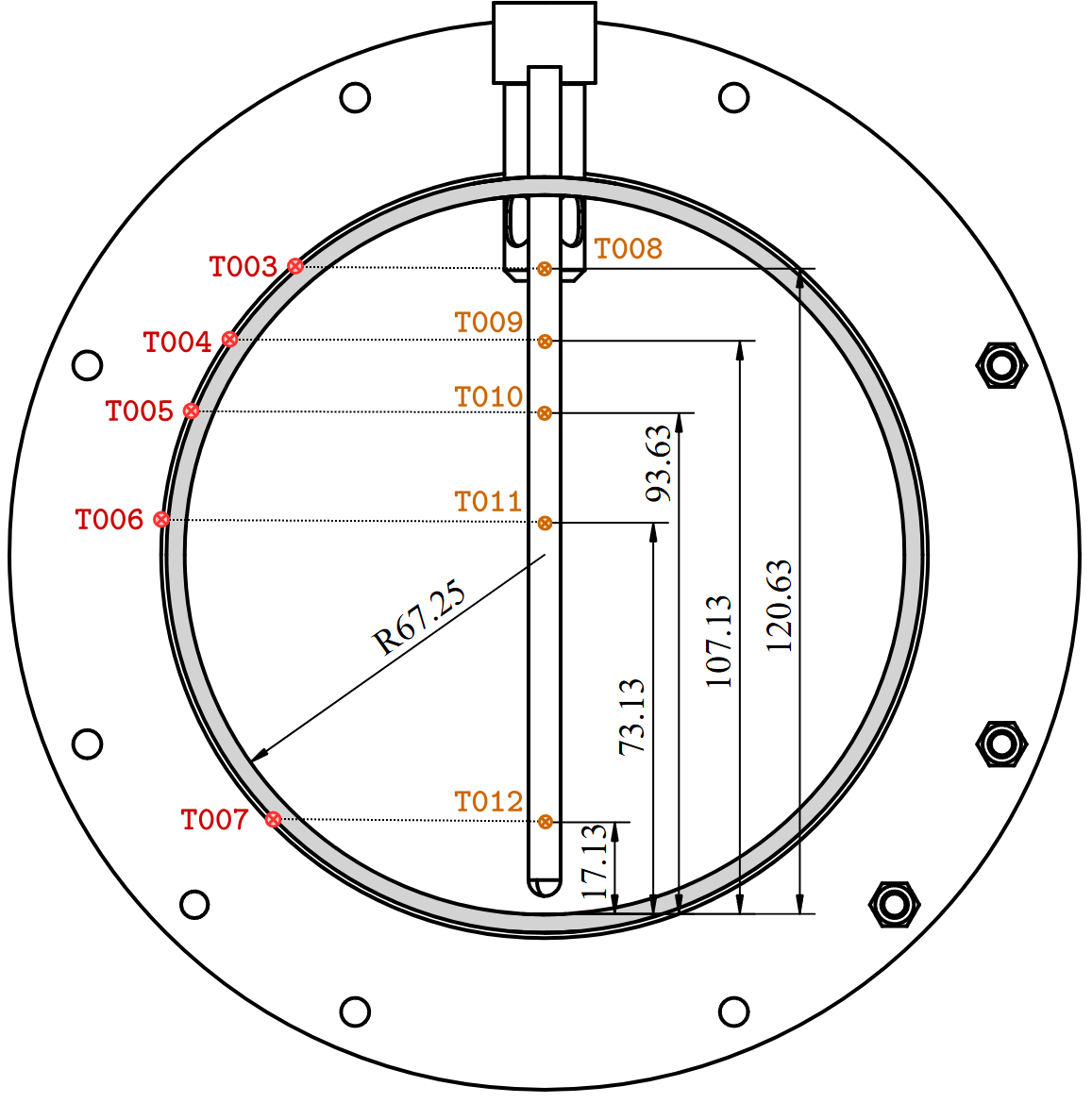}
			\caption{}
			\label{fig:sloshing-cell-temperature-probe}
		\end{subfigure}
		\caption{Sloshing cell schematic. (a) Instrumentation layout: (1) relief valve; (2) flexible surface heaters; (3) vapor diffuser; (4) Type-T thermocouple \texttt{T002}; (5) ullage pressure transducer; (6) triaxial accelerometer; (7) differential-pressure transducer (fill level); (8) Type-K thermocouples \texttt{T003--T012} (center probe and outer wall); (9) triaxial load cells. (b) Longitudinal mid-plane section showing the temperature probe and the vertical distribution of thermocouples \texttt{T003--T012}; elevations are listed in \autoref{tab:sensor-loc}.}
		\label{fig:sloshing-cell-merged}
	\end{figure}

	The sloshing cell (detail in \autoref{fig:sloshing-cell}) is a 3.5 mm-thick AISI~316 stainless-steel cylinder fitted with a spring-loaded safety valve (1) set to 2 bar absolute. Thermal stratification of the walls is imposed by four polyimide surface heaters (2) (\textit{Omega KHLVA-104/10}, 10W/$\text{inch}^{2}$) located on top of the cylindrical body. Warm vapor from the pressurizing tank is injected via a diffuser (3). A dedicated type-T thermocouple \texttt{T002} placed immediately upstream monitors the pressurant temperature. The ullage pressure is measured with a \textit{Swagelok PTI-S-AC3-32AQ-B} transducer (5), absolute range \(0\)–4 bar, accuracy \(\pm 0.25\%\) FS. A triaxial variable-capacitance accelerometer \textit{Endevco 7298-2} (6, \(\pm 2\,g\) range) is rigidly mounted on the dome to record the imposed excitation. A differential-pressure transducer \textit{PMI CS14-AA00001PD4A-3} (7, range \(0\)–68mbar, accuracy \(\pm 0.5\%\) FS) provides the fill-level measurement. A custom probe (8) at the centerline houses five type-K junctions (\texttt{T008}–\texttt{T012}) distributed along the vertical axis. Five additional type-K thermocouples (\texttt{T003}–\texttt{T007}) are mounted on the external wall at matching heights to assess circumferential stratification. The vertical position of the sensors is detailed in Table~\ref{tab:sensor-loc} and a schematic view is provided in \autoref{fig:sloshing-cell-temperature-probe}. The cell body is mounted on two aluminum feet bolted to a base plate that interfaces with two triaxial load cells (9) (\textit{PMI K3D120-1KN} $\pm1000N$ range, $0.5\,\%$~FS accuracy). All analog signals were acquired at 10Hz. \autoref{tab:OpCond} summarizes the key quantities of the non-isothermal tank.

	\begin{table}[htbp]
		\centering
		\caption{Vertical locations ($z/D$) of the thermocouples on the central probe and the outer tank wall.}
		\label{tab:sensor-loc}
		\begin{tabular}{ccc}
			\toprule 
			\textbf{Vertical Position} & \textbf{Central Probe} & \textbf{Outer Wall}  \\
			\midrule
			Pressurant Inlet & \multicolumn{2}{c}{\texttt{T002}} \\
			\midrule
			$z/D = 90\%$ & \texttt{T008} (Vapor) & \texttt{T003} (Vapor) \\ 
			$z/D = 80\%$ & \texttt{T009} (Vapor) & \texttt{T004} (Vapor) \\
			$z/D = 70\%$ & \texttt{T010} (Vapor) & \texttt{T005} (Vapor) \\
			$z/D = 54\%$ & \texttt{T011} (Vapor) & \texttt{T006} (Vapor) \\
			$z/D = 13\%$ & \texttt{T012} (Liquid) & \texttt{T007} (Liquid) \\
			\bottomrule
		\end{tabular}
	\end{table}

	\begin{table}[htbp]
		\centering
		\caption{Dimensions and wall thermophysical properties of the non-isothermal sloshing cell.}
		\label{tab:OpCond}
		\begin{tabular}{llc}
			\toprule
			\textbf{Quantity} & \textbf{Symbol} & \textbf{Value} \\
			\midrule
			Internal Radius & $R$  & 67.25 mm \\
			Cylinder Length & $L$ & 336 mm \\
			Dome Radius & $R_d$ & 100 mm \\
			Wall Thickness & $\delta_w$ & 3.5 mm \\
			Total internal Volume & $V_t$   & 5.2 l \\
			\midrule
			Wall Mass & $m_w$    & 11.2 kg \\
			Wall Specific Heat & $c_w$ & 0.5 J/g/K \\
			Wall Conductivity & $\kappa_w$ & 15 W/m/K \\
			\bottomrule
		\end{tabular}
	\end{table}
	
	\subsection{Experimental Procedure for Non-isothermal Experiments}\label{sec5p4}

	Each non-isothermal test follows a multi-stage procedure to establish single-species conditions and controlled initial thermal stratification. 
	
	A critical prerequisite is ensuring the working fluid is free of dissolved non-condensable gases, which are known to alter the thermodynamic response \citep{arndt}.  To achieve this, the HFE-7000 is degassed using the pressure reduction method \citep{luyckx}, until the measured $(p, T)$ lie on the saturation curve, which confirms single-species conditions. 
	
	Next, the sloshing cell is vacuumed before being filled with degassed, ambient-temperature liquid from the pressurizing tank by pressure-feeding. The fill level is set slightly below the target to account for the volume increase due to vapor condensation during subsequent pressurization
	
	Thermal conditioning follows. Heater belts on the pressurizing tank build up its ullage pressure and temperature, while the pressurizing line and the sloshing cell are preheated to limit condensation of injected superheated vapor along its pathway. Pre-heating is crucial to mitigate the unavoidable condensation of the injected super-heated vapor along its cooler pathway to the sloshing cell. Excess condensation would significantly increase the fill level within the sloshing cell and limit the pressure build-up required to set the initial conditions. Lastly, the differential-pressure piping is heated to avoid spurious vapor condensation that could bias the level measurement. All the heating systems are controlled by separate PID controllers to prevent temperature overshoot that could damage the system. 
	
	As the set point temperatures are reached, active pressurization is initiated by injecting warm vapor from the pressurizing tank until the sloshing cell reaches the prescribed pressure of $p \approx 170$ kPa. 
	Then a relaxation period follows, allowing the system to stabilize and reach the prescribed initial pressure within the sloshing cell reported in Table~\ref{tab:ini_cond}. This relaxation time is essential to ensure the development of thermal stratification in both fluid phases, and thereby ensure repeatability of the experiments.
	
	\begin{figure}[htbp]
		\centering
		\includegraphics[width=0.7\linewidth]{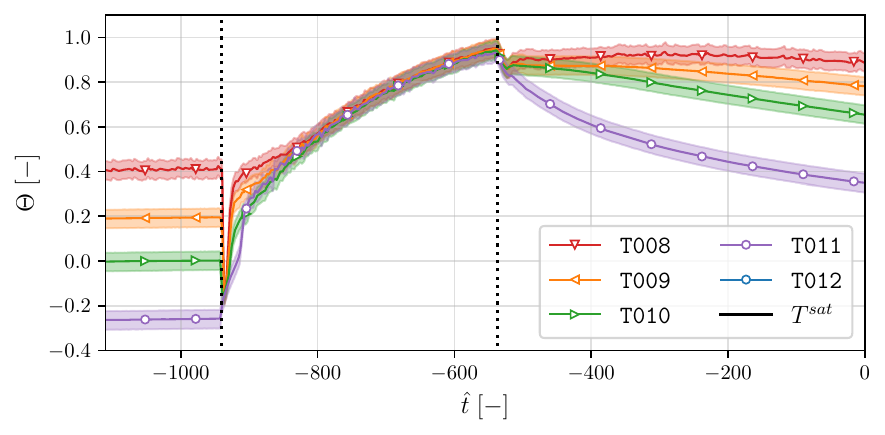}
		\caption{Non-dimensional temperature evolution at the 5 measurement points within the sloshing cell during active pressurization and relaxation of test $S0$. T012 is located in the bulk liquid, T011 near interface and T008-T010 is the vapor phase. The vertical lines denote the start and end of the warm vapor injection, $\hat t=0$ marks the initiation of the sloshing signal.}
		\label{fig:stratification}
	\end{figure}
	
	\autoref{fig:stratification} shows the temporal evolution of the temperature field within the sloshing cell prior to the sloshing excitation. The two vertical lines mark the onset and termination of the warm-vapor injection, and \(\hat t = 0\) designates the start of the sloshing event. Because the tank walls are initially externally heated, the ullage is already stratified before pressurization begins. During the active pressurization phase, the ullage rapidly approaches an isothermal state.    
	
\begin{figure}[htbp]
		\centering
		\includegraphics[width=0.7\linewidth]{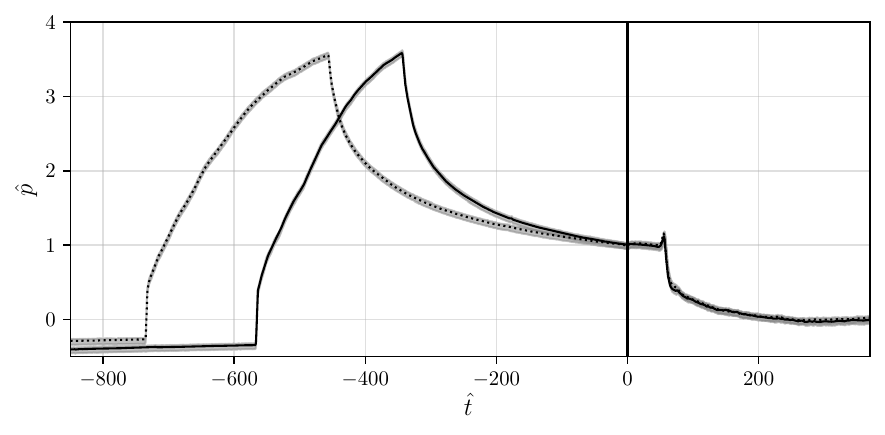}
		\caption{Non-dimensional pressure evolution along the testing procedure for two runs of test $S4$ tests performed on different days, showing the impact of ambient temperature on the repeatability of the relaxation. $\hat{t}=0$ marks the initiation of the sloshing signal highlighted by a vertical line.}
		\label{fig:relaxation}
	\end{figure}

	Once the injection stops and the system relaxes, a thermal stratification progressively develops in the ullage gas: the near-wall region remains superheated due to the top heaters, whereas the temperature in the vicinity of the liquid-vapor interface decreases as condensation occurs, closely following the saturation temperature. Throughout the entire process, the bulk liquid temperature remains essentially constant, while the interface follows the saturation curve. During relaxation, the pressure decays as the superheated vapor condenses onto the colder liquid free-surface and walls. This behavior is well-described by an empirical exponential decay, such as
	\begin{equation}
		\hat{p}(\hat{t}) =
		1+(\hat{p}_{\max}-1)\,
		\exp\!\left[\frac{\hat t_{\text{inj}}-\hat{t}}{\hat \tau_r}\right] \text{ for } \hat t\in[\hat t_{\text{inj}},0],
		\label{eq:baseline_decay}
	\end{equation}

	where $\hat t_{\mathrm{inj}}$ is the non-dimensional time at the end of injection, 
	$\hat p_{\max}=\bigl(p(t_{\mathrm{inj}})-p_{eq}\bigr)/\Delta p$ is the dimensionless pressure at $\hat t=\hat t_{\mathrm{inj}}$. The relaxation time $\hat \tau_r$ is obtained by fitting \eqref{eq:baseline_decay} to the measured pressure during the relaxation phase. Typical values are $\mathcal{O}(10^{2})$ while sensitive to ambient boundary conditions, i.e. increasing with higher ambient temperature (see \autoref{fig:relaxation}). The fitted $\hat \tau_r$ and $p(t_{\mathrm{inj}})$ are reported in \autoref{tab:ini_cond}.
	
	\subsection{Test Matrix}\label{sec5p5}

	The experimental test matrix was designed to investigate the effect of forcing amplitude and fill level on a reference full-scale, 1 m radius tank, under vertical harmonic excitation at 1 Hz.
	This choice is motivated by its relevance to flight conditions, as significant vibrational energy is measured in the 0.8-1.2 Hz range during gust-load and maneuver segments \citep{catherines_vibrations_1975}.
	Furthermore, for a representative full-scale horizontal tank of radius 1m, this frequency range coincides with the instability region of the antisymmetric (2,0) “middle-jet” mode (see \autoref{fig:stab_bound}). This sloshing regime is expected to be particularly effective in inducing thermal mixing and destratification in wave-breaking conditions \citep{Das2009-vertical}.
	
	The testing excitation frequency was set to correspond to this reference case by conserving the dimensionless forcing frequency $\omega_e/\omega_{1,0}$, amplitude $z_e/R$ and acceleration $\hat a_v$. 
	The excitation and initial thermodynamic conditions for all test cases are summarized in \autoref{tab:ini_cond}. The initial volume-averaged temperature of each control volume is also reported; it is evaluated as the weighted average of the instantaneous thermocouple readings, with weights adjusted to the vertical sensor location. The relative locations of the testing conditions in the vertical stability diagram are shown in \autoref{fig:stab_bound}. Test $S0$ is predicted to lie outside the instability envelope of the $(2,0)$ sloshing mode, whereas $S1$ is situated on its theoretical boundary and $S2$ resides well within this instability region. In opposition all points at 70\% filling ($S3, S4$ and $S5$) are located inside the instability region of mode $(2,0)$, with $S3$ sitting at the theoretical boundary. Finally, the corresponding values of the dimensionless parameters for each configuration are evaluated and reported in \autoref{tab:DimlessValues}

	\begin{table}[htbp]
		\centering
		\begin{tabular}{llcccccc}
			\toprule
			& Units & \textbf{S0} & \textbf{S1} & \textbf{S2} & \textbf{S3} & \textbf{S4} & \textbf{S5} \\
			\midrule
			$H/2R$ & [-] & 0.50 & 0.52 & 0.50 & 0.69 & 0.67 & 0.67 \\
			\midrule
			$p(t^{\text{inj}})$ & [kPa] & 169.9 & 170.1 & 170.1 & 170.6 & 168.3 & 170.2 \\
			$\hat \tau_r$ &  [-] & 150.6 & 167.2 & 237.4 & 82.0 & 270.7 & 120.0 \\
			\midrule
			$z_e$ & [mm] & 3.3 & 6.3 & 8.9 & 3.3 & 6.3 & 8.9 \\
			$\omega_e$ & [rad/s] & 23.2 & 23.2 & 23.2 & 23.2 & 23.2 & 23.2 \\
			\midrule
			$p(0)$ & [kPa] & 105.1 & 105.0 & 105.0 & 105.0 & 105.1 & 105.0 \\
			$\overline{T}_l(0)$ &  [K] & 294.8 & 295.4 & 292.1 & 292.8 & 298.1 & 296.1 \\
			$\overline{T}_v(0)$ & [K] & 315.2 & 314.4 & 315.2 & 310.8 & 312.1 & 312.7 \\
			$\overline{T}_w(0)$ & [K] & 309.9 & 309.8 & 309.1 & 305.5 & 309.5 & 308.3 \\
			$T_{eq}$ &  [K] & 303.2 & 303.4 & 301.6 & 299.9 & 304.5 & 302.9 \\
			\bottomrule
		\end{tabular}
		\caption{Thermodynamic state at the end of the active pressurization ($t^{\text{inj}}$) and at the start of sloshing ($t = 0$), together with the imposed vertical forcing. $H/2R$ is the liquid fill level; $p(t^{\text{inj}})$ and its decay constant $\hat \tau_r$ describe the exponential pressure relaxation between the two instants. The excitation is set by the dimensionless acceleration $\hat a_v$ and angular frequency $\omega_e$. At $t = 0$ the pressure $p(0)$ and the volume-averaged liquid, vapor, and wall temperatures $\overline T_l(0)$, $\overline T_v(0)$, $\overline T_w(0)$ are recorded; $T_{eq}$ is the equilibrium temperature used for non-dimensionalization.}
		\label{tab:ini_cond}
	\end{table}

	\begin{table}[htbp]
		\centering
		\begin{tabular}{llcccccc}
			\toprule
			&&  \textbf{S0} & \textbf{S1} & \textbf{S2} & \textbf{S3} & \textbf{S4} & \textbf{S5} \\
			\midrule
			$\hat{a}_v$ & [\scriptsize$\times10^{-1}$] & 1.8 & 3.5 & 4.9 & 1.8 & 3.5 & 4.9 \\
			$\mathrm{Re}_l$ & [\scriptsize$\times10^{4}$] & 1.6 & 3.1 & 4.3 & 1.6 & 3.3 & 4.5 \\
			$\mathrm{Ri}_l$ & [\scriptsize$\times10^{0}$] & 4.5 & 1.1 & 0.7 & 3.9 & 0.8 & 0.5 \\
			$\mathrm{Pr}_l$ & [\scriptsize$\times10^{0}$] & 8.4 & 8.3 & 8.6 & 8.6 & 8.1 & 8.3 \\
			$\mathrm{Pr}_v$ & [\scriptsize$\times10^{-1}$] & 8.4 & 8.4 & 8.4 & 8.5 & 8.5 & 8.4 \\
			$\Pi^{Nu}_{w,l}$ & [\scriptsize$\times10^{-6}$] & 6.9 & 6.8 & 7.0 & 6.1 & 6.1 & 6.1 \\
			$\Pi^{Nu}_{w,v}$ & [\scriptsize$\times10^{-4}$] & 2.5 & 2.5 & 2.5 & 3.2 & 3.1 & 3.1 \\
			$\Pi^{Nu}_{l,i}$ & [\scriptsize$\times10^{-6}$] & 3.7 & 3.5 & 3.7 & 2.3 & 2.4 & 2.4 \\
			$\Pi^{Nu}_{v,i}$ & [\scriptsize$\times10^{-4}$] & 1.3 & 1.4 & 1.3 & 2.2 & 2.1 & 2.1 \\
			$\Pi^{\mathrm{sat}}_l$ & [\scriptsize$\times10^{-3}$] & 4.0 & 3.7 & 4.2 & 1.9 & 1.8 & 1.8 \\
			$\Pi^{\mathrm{sat}}_v$ & [\scriptsize$\times10^{-1}$] & 3.3 & 3.2 & 3.0 & 1.4 & 2.7 & 2.6 \\
			$\mathrm{Ja}/\gamma$ &  [\scriptsize$\times10^{-1}$] & 1.4 & 1.3 & 1.6 & 1.2 & 0.9 & 1.1 \\
			$\Pi_{p\rho}$ & [\scriptsize$\times10^{-1}$] & 3.7 & 3.7 & 3.8 & 4.3 & 3.9 & 4.0 \\
			$\Pi_{pT}$ & [\scriptsize$\times10^{0}$] & 1.3 & 1.2 & 1.1 & 0.6 & 1.1 & 1.0 \\
			$\mathrm{Fo}$ & [\scriptsize$\times10^{-2}$] & 3.3 & 3.3 & 3.3 & 3.3 & 3.3 & 3.3 \\
			$\varepsilon$&  [\scriptsize$\times10^{0}$] & 2.0 & 1.9 & 1.9 & 1.1 & 1.3 & 1.3 \\
			\bottomrule
		\end{tabular}
		\caption{Dimensionless parameters of the sloshing response of the vertically excited sloshing cell estimated from the initial thermodynamic state and forcing condition listed in Table~\ref{tab:ini_cond}}
		\label{tab:DimlessValues}
	\end{table}
	
	\section{Data analysis}\label{sec6}
	\subsection{Free-decay Analysis via POD}\label{sec6p1}
	
	To determine the natural frequencies and damping ratios of the sloshing modes, free-decay tests were conducted. The motion was induced by a brief, step-like longitudinal acceleration ($0.62\,g$ for $T_s=0.18$s) that displaces the free surface from its equilibrium position. After the pulse, the interface is left to oscillate freely. While far from an ideal impulse, this finite pulse provides broadband spectral energy sufficient to excite the lowest sloshing modes, with its first zero-crossing occurring at $\omega = 2\pi/T_s \approx 34.9,$rad/s. In a linear framework, the subsequent free-decay motion can be characterized by treating the time evolution of each mode as a second order damped system \citep{Ibrahim_2005}, that is:
	
	\begin{equation}
		\eta(\mathbf{x}, t) = \sum \eta_{n,m}(\mathbf{x})\cos{(\omega_{d,nm}t+\varphi_{n,m})}e^{-\varsigma_{n,m}t}\label{eq:free_decay},
	\end{equation}
	where $\omega_{d,nm}$ is the damped frequency of mode $(n,m)$, $\varsigma_{n,m}=\zeta_{n,m}\,\omega_{n,m}$  the modal damping rate, with $\zeta_{n,m}$ the damping ratio, and $\omega_{n,m}$ the natural frequency of each mode. We have:
	\begin{equation}\label{eq:nat_freq_decay_fitting}
		\omega_{n,r}^2=\omega_{d,r}^2+\varsigma_r^2.
	\end{equation}
	
\begin{figure}[htbp]
		\centering
		\begin{subfigure}[b]{0.49\linewidth}
			\centering
			\includegraphics[width=\linewidth]{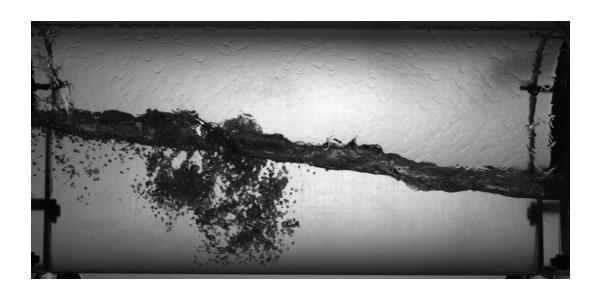}
		\end{subfigure}
		\begin{subfigure}[b]{0.49\linewidth}
			\centering
			\includegraphics[width=\linewidth]{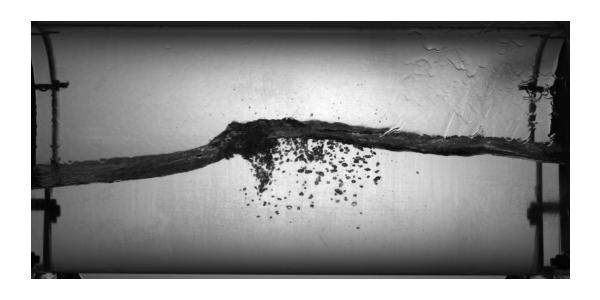}
		\end{subfigure}
		\begin{subfigure}[b]{0.49\linewidth}
			\centering
			\includegraphics[width=\linewidth]{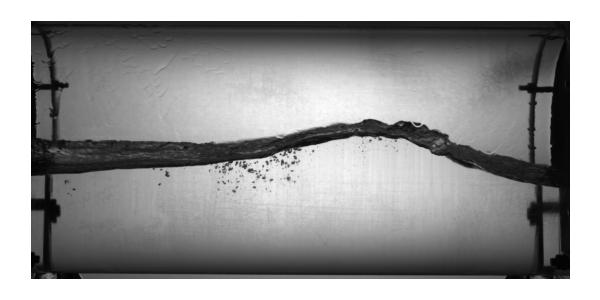}
		\end{subfigure}
		\caption{Representative raw frames from the lateral high-speed camera during the free-decay at 50\% fill, recorded after a longitudinal step-like acceleration (0.62\,g for $T_s=0.18$\,s). Panels correspond to $t=\,$[0.20, 0.70, 1.30]\,s after the end of the pulse.}
		\label{fig:free-decay-50-raw}
	\end{figure}

	The analysis of free decay subsequent to the short step therefore enables the estimation of the natural frequencies and damping rates. Representative frames of the video recording of the free-damping at 50\% are shown in \autoref{fig:free-decay-50-raw}. Videos of these tests are provided as supplementary material.
	
	To extract the dominant flow patterns and their frequencies from the videos, a combination of Proper Orthogonal Decomposition (POD) of the high-speed video recordings and nonlinear fitting was carried out. Denoting by $g(\mathbf{x}_i,t_k)$ the gray scale intensity at pixel location $\mathbf{x}_i$ and time $t_k=k\Delta t$, the POD expresses the video sequence with a decomposition of the form
	
	\begin{equation}
		\label{POD}
		g(\mathbf{x},t)=\sum^{n_R}_{r=1} \sigma_r \phi_r(\mathbf{x}_i)\psi_r(t_k)\,,
	\end{equation} where $\sigma_r$ is the amplitude of each mode, having spatial structure $\phi_r$ and temporal structure $\psi_r$ and $n_R=\min(n_s,n_t)$, with $n_s$ the number of pixels in each image and $n_t$ the number of images in each video. These structures are the eigenvectors of the temporal and spatial covariance matrices of the data respectively-- this definition results from the optimality condition that leads to the maximization of the amplitudes $\sigma_r$ or, equivalently, the minimization of the least square error produced by truncating \eqref{POD} at any index $r<n_R$ \citep{Dawson2023,Mendez2023,Poletti2024}.
	
	\begin{figure}[htbp]
		\centering
		\includegraphics[width=0.7\linewidth]{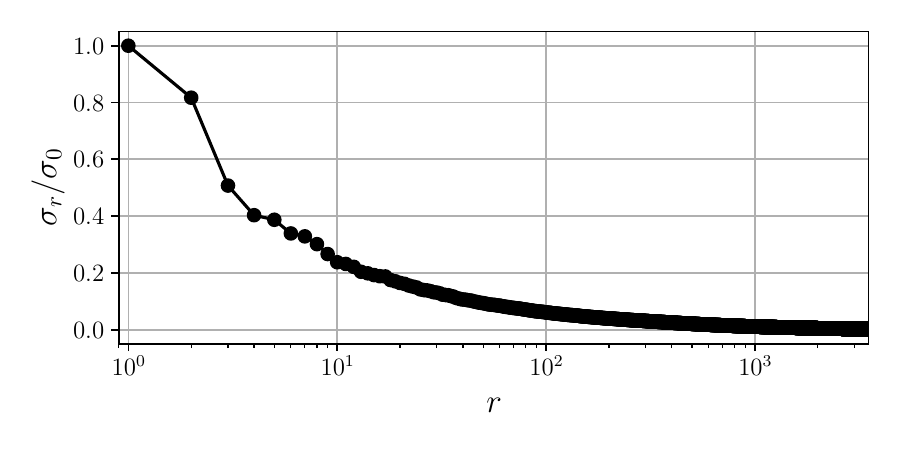}
		\caption{Normalized POD singular values $\sigma_r/\sigma_0$ versus mode index $r$ of the high-speed video recording of the free-damping following the step-input test at 50\% filling.}
		\label{Sigma_decay}
	\end{figure}
	
	While POD is fully data-driven and does not enforce a one-to-one correspondence between modes and the theoretical eigenfunctions in \eqref{eq:free_decay}, its optimal reconstruction provides strong filtering, with only a handful of modes needed to capture the energy decay, as shown in Figure \ref{Sigma_decay}. For this reason, POD was employed as a first step toward identifying the eigenfunctions in \eqref{eq:free_decay} directly from the video data. The temporal structures extracted by POD were then used to construct a characteristic time series, $\xi(t)$, from which the decaying exponential functions were sought:
	
	\begin{equation}
		\label{xi_eq}
		\xi(t)=\sum^{n^*_R}_{r=1} \sqrt{\sigma_r}\,\psi_r(t_k)
		\;\;\approx\;\;\sum^{n_D}_{r=1} a_r \cos\!\big(\omega_{d,r}t+\varphi_{r}\big)\,e^{-\varsigma_{r}t}.
	\end{equation}
	
	The parameters $a_{r}$ $\varsigma_{r}$ and $\varphi_{r}$ are then identified from $\xi(t)$ using nonlinear least square, with initial guess for $\omega_{d,r}$ set from the Fast Fourier Transform (FFT) of the temporal structures \(\psi_r(t_k)\). Once these are identified, a set of $n^*_D$ eigenfunction evolution $$\psi_{D,r}(t_k)=\cos{(\omega_{d,r}t+\varphi)}e^{-\varsigma_{r}t_k}$$ is used as a temporal basis to identify the spatial basis using traditional least squares. That is, given $\bm{\Psi}_D\in\mathbb{R}^{n_t\times n^*_D}$ the matrix collecting these bases along the columns, the least square solution for the spatial structures can be written as \citep{Mendez2023}:
	
	\begin{equation}
		\label{phiD}
		\bm{\Phi}_D \bm{\Sigma}_D= \mathbf{G} \bm{\Psi}_D \bigl(\bm{\Psi}^T_D  \bm{\Psi}_D \bigr)^{-1}\,,
	\end{equation} where $\mathbf{G}\in\mathbb{R}^{n_p\times n_t}$ is the matrix collecting the $n_t$ images of the videos, each image reshaped into a column vector storing the $n_p$ gray scale values in each pixel, $\bm{\Phi}_D \in\mathbb{R}^{n_p\times n_D}$ the matrix of spatial structure also reshaped column wise and $\bm{\Sigma}_D$ is a diagonal matrix containing the normalization values such that the $l_2$ norm of each column of $\bm{\Phi}_D$ is unitary. The spatial structures $\phi_{D,r}(\mathbf{x}_i)$ retreived from the columns of $\bm{\Phi}_D$ are expected to be the gray-scale image counter parts of the spatial structures $\eta_{n,m}(\mathbf{x})$ in \eqref{eq:free_decay}.

	It is worth noting that the final decomposition is essentially a Dynamic Mode Decomposition (DMD, \cite{Schmid,Rowley2}) of the grayscale videos. However, none of the standard DMD algorithms (see \cite{Tu2014jcd}), nor the variants first proposed in the climatology community such Principal Oscillation Patterns (POP, \cite{Hasselmann1988,Storch1990}) and Linear Inverse Modeling (LIM, \cite{Penland1996,LIM1}), as implemented in the open-source package MODULO \citep{Poletti2024}, yielded satisfactory results. This motivated the present custom adaptation, where nonlinear fitting is applied to the characteristic time series \eqref{xi_eq} and then projected according to \eqref{phiD}.

	\subsection{Parameter inference via Augmented Extended Kalman Filter (AEKF)}
	\label{sec6p2}
	
	An augmented and extended Kalman filter was used to estimate the heat transfer coefficient from the available measurements. The Kalman filter provides a recursive framework to combine predictions from a dynamic model with noisy measurements, producing improved estimates of the system state at each sampling time. For nonlinear dynamics, the Extended Kalman Filter (EKF) achieves this by linearizing both the model equations and the observation function around the current estimate \citep{emery_2004,bonini_2024}. The augmented EKF (AEKF) \citep{BALDASSARRE2024e27343} employed in this work includes the unknown heat transfer coefficients as additional state variables. This augmentation allows the filter to adjust parameter values dynamically so that the model predictions remain consistent with the experimental data.
	
	The dynamic model employed in the one introduced in Section \ref{sec3}, assuming constant properties obtained from the NIST database \citep{LEMMON-RP10} at the initial thermodynamic state for each control volume. The external heat supplied to the tank is set to a constant $Q_{ext}=30\,\text{W}$ and the volume of each control volume is assume to remain constant $dV_l/dt=dV_v/dt=0$. The augmented state space, including as well the unknown heat-transfer coefficients, reads 
	
	\begin{equation}
		\mathbf{s}
		\;=\;
		\bigl[
		\overline T_l,\,
		\overline T_v,\,
		m_l,\,
		p,\,
		T_w(\mathbf{x}),\,
		h_{wl},\, h_{wv},\, h_{li},\, h_{vi}
		\bigr]^{\mathsf T},
	\end{equation} where $\mathbf{x}$ is the discretization in 10 nodes within the wall thickness. The augmented thermodynamic model can then be written as 
	
	\begin{equation}
		\dot{\mathbf{s}} = f(\mathbf{s}),
		\label{eq:dyn_syst}
	\end{equation} with $f$ defined by \eqref{eq:massVariation}–\eqref{eq:p_state} and the time derivatives of  the heat transfer coefficients taken as 
	
	\begin{equation}
		\frac{d h_{ij}}{dt}=0.
		\label{eq:htc_model}
	\end{equation}
	
	The observation, within the traditional EKF approach, are written as 
	
	\begin{equation}
		\label{Obs}
		\mathbf{y}_k
		=
		\bigl[\overline{T}_{l,k},\, \overline{T}_{wo,k},\, \overline{T}_{v,k},\, p_k\bigr]^{\mathsf T}
		=
		\mathcal{g}(\mathbf{s}_k)+\mathbf{v}_k,
		\qquad
		\mathbf{v}_k\sim\mathcal{N}(\mathbf{0},\mathbf{R}),
	\end{equation}
	where the subscript denotes the time step $k$ within the sequence of measurements $k=[1\dots n_t]$ associated to the time steps $t_k=k \Delta t$, $\mathcal{g}$ denotes the measurement function, $\mathcal{N}(\mathbf{\mu},\mathbf{R})$ denotes a multivariate Gaussian distribution with mean vector $\mathbf{\mu}$ and covariance matrix $\mathbf{R}$. Here, the measurement-noise covariance associated to the sensors uncertainty $\mathbf{R}$ in \eqref{Obs} is taken as a diagonal matrix $\mathbf{R}=\mbox{diag}(\mathbf{r})$, with entries given by the measurement uncertainty of each variable as described in section \ref{sec5}. 
	
	For each of the measurements $\mathbf{y}_k$ and associated covariance matrix $\mathbf{R}$, and given the corresponding model predictions of the state $\tilde{\mathbf{s}}_{k}^{-}$, and its covariance $\mathbf{P}_{k}^{-}$, the Kalman filter provides an improved estimate of the state, denoted as $\tilde{\mathbf{s}}_{k}$, and its covariance matrix $ \mathbf{P}_{k}$, using the update (assimilation step):
	
	\begin{align}
		\label{Kalman_update}
		\tilde{\mathbf{s}}_{k}
		&= \tilde{\mathbf{s}}_{k}^{-} + \mathbf{K}_k\!\left(\mathbf{y}_k - \mathcal{g}(\tilde{\mathbf{s}}_{k}^{-})\right),\\
		\mathbf{P}_{k}
		&= (\mathbf{I}-\mathbf{K}_k\mathbf{H}_k)\,\mathbf{P}_{k}^{-}\,(\mathbf{I}-\mathbf{K}_k\mathbf{H}_k)^{\mathsf T}
		+ \mathbf{K}_k\,\mathbf{R}\,(\mathbf{K}_k)^{\mathsf T}.
	\end{align} where 
	
	\begin{equation}
		\label{Kalman_gain}
		\mathbf{K}_k = \mathbf{P}_{k}^{-}\,\mathbf{H}_k^{\mathsf T}\,(\mathbf{H}_k\,\mathbf{P}_{k}^{-}\,\mathbf{H}_k^{\mathsf T} + \mathbf{R}))^{-1}, 
	\end{equation} is the Kalman gain and 
	
	\begin{equation}
		\mathbf{H}_k = \left.\frac{\partial \mathcal{g}}{\partial \mathbf{s}}\right|_{\tilde{\mathbf{s}}_{k}^{-}},
	\end{equation} is the Jacobian of the observation function evaluated at the current model prediction $\tilde{\mathbf{s}}_{k}^{-}$. 
	
	The model propagation of the predicted states $\tilde{\mathbf{s}}_{k}^{-}$ was carried via Crank–Nicolson (CN),
	\begin{equation}
		\tilde{\mathbf{s}}_{k}^{-}
		=
		\tilde{\mathbf{s}}_{k-1}
		+
		\tfrac{\Delta t}{2}\Big[
		f\!\bigl(\tilde{\mathbf{s}}_{k-1}\bigr)
		+
		f\!\bigl(\tilde{\mathbf{s}}_{k}^{-}\bigr)
		\Big],
		\label{eq:cn_predict}
	\end{equation}
	which was solved at each step with a small number of Newton iterations. The CN method was chosen for its unconditional stability and second-order accuracy, allowing the use of relatively large time steps $\Delta t$, that match the measurement availability without compromising stability.
	The associated covariance matrix update was computed via linear propagation as 
	
	\begin{equation}
		\label{P_update}
		\mathbf{P}_{k}^{-}  =  \mathbf{A}_k\,\mathbf{P}_{k-1}\,\mathbf{A}_k^{\mathsf T}  + \mathbf{Q}_k\,,
	\end{equation} where
	
	\begin{equation}
		\mathbf{A}_k  =  \Bigl(\mathbf{I}-\tfrac{\Delta t}{2}\,\mathbf{J}_f\big|_{\tilde{\mathbf{s}}_{k}^{-}}\Bigr)^{-1}  \Bigl(\mathbf{I}+\tfrac{\Delta t}{2}\,\mathbf{J}_f\big|_{\tilde{\mathbf{s}}_{k-1}}\Bigr)
	\end{equation} is the associated linear propagator, $\mathbf{J}_f=\partial f/\partial\mathbf{s}$ is the Jacobian of the thermodynamic model and $\mathbf{Q}_k$ is the covariance matrix related to the reliability of the model at step $k$. This matrix is user defined and was also taken as diagonal $\mathbf{Q}_k=\sigma^2_{Q,k}\mathbf{I}$.
	
	The inference according to \eqref{Kalman_update} strongly depends on the relative weight between $\mathbf{P}_{k}^{-}$ and $\mathbf{R}$ in \eqref{Kalman_gain}: if the entry on the model covariance $\mathbf{P}_{k}^{-}$ are much smaller than those in $\mathbf{R}$, the filter gives more credit to the model prediction than to the measurements. The opposite is true if $\mathbf{P}_{k}^{-}$ has much larger entries than those in $\mathbf{R}$. Therefore, in addition to the model dynamics, a major role in the filtering is played by the model covariance $\mathbf{Q}_k$ in \eqref{P_update}.
	
	The main control parameters in the filtering is the ratio of entries between $\mbox{diag}(\mathbf{Q}_k)$ and $\mbox{diag}(\mathbf{R})$. These were dynamically adapted to allow the filter to mildly release the assumption in \eqref{eq:htc_model} and allow the heat transfer coefficient to vary. More specifically, during quasi-steady phases, small variances ($\mathcal{o}(1)$) are given to the entries in $\mathbf{Q}_k$ corresponding to the heat transfer coefficients. At the onset of mixing, detected as the pressure variation exceeds a user defined threshold  $|p_k-p_{k-1}|>\Delta p_{\Delta}$, the variance in the same entries are increased by six orders of magnitude, thus allowing the filter to ignore \eqref{eq:htc_model} and let the heat transfer coefficients adjust as best required to match the measurements.
	
	The AEKF inference was carried out over the non-dimensional window $\hat t\in[-100,200]$, starting 100 sloshing periods before excitation to capture the end of the initial relaxation. To reduce sensitivity to initialization, 100 EKF runs with randomized initial $h_{ij}$ were performed and the solution minimizing the root-mean-square error between predicted and measured signals wass retained. The outcome is a time-resolved estimate $\tilde{\mathbf{s}}_k$, including the inferred heat-transfer coefficients $h_{ij}(t)$.

	\section{Results}\label{sec7}
	
	This section is divided into two main subsections. Section \ref{sec7p1} focuses on the sloshing dynamics, analyzed in isothermal conditions, while Section reports on its thermodynamics effects in \ref{sec7p2}.
	
	\subsection{Isothermal sloshing characterization}\label{sec7p1}
	
	\subsubsection{Natural frequency estimation}
	
\begin{figure}[htbp]
		\centering
		\begin{subfigure}[b]{0.75\textwidth}
			\includegraphics[width=\textwidth]{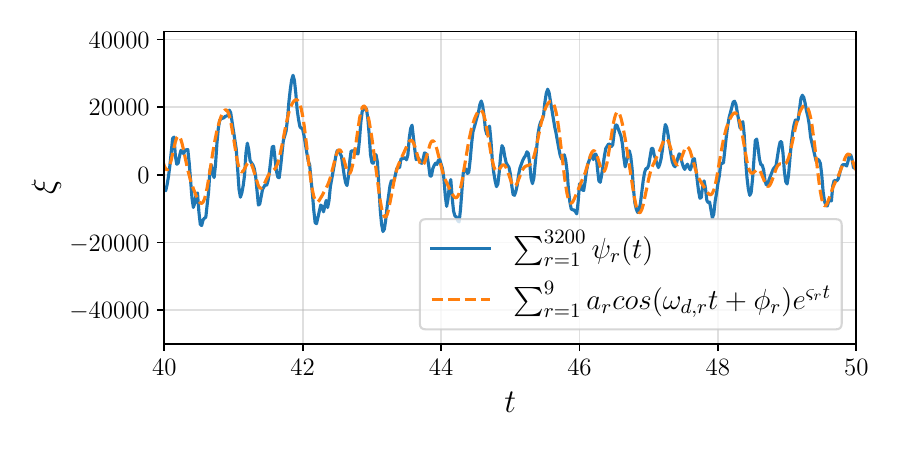}
			\caption{}
		\end{subfigure}
		\begin{subfigure}[b]{0.75\textwidth}
			\includegraphics[width=\textwidth]{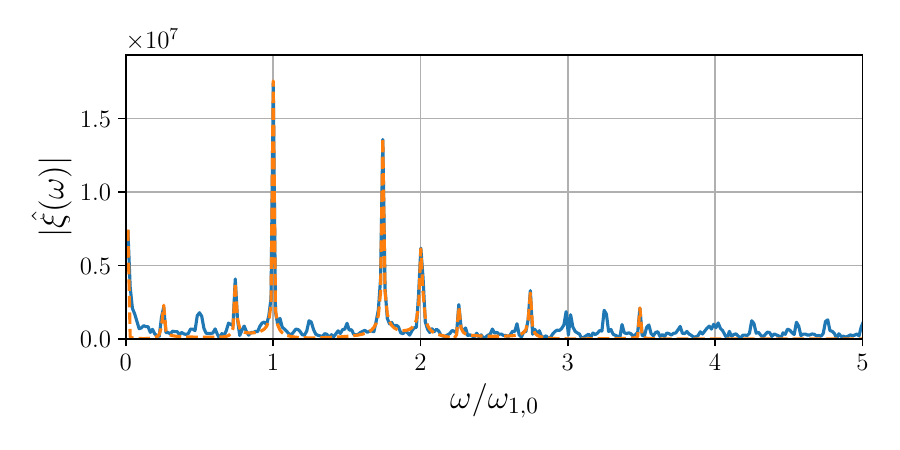}
			\caption{}
		\end{subfigure}
		\caption{Free-decay identification at 70\% fill. Top: characteristic time series $\xi(t)$ (solid) and fit $\sum_{r=1}^{n_D} a_r \cos(\omega_{d,r} t + \varphi_r)e^{-\varsigma_r t}$ (dashed). Bottom: corresponding normalized frequency spectrum $|\hat{\xi}(\omega)|$.}
		\label{fig:POD_psi_fitting_70}
	\end{figure}

\begin{figure}[htbp]
		\centering
		\begin{subfigure}[b]{0.75\textwidth}
			\includegraphics[width=\textwidth]{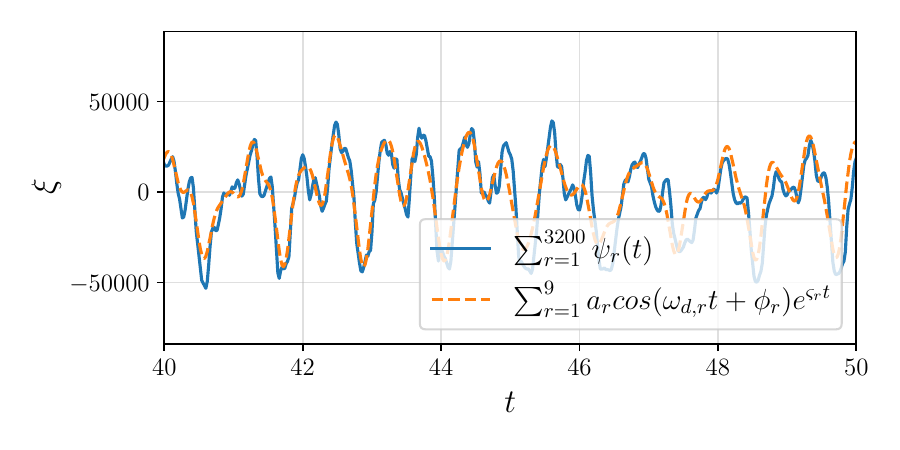}
			\caption{}
		\end{subfigure}
		\begin{subfigure}[b]{0.75\textwidth}
			\includegraphics[width=\textwidth]{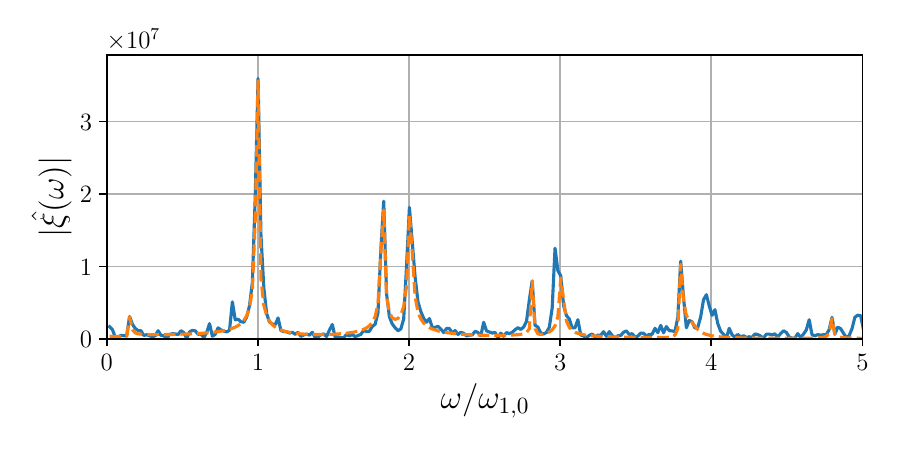}
			\caption{}
		\end{subfigure}
		\caption{Free-decay identification at 50\% fill. Top: characteristic time series $\xi(t)$ (solid) and fit $\sum_{r=1}^{n_D} a_r \cos(\omega_{d,r} t + \varphi_r)e^{-\varsigma_r t}$ (dashed). Bottom: corresponding normalized frequency spectrum $|\hat{\xi}(\omega)|$.}
		\label{fig:POD_psi_fitting_50}
	\end{figure}

	We analyze the step–like free response following the procedure of Section \ref{sec6p1}. For each fill level, we retain the leading $n_R^\ast=3200$ POD modes, capturing $>99.99\%$ of the energy. The characteristic series
	$\xi(t)=\sum_{r\le n_R^\ast}\sqrt{\sigma_r}\,\psi_r(t)$
	is fitted over a $50\,\mathrm{s}$ free–decay window using a linear combination of $n_D=9$ damped cosines in \eqref{xi_eq}.
	The $\xi$–fit scores are $R^2=0.73$ (50\%) and $R^2=0.83$ (70\%). \autoref{fig:POD_psi_fitting_70} and \autoref{fig:POD_psi_fitting_50} shows the fit over the last $10\,\mathrm{s}$ of each record for readability and the associated frequency spectrum at 70 and 50\% filling respectively. At 70\% fill, the response is dominated by two damped frequencies, $\omega_{d,1}$ and $\omega_{d,2}\!\approx\!1.70\,\omega_{d,1}$, with harmonic content at integer combinations $n\,\omega_{d,1}+ m\,\omega_{d,2}$. The strongest secondary peak occurs at $2\,\omega_{d,1}$, corresponding to the first harmonic of the leading component. At 50\% fill, the two first peaks are found at $\omega_{d,1}$ and $\omega_{d,2}\!\approx\!1.8\,\omega_{d,1}$. We note that for this fill level the third peak, which corresponds to the first harmonic of the leading component $\omega_{d,3}=2\,\omega_{d,1}$, shows an amplitude equivalent to the one of $\omega_{d,2}$. 
	
	\begin{figure*}[htbp]
		\centering
		\begin{subfigure}[b]{0.49\linewidth}
			\centering
			\includegraphics[width=\linewidth]{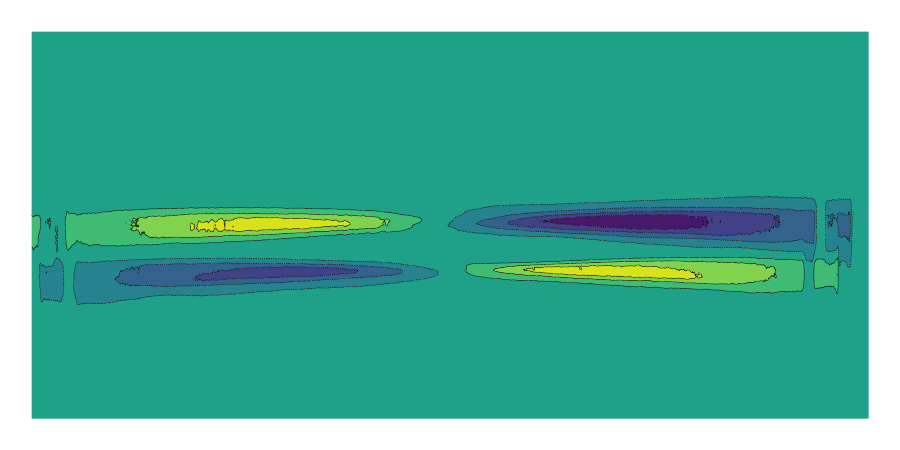}
			\caption{}
		\end{subfigure}
		\begin{subfigure}[b]{0.49\linewidth}
			\centering
			\includegraphics[width=\linewidth]{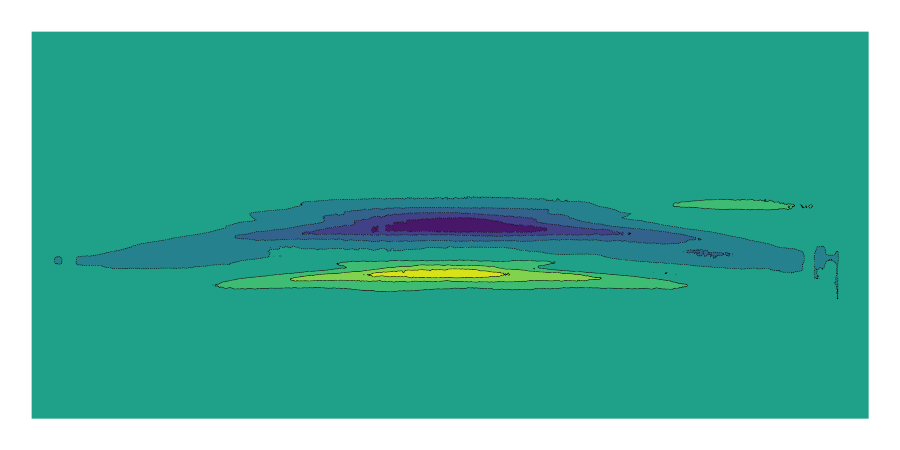}
			\caption{}
		\end{subfigure}
		\begin{subfigure}[b]{0.49\linewidth}
			\centering
			\includegraphics[width=\linewidth]{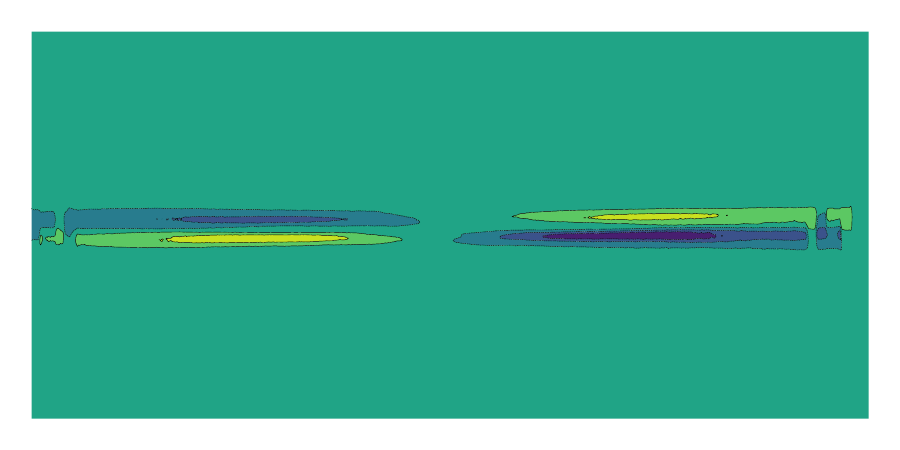}
			\caption{}
		\end{subfigure}
		\begin{subfigure}[b]{0.49\linewidth}
			\centering
			\includegraphics[width=\linewidth]{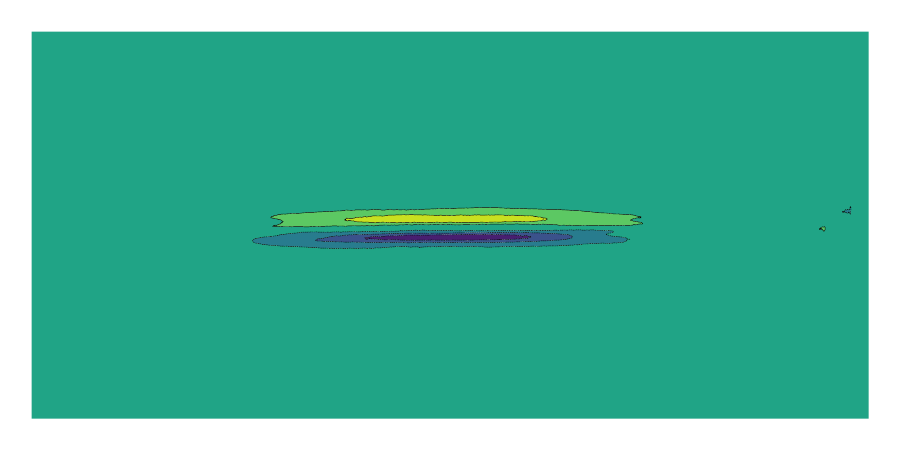}
			\caption{}
		\end{subfigure}
		\caption{Spatial structures at 50\% filling (top row) and 70\% filling (bottom row) obtained by least–squares projection of the image stack onto the fitted temporal basis $\mathbf{\Psi}_D$. Structures associated with \(\omega_{d,1}\) (left) and  with \(\omega_{d,2}\) (right)}
		\label{fig:Phi_fitting}
	\end{figure*}
	
	The fitting as per \eqref{eq:free_decay} yields $\zeta_n \ll 1$, we therefore find a negligible correction of the damped frequencies so that $\omega_d\approx\omega_n$.
	
	The spatial structures $\psi_{D,1}$ and $\psi_{D,2}$, obtained by least–squares projection of the grayscale snapshot matrix $\mathbf{G}$ onto the fitted temporal basis $\mathbf{\Psi}^D$ (see \eqref{phiD}), are shown in \autoref{fig:Phi_fitting} for 70 and 50\% filling respectively. For both fill levels, $\omega_{d,1}$ maps to an antisymmetric planar–wave pattern consistent with the expected first longitudinal mode $(1,0)$, while $\omega_{d,2}$ maps to a symmetric central standing-wave pattern consistent with the second longitudinal mode $(2,0)$. No clear two–crest pattern, which could be attributable to $(3,0)$, is observed above noise in the remaining peaks. 
	
	The associated non-dimensional natural frequencies and their damping ratios are summarized in \autoref{tab:dimless_nat_freq}.
	
	\begin{table}[htbp]
		\centering
		\begin{tabular}{ccccccc}
			\toprule
			
			$H/2R$& {$\Omega_{1,0}^2$} & {$\zeta_{1,0}$} 
			& {$\Omega^2_{2,0}$} & {$\zeta_{2,0}$} \\[2pt]
			\midrule
			0.50 &  0.26 & $2.6\times10^{-3}$&  0.80 & $4.7\times10^{-3}$   \\
			0.70 &  0.33 & $1.8\times10^{-3}$   & 0.99 &  $2.1\times10^{-3}$  \\
			\bottomrule
		\end{tabular}
		\caption{Values of dimensionless natural frequencies $\Omega^2_{n,m}=\omega^2_{n,m}R/g$ and damping ratio $\zeta_{n,m}$ obtained by analysis of the sloshing response under a longitudinal step-like excitation of the tank}
		\label{tab:dimless_nat_freq}
	\end{table}
	
	By inserting the experimentally determined frequencies into the parametric‐stability diagram (see \autoref{fig:stab_bound}), the instability regions shift noticeably compared to the potential‐flow prediction of the flat-ended geometry. In the 50\% fill case, $S0$ lies outside any unstable region, whereas both $S1$ and $S2$ fall into the (2,0) instability area. At 70 \% fill, the condition ${\omega_e}/{2} = 0.96\,\omega_{(2,0)}\,$ places $S3$, $S4$ and $S5$ within the primary (2,0) instability region. 
	\begin{figure*}[htbp]                 
		\centering
		\begin{subfigure}[b]{0.32\linewidth} 
			\includegraphics[width=\linewidth]{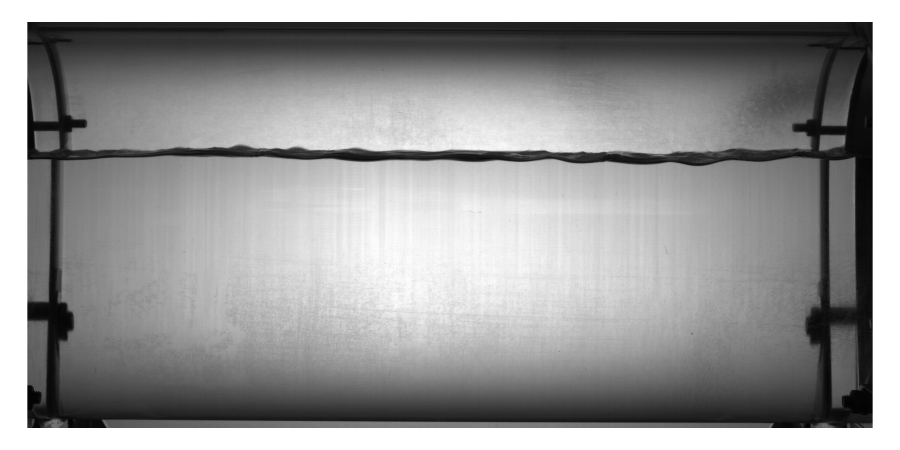}
			\caption{}
			\label{fig:iso_ripples_70}
		\end{subfigure}  
		\begin{subfigure}[b]{0.32\linewidth} 
			\includegraphics[width=\linewidth]{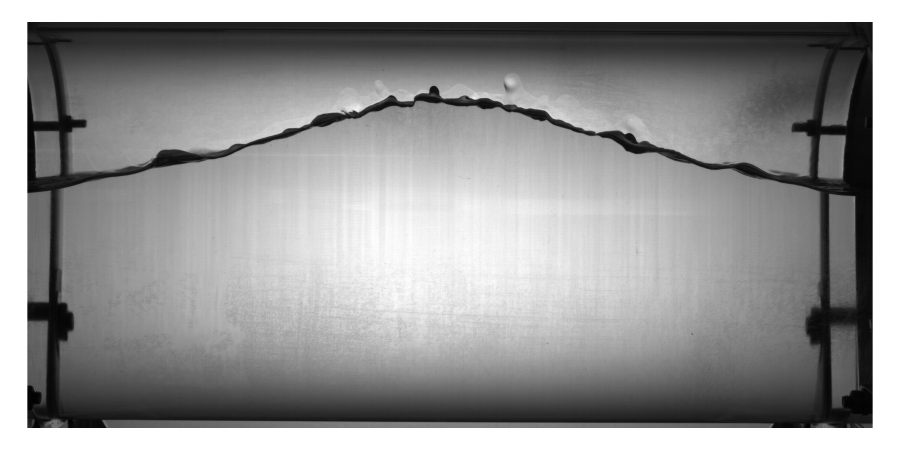}
			\caption{}
			\label{fig:iso_mode20_70_satured}
		\end{subfigure}
		\begin{subfigure}[b]{0.32\linewidth}
			\includegraphics[width=\linewidth]{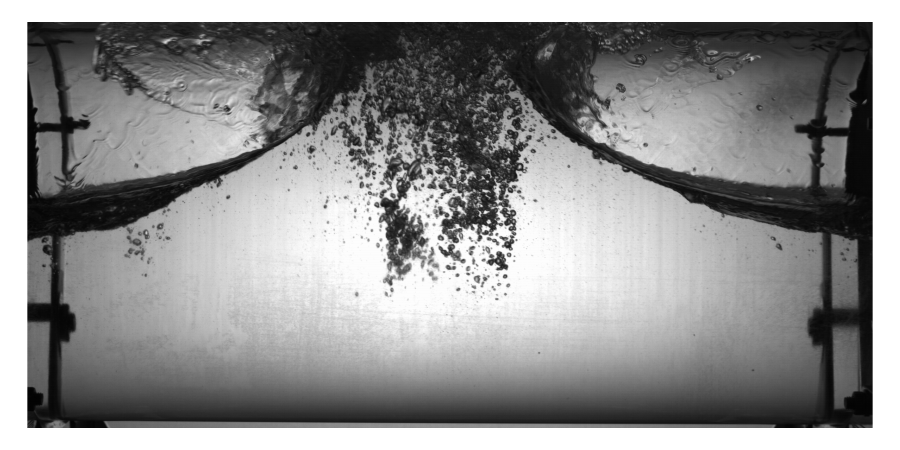}
			\caption{}
			\label{fig:iso_mode_20_70}
		\end{subfigure}
		\begin{subfigure}[b]{0.32\linewidth} 
			\includegraphics[width=\linewidth]
			{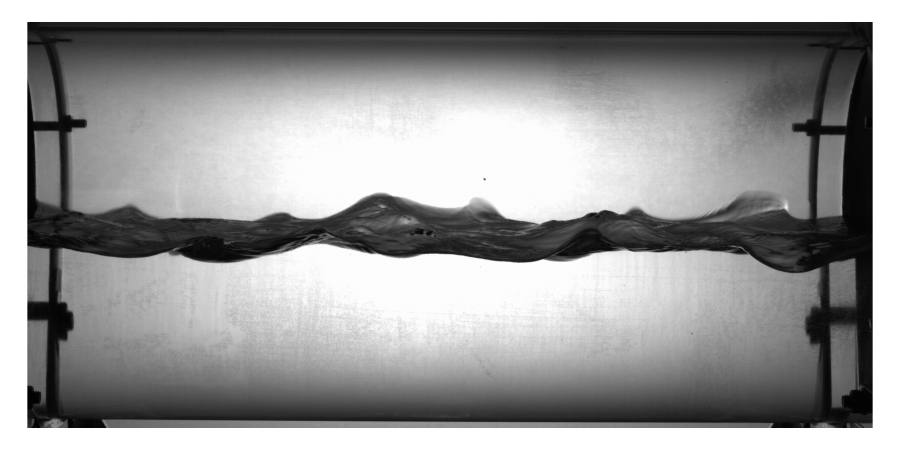}
			\caption{}
			\label{fig:iso_ripples_50}
		\end{subfigure}  
		\begin{subfigure}[b]{0.32\linewidth} 
			\includegraphics[width=\linewidth]
			{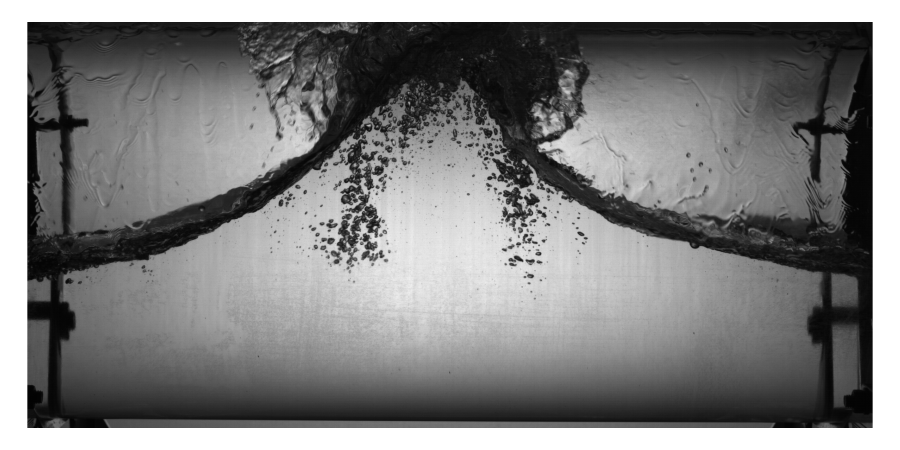}
			\caption{}
			\label{fig:iso_mode20}
		\end{subfigure}
		\begin{subfigure}[b]{0.32\linewidth}
			\includegraphics[width=\linewidth]
			{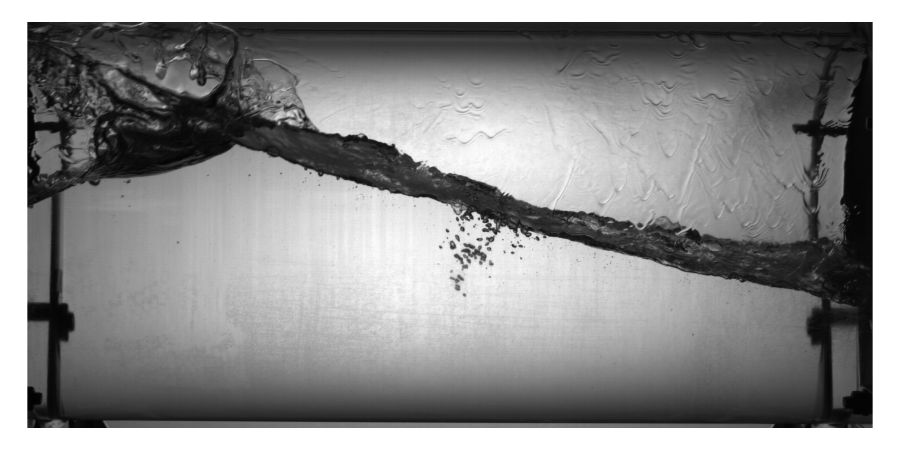}
			\caption{}
			\label{fig:iso_mode_10}
		\end{subfigure}\\
		
		\caption{Back-lit lateral snapshots of the liquid free surface under vertical vibration. Top row: bulk liquid motion at 70\% filling: 
			(a) intial interface ripples at test condition $S3$; 
			(b) saturated standing wave corresponding to modal shape $(2,0)$ in $S3$; 
			(c) strong central-jet impinging the tank ceiling corresponding to modal shape $(2,0)$ at $S4$. 
			Bottom row: bulk liquid motion at 50\% filling:  
			(a) saturated Small-amplitude ripples at test condition $S0$; 
			(b) central jet impinging the roof corresponding to modal shape $(2,0)$ at $S1$; 
			(c) quasi-planar wave corresponding to modal shape $(1,0)$ at $S1$.}
		\label{fig:iso_screen}
	\end{figure*}
	
	\subsubsection{Sloshing dynamics and mode competition}
	
	We report here on the response of the liquid interface for the isothermal experimental conditions detailed in \autoref{tab:ini_cond}. 
	For both fill levels, the observed liquid response is in line with the stability diagrams in \autoref{fig:stab_bound_exp_50_percent} and \autoref{fig:stab_bound_exp_70_percent}.  
	
	For the case at \(\,70\%\) fill level, because the excitation frequency is set very close to \(2\omega_{(2,0)}\), every test condition develops a central standing wave, corresponding to the modal response \((2,0)\). The onset of the motion is characterized by ripples on the interface as shown in \autoref{fig:iso_ripples_70}, followed by a brief growth phase of a central standing wave until the free surface steepens into a single axis-centered jet illustrated in \autoref{fig:iso_mode20_70_satured}. In the weakest-forcing case (\(S3\)), the jet saturates at a moderate amplitude and stabilizes without reaching the tank roof. This behavior suggests that, while the acceleration level is sufficient to trigger the mode $(2,0)$, it only marginally exceeds the instability threshold. For stronger accelerations, the jet impacts the tank roof, producing interface break-up and splashing, while retaining a clear \((2,0)\) modal shape throughout the entire forcing as illustrated in \autoref{fig:iso_mode_20_70}.  
	
\begin{figure*}[htbp]                
		\centering
		\begin{subfigure}[b]{0.49\linewidth} 
			\includegraphics[width=\linewidth]{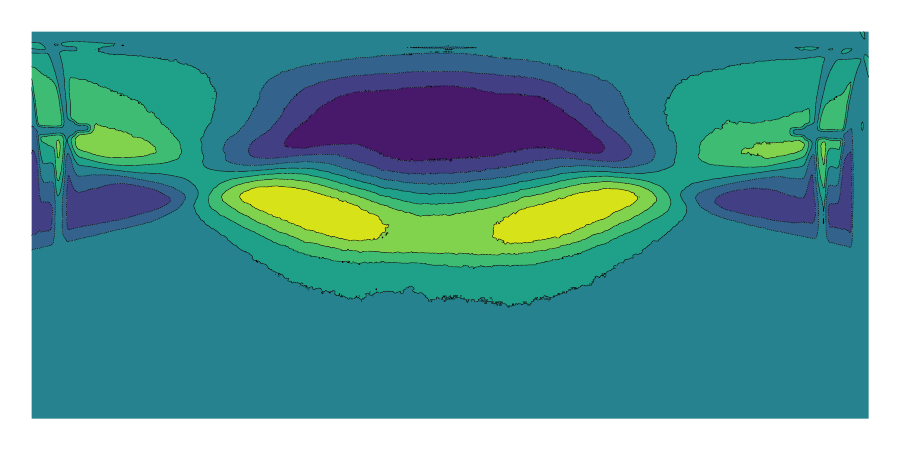}
			\caption{}
			\label{fig:S3_pod_0}
		\end{subfigure}  
		\begin{subfigure}[b]{0.49\linewidth} 
			\includegraphics[width=\linewidth]{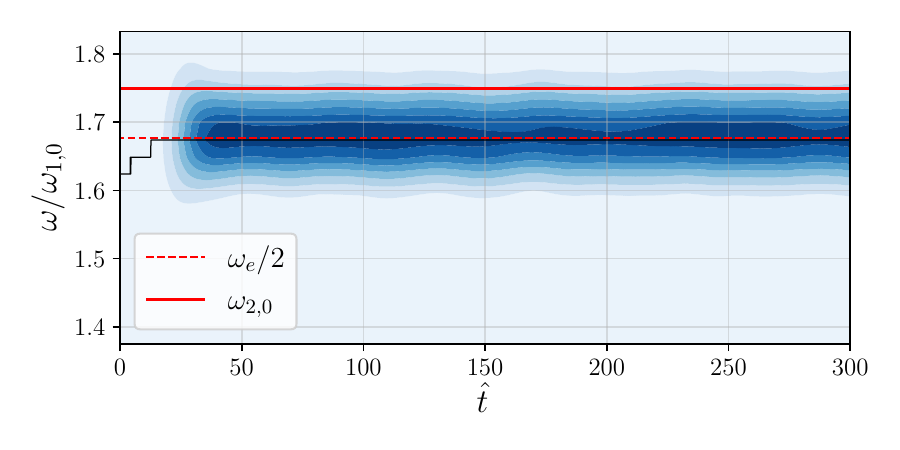}
			\caption{}
			\label{fig:S3_pod_0_cwt}
		\end{subfigure}  
		\caption{Leading POD mode and its time–frequency content for case $S4$ (70\% fill, vertical harmonic excitation at $\hat a_v=0.35$). 
			(a) Spatial structure $\phi_1$ and (b) CWT scalogram of the corresponding temporal coefficient $\psi_1(\hat t)$.}
	\end{figure*}

	Decomposing high-speed video recordings with POD provides insights into the flow dynamics, revealing an axisymmetric jet as the most energetic spatial structure across all cases (see \autoref{fig:S3_pod_0} for test $S3$). The associated temporal coefficients $\psi_r(\hat t)$ are then analyzed using CWT to determine the frequencies involved.
	The analysis employs a complex Morlet wavelet (center frequency 6.0 Hz, bandwidth 2.5 Hz), with scales sampled on a logarithmic grid using 48 voices per octave, to resolve the evolution of modal frequencies over time. The leading temporal coefficient was found to be characterized by a fixed frequency at half the forcing \(\omega_e/2\simeq\omega_{(2,0)}\), as expected in the primary response to parametric excitation as depicted in the scaleogram in \autoref{fig:S3_pod_0_cwt}. The following modes exhibit the same frequency or higher harmonics. Interestingly, several lower-order POD modes display a subharmonic at \(\omega_{(2,0)}/2\); however, this frequency does not coincide with any eigenmode at \(70\%\) filling and its energy content is more than an order of magnitude smaller than that of the primary jet mode, implying minor impacts on the global dynamics. 

	At 50\% filling, the test $S0$ shows limited bulk liquid motion when subjected to the low-level excitation, in line with the stability diagram in \autoref{fig:stab_bound_exp_50_percent} that collocates this test case outside instability tongues. Only small wave ripples are visible on the free-surface and their amplitude remains constant over time, as depicted in \autoref{fig:iso_ripples_50}.  
	By contrast, tests $S1$ and $S2$ appear to yield to a strong \((2,0)\) modal response. The liquid motion follows the same pattern than the 70\% filling, characterized by small initial ripples, followed by a standing wave growth until the formation of a strong single central vertical jet.  The jet rapidly impinges on the tank roof, producing a breakup and splashing.  During this stage the global flow motion is dominated by the jet-like motion, depicted in \autoref{fig:iso_mode20}.
	
	After several cycles of jet impacts on the tank roof, both $S1$ and $S2$ undergo a reorganization of the bulk liquid motion. First, the splashing is observed to off center the jet, then bulk motion quickly reorganizes into a longitudinal planar wave characteristic of the \((1,0)\) sloshing mode, illustrated in \autoref{fig:iso_mode_10}. 
	In case $S1$, this planar motion dominates the flow dynamics for the rest of the experiment, with only intermittent, low-amplitude standing wave patterns. In $S2$ ($\hat a_v$=0.49), the bulk liquid motion continuously alternates between planar and standing waves. This behavior repeats quasi-periodically between several cycles (typically $5$-$15$) of the \((2,0)\) mode and a shorter sequence ($1$-$3$ cycles) of the \((1,0)\) mode, suggesting mode interaction. 
	
	These transitions in $S1$ and $S2$ are associated with subharmonic resonance: the planar wave oscillates at about half the frequency of the central jet. This frequency halving is the classical period-doubling phenomenon \citep{Ciliberto, Bardazzi_2024, aston, Ibrahim_2015_review}, arising from nonlinear sloshing effects. In the present configuration the forcing satisfies $\omega_e \approx 2\omega_{2,0} \approx 4\omega_{1,0}$, so the response at $\omega_e/2$ excites mode $(2,0)$; quadratic couplings then enable a $2{:}1$ internal resonance of mode $(1,0)$, which drives symmetry breaking and energy transfers between the jet-like and planar patterns observed in $S1$ and $S2$.

\begin{figure*}[htbp]                
		\centering
		\begin{subfigure}[t]{0.49\linewidth} 
			\includegraphics[width=\linewidth]{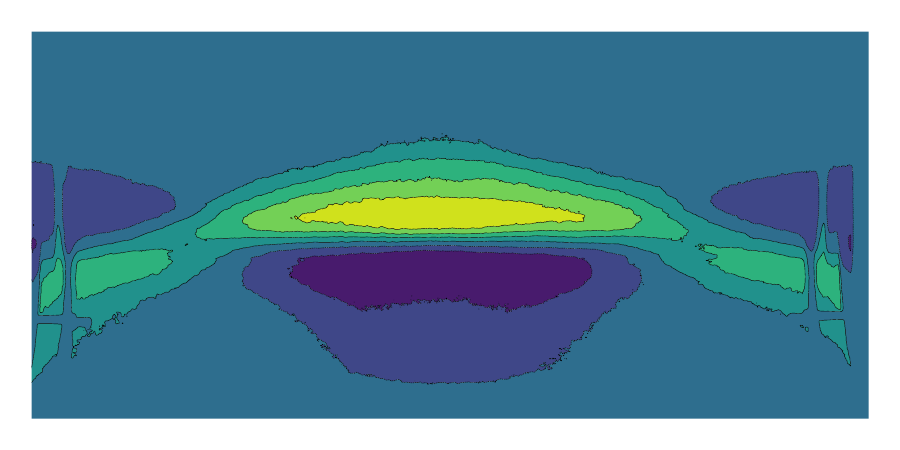}
			\caption{}
			\label{fig:S1_pod_0}
		\end{subfigure}  
		\hfill
		\begin{subfigure}[t]{0.49\linewidth} 
			\includegraphics[width=\linewidth]{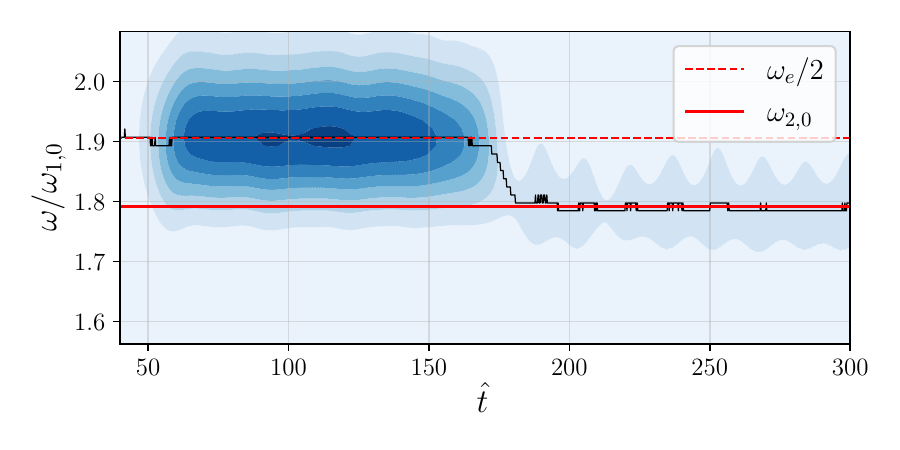}
			\caption{}
			\label{fig:S1_pod_0_cwt}
		\end{subfigure}  
		\begin{subfigure}[t]{0.49\linewidth} 
			\includegraphics[width=\linewidth]{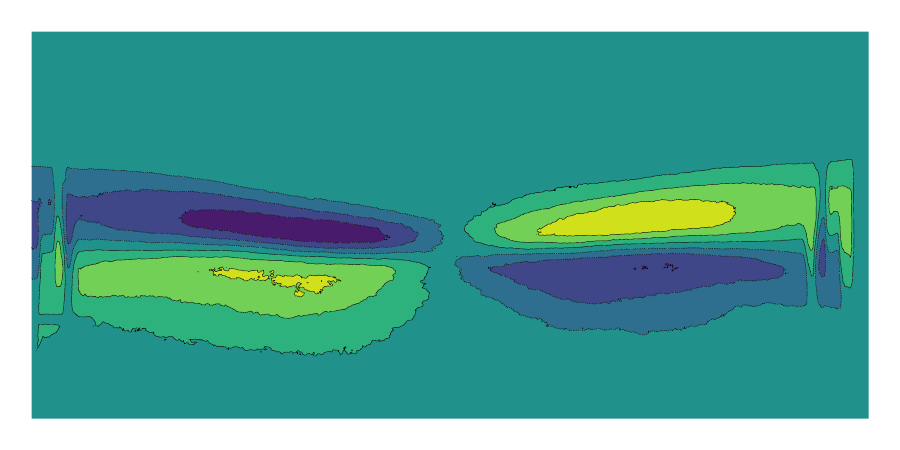}
			\caption{}
			\label{fig:S1_pod_2}
		\end{subfigure}  
		\hfill
		\begin{subfigure}[t]{0.49\linewidth} 
			\includegraphics[width=\linewidth]{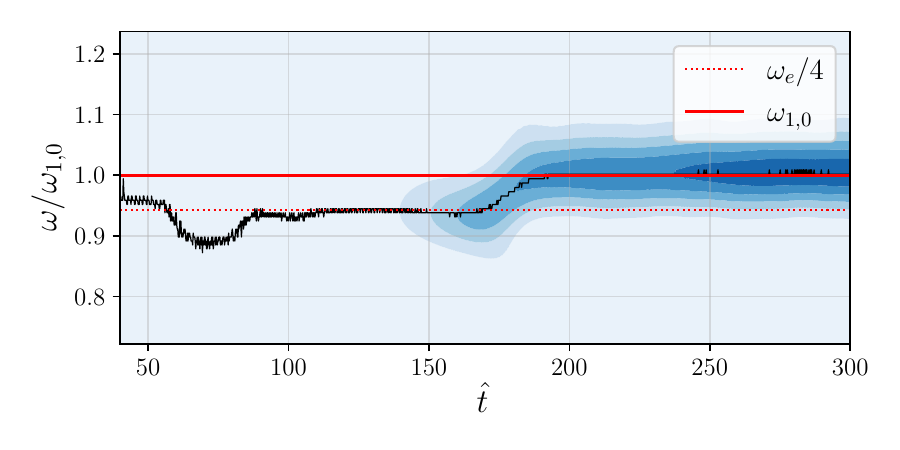}
			\caption{}
			\label{fig:S1_pod_2_cwt}
		\end{subfigure}  
		\caption{POD–CWT analysis for $S1$. (a,c) Spatial modes $\phi_1,\phi_3$. 
			(b,d) Scalograms of $\psi_1,\psi_3$: the solid black line indicates the instantaneous ridge frequency, the dashed vertical lines indicate subharmonic of the excitation frequency and solid ones the natural frequencies of the tank.}
		
		\label{fig:S1_pod_modes}
	\end{figure*}

\begin{figure*}[htbp]                
		\centering
		\begin{subfigure}[b]{0.49\linewidth} 
			\includegraphics[width=\linewidth]{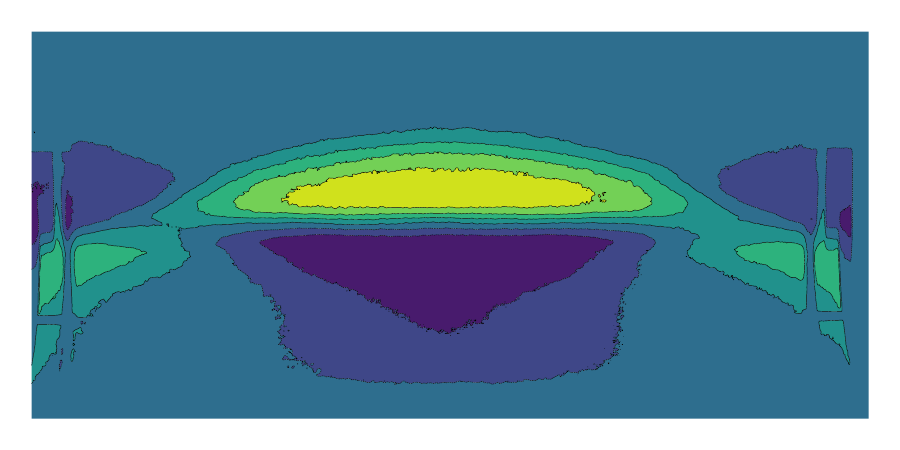}
			\caption{}
			\label{fig:S2_pod_0}
		\end{subfigure}  
		\hfill
		\begin{subfigure}[b]{0.49\linewidth} 
			\includegraphics[width=\linewidth]{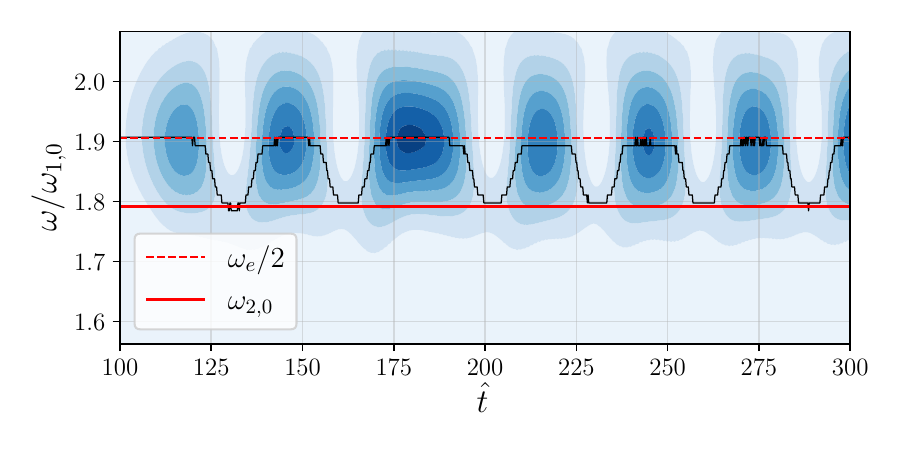}
			\caption{}
			\label{fig:S2_pod_0_cwt}
		\end{subfigure}  
		\begin{subfigure}[b]{0.49\linewidth} 
			\includegraphics[width=\linewidth]{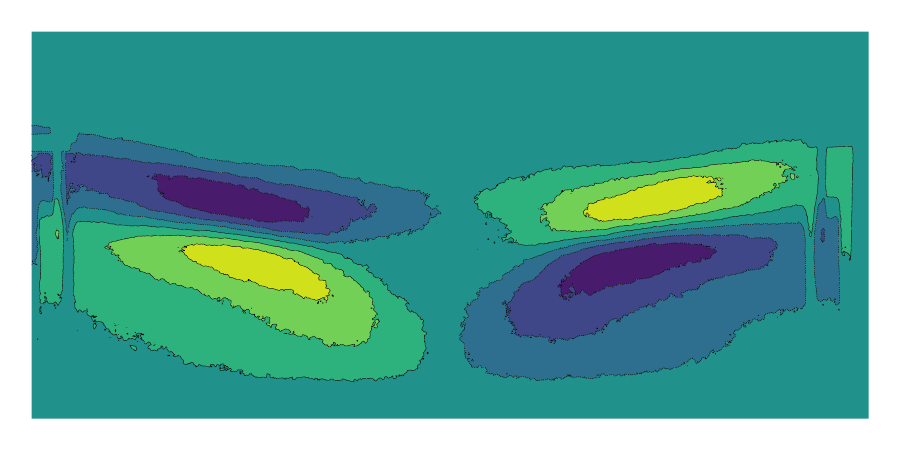}
			\caption{}
			\label{fig:S2_pod_2}
		\end{subfigure}  
		\hfill
		\begin{subfigure}[b]{0.49\linewidth} 
			\includegraphics[width=\linewidth]{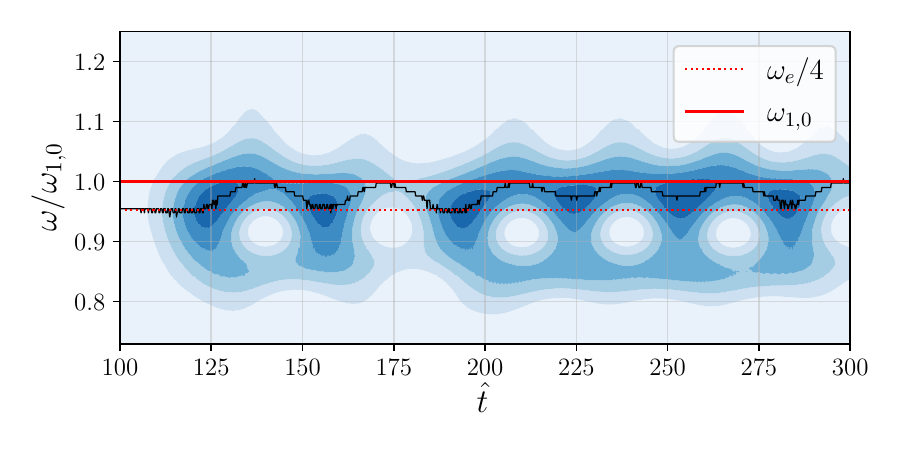}
			\caption{}
			\label{fig:S2_pod_2_cwt}
		\end{subfigure}  
		\caption{POD–CWT analysis for $S2$. (a,c) Spatial modes $\phi_1,\phi_3$. 
			(b,d) Scalograms of $\psi_1,\psi_3$: the solid black line indicates the instantaneous ridge frequency, the dashed vertical lines indicate subharmonic of the excitation frequency and solid ones the natural frequencies of the tank.}
		\label{fig:S2_pod_modes}
	\end{figure*}

	To analyze the transition mechanisms in $S1$ and $S2$, the grayscale snapshot matrix of the high-speed video recordings is decomposed using POD. In both tests the spatial structures $\phi_r$ exhibit similar patterns, with the first and second POD modes capturing the axisymmetric jet (see Figure~\ref{fig:S1_pod_0} and \ref{fig:S2_pod_0}), and the third mode representing the longitudinal \((1,0)\) wave as depicted in Figure~\ref{fig:S1_pod_2} and \ref{fig:S2_pod_2}.

	A CWT analysis reveals that, for the test case $S1$, the jet-like mode ($r=0$) initially locks in the subharmonic $\omega_e/2$, as expected for the Mathieu-type parametric response (see \autoref{fig:S1_pod_0_cwt}). During this phase, the planar wave mode ($r=3$), shows a low amplitude energy and a stable frequency at the second subharmonic $\omega/4$ (see \autoref{fig:S1_pod_2_cwt}). Around $\hat t\simeq180$, the bulk dynamics change establishes: the jet-like mode loses energy and its instantaneous ridge drifts toward the estimated $(2,0)$ natural frequency. Simultaneously, the energy level of mode $3$ increases and its frequency shifts toward the natural planar response $\omega_{1,0}$. 
	
	In test $S2$, the onset of the motion is also characterized by a symmetric motion occurring at $\omega_e/2$, as captured by the mode 0 (see \autoref{fig:S2_pod_0_cwt}). During this initial transient, the planar mode holds a low energy level, at the second subharmonic of the excitation frequency $\omega_e/4$. The transition from standing waves to planar waves takes place within a shorter time than in  $S1$, namely approximately 15 periods. This change in sloshing dynamics is associated with a frequency shift in the leading POD mode toward its natural frequency, $\omega_{2,0}$ and $\omega_{1,0}$ for mode 0 and 2, respectively. After a few periods of planar wave, the bulk dynamics revert to a standing wave, associated with frequencies shifting to subharmonic of the forcing. This behavior repeats quasi-periodically with a period $T_{\mathrm{alt}}\approx 30\pi/\omega_e$.

	\subsection{Non-isothermal sloshing characterization}\label{sec7p2}
	
\begin{figure*}[t]                
		\centering
		\begin{subfigure}[b]{\textwidth} 
			\includegraphics[width=\textwidth]{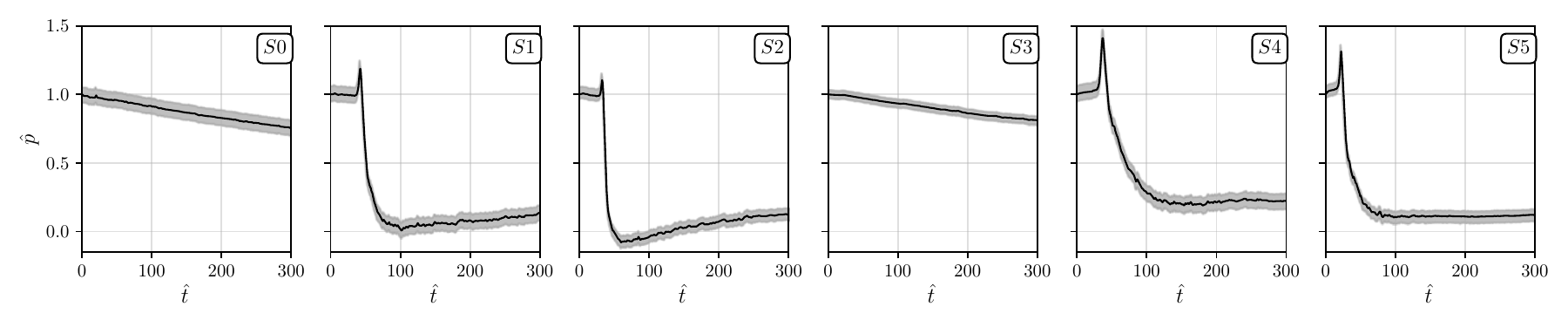}
			\caption{}
			\label{fig:P-nd}
		\end{subfigure}  
		\begin{subfigure}[b]{\textwidth} 
			\includegraphics[width=\textwidth]{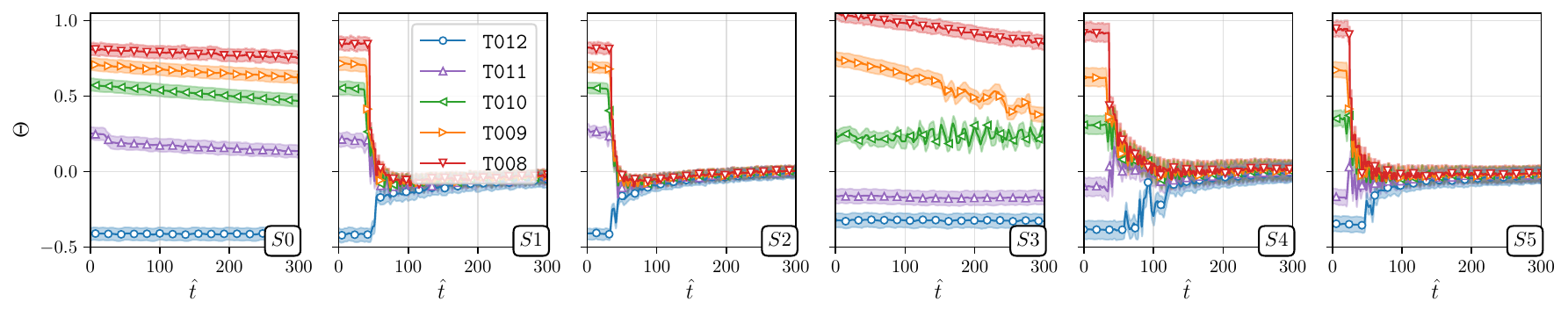}
			\caption{}
			\label{fig:T-nd}
		\end{subfigure}
		\begin{subfigure}[b]{\textwidth}
			\includegraphics[width=\textwidth]
			{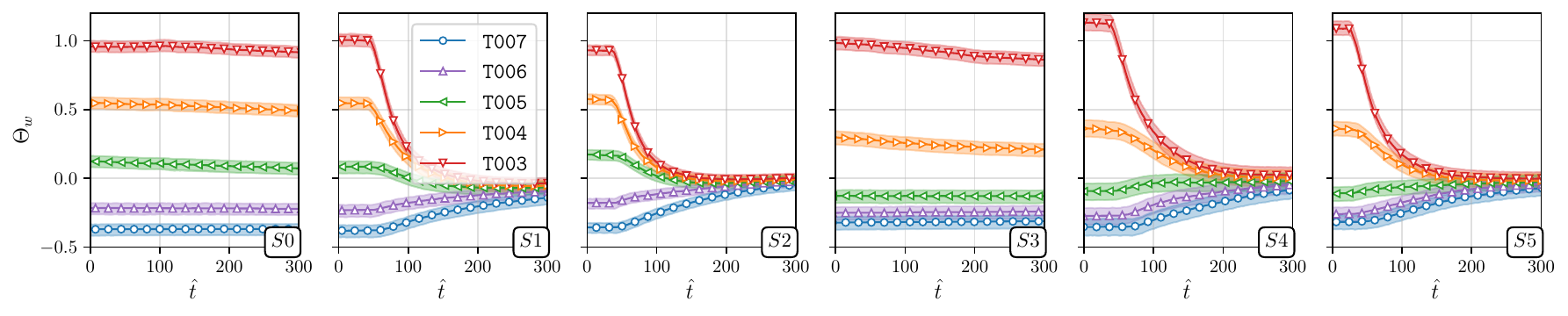}
			\caption{}
			\label{fig:Tw-nd}
		\end{subfigure}
		\caption{Non-dimensional thermodynamic evolution within the sloshing cell for the different testing conditions reported in \autoref{tab:ini_cond}. (a) Pressure, (b) fluid temperature, (c) wall temperature.
			Thermocouple locations are listed in \autoref{tab:sensor-loc}.}
		\label{fig:thermo_noniso}
	\end{figure*}

	Pressure and temperature histories from the sloshing experiments are shown in dimensionless form, normalized as described in Sec.~\ref{sec4}.  In the following, the results are presented from the start of the sloshing event, marked by $\hat{t}=0$. \autoref{fig:P-nd} depicts the scaled absolute pressure, whereas \autoref{fig:T-nd} and \ref{fig:Tw-nd} show the temperature readings of the fluid and the walls, respectively, from the thermocouples located in the tank centerline from the bulk liquid up to close to the roof. The vertical locations of all sensors are listed in Table~\ref{tab:sensor-loc}.

	The isothermal study showed that at 50\% filling and under the lowest acceleration ($S0$), the fluid response was limited to small amplitude free-surface ripples. Throughout the entire forcing event, the pressure decay remains comparable to that observed in the relaxation phase (i.e., at $\hat t<0$). The ullage temperature follows the same trend and decreases slowly, while the bulk-liquid temperature remains quasi-constant. A short inflection on the readings of \texttt{T011}, located in the vicinity of the interface, around 30 periods, suggests light liquid splashing at $H/2R=0.54$.
	Overall, no sloshing-induced thermal mixing or pressure drop is observed. This indicates that the small-amplitude surface waves do not enhance either wall-liquid heat transfer or vapor-liquid exchanges. 
	
	For $S1$, the pressure signal remains quasi-steady for 42 forcing periods, corresponding to the onset of the bulk motion. Then, a pressure burst occurs, followed by pressure decay that can be well approximated by an exponential law. Denoting by $t^{\star}$ the onset of this decay, the pressure evolution is well described by  
	\begin{equation}
		\hat{p}(\hat{t}) = p^\star\exp\!\left[\frac{t^{\star}-\hat{t}}{\hat \tau_p}\right],
		\label{eq:pressure_model}
	\end{equation}
	with $p^\star$ the peak pressure value and $\hat \tau_p$ characteristic dimensionless timescale of the pressure drop, fitted here to $\hat \tau_p = 10$.  After roughly $3\hat \tau_p$, corresponding here to 30 forcing periods, the pressure is within $\sim5\%$ of the final equilibrium value. 
	
	The temperature readings in the ullage fall to the saturation temperature before the actual start of the pressure drop $t^\star$. In particular, \texttt{T010} and \texttt{T009} show a delay of 4 and 2 forcing periods, respectively, corresponding to the 1 and 2 response periods of the induced jet. This indicates the progressive growth of the jet amplitude, which takes 3 periods to reach the tank roof and trigger the pressure drop. Once the temperature reaches the saturation temperature, it follows the same exponential trend as the pressure. 
	
	The outer-wall temperature readings start to evolve towards equilibrium five forcing periods after the initiation of pressure drop, i.e.\ at $\hat t \approx t^{\star}+5$.
	The time delay is attributed to the diffusive effects due to the propagation through the wall thickness. The wall temperature evolution after the mixing onset is well described by an exponential of the form:
	\begin{equation}
		\Theta_w(\hat t)=
		\Theta_w^0\,
		\exp\!\left[\frac{t^{\star}+\hat \tau_{\text{diff}}- \hat t}{\hat \tau_w}\right],
	\end{equation}
	with $\hat \tau_w$ the characteristic time scale of the wall temperature evolution and $\hat \tau_{\text{diff}}$ the diffusive lag. We fit here $\hat \tau_w=35$, as reported in \autoref{tab:timescales}. 
	
	Once the pressure has already fallen by about 50\%, i.e., roughly 10 periods, the bulk liquid temperatures start to evolve towards equilibrium too, suggesting wall-liquid heat exchanges due to strong liquid motion. As the pressure reaches equilibrium (i.e. $\hat p=0$), the liquid is at equilibrium temperature, indicating that the entire fluid is at saturation. The temperature of the wall continues evolving towards equilibrium, and a slow pressure recovery is observed, attributed to the exchange of energy between the saturated fluid mixture and the walls. Once the entire system has reached thermal equilibrium, a slow pressure recovery is still measured, attributed to the additional energy passed by the heaters located on the tank upper wall.
	
	The observed thermodynamic evolution induced by the sloshing event can be described by several phases:
	\begin{enumerate}
		\item Onset of the mode $(2,0)$: the standing wave grows and reaches the tank roof within two forcing periods, leading to a pressure burst to $\hat p = p^\star$ at $\hat t = t^\star$,
		\item Ullage destratification and pressure drop: for $\hat t> t^\star$, the pressure and ullage temperature follow an exponential decay toward equilibrium with characteristic time $\hat \tau_p$. The characteristic time of pressure decay is one order of magnitude shorter than in relaxation, suggesting that the pressure drop is here associated with a much faster mechanism, namely the sloshing induced mixing.
		\item  Outer-wall conduction response: after a delay $\hat \tau_{\mathrm{diff}}\approx5$, the external wall temperature starts to evolve toward equilibrium with a slower exponential trend of characteristic time $\hat \tau_w\approx35$,
		\item Wall-liquid heat exchange: at $\hat t\approx t^\star+\hat \tau_p$, the liquid temperature start to rise towards equilibrium. 
		\item Complete fluid mixing: at $\hat t \approx t^\star + 3 \hat \tau_p$, the pressure and fluid temperature reached the equilibrium point (i.e. $\hat p = \Theta =0$), and both vapour and liquid reach saturated conditions.
		\item Pressure recovery: the saturated fluid mixture receives heat from the top heaters, leading to a small pressure recovery. 
	\end{enumerate}

	In test $S2$, the dimensionless vertical acceleration is increased to $\hat a_v = 0.49$.  The same six-stage thermodynamic sequence described above is reproduced, but on a slightly shorter timescale.  The pressure drop starts at $t^{\star}=33$ and proceeds more strongly, with a fitted exponential timescale of $\hat \tau_{p}=4$. The thermal destratification is nearly complete in 12 forcing periods.  The ullage thermocouple \texttt{T010} reaches saturation only two forcing periods before $t^{\star}$, while \texttt{T009} does so exactly at $t^{\star}$. This indicates a faster growth of the middle-jet amplitude, reaching the tank roof in only 2 forcing period. These faster processes of thermal mixing suggest an enhanced ullage/liquid heat transfer compared to $S1$.  Conversely, the external wall temperature begins to evolve towards equilibrium after the same diffusive delay, $\hat \tau_{\text{diff}}\approx 5$, and its subsequent evolution is again well described by an exponential trend with a dimensionless characteristic time $\hat \tau_{w}=35$. Consequently, while the ullage–liquid heat exchange appears to be enhanced by the forcing amplitude, the wall–mixture heat transfer remains constant, suggesting that it is effectively independent of the imposed acceleration within the investigated conditions. In other words, the thermal response of the solid wall seems governed primarily by its own intrinsic diffusive timescale rather than by the dynamics of the fluid forcing. Normalizing the time constant $\hat \tau_w=\tau_w\,\omega_e/2\pi$ by the diffusive scales gives an alternative Fourier number $\widetilde{\mbox{Fo}}=\mathrm{Fo}\hat \tau_w=\alpha_w\tau_w/\delta_w^2$. The values of the modified Fourier number are reported in \autoref{tab:timescales}. We report $\widetilde{\textrm{Fo}}\approx1$,  confirming that the heat diffusion across the thickness $\delta_w$ sets the thermal response of the wall.

	\begin{table}[htbp]
		\centering
		\begin{tabular}{cccccc}
			\toprule
			\textbf{ID} & $t^{\star}$ [-] & $p^\star$ [-]& $\hat \tau_{p}$ [-] & $\hat \tau_{w}$ [-] & $\widetilde{\mbox{Fo}}=\mathrm{Fo}\hat \tau_w$ [-] \\
			\midrule
			$S0$ & -  & - & - & - & - \\
			$S1$ & 43 & 1.19 & 10 & 35 & 1.16 \\
			$S2$ & 34 & 1.10 & 4 & 35 & 1.16 \\
			$S3$ & -  & - & - & - & - \\
			$S4$ & 37 & 1.51 & 23 & 37 & 1.22 \\
			$S5$ & 22 & 1.37 & 10 & 34 & 1.12 \\
			\bottomrule
		\end{tabular}
		\caption{Characteristic dimensionless time scales of the non-isothermal sloshing response. $t^{\star}$ is the onset of the pressure drop (in forcing periods), $p^{\star}$ the pressure burst amplitude, and $\hat\tau_{p}$ and $\hat\tau_{w}$ the exponential decay time scales for pressure and wall temperature, respectively. Modified Fourier number $\widetilde{\mathrm{Fo}}=\mathrm{Fo}\,\hat\tau_{w}=\alpha_w\tau_w/\delta_w^2$. A dash indicates that no thermal destratification was observed.}
		
		\label{tab:timescales}
	\end{table}
	
	At the 70\% fill level, the lowest acceleration case ($S3$) reveals evidence of splashing or low amplitude middle-jet development. Indeed, the lower ullage thermocouples (\texttt{T009} and \texttt{T010}) show temperature oscillations as off $\hat t \approx 110$, suggesting intermittent wetting from the bulk liquid motion. This observation is not associated with any inflection on the pressure, or liquid temperature readings (\texttt{T011} and \texttt{T012}). These quantities, along with the highest ullage thermocouple (\texttt{T008}) continue to evolve at the same rate than during the pre-sloshing relaxation phase, suggesting that the bulk liquid did not reach the tank upper wall and that the liquid motion did not impact the thermal stratification. These observations are in line with the findings of \cite{Das2009-vertical}, who showed that stable jet formation in an upright cylinder had a negligible impact on thermal stratification. 
	
	At higher accelerations, the destratification process at 70\% fill (tests $S4$ and $S5$) closely follows the six-step sequence observed at 50\% fill. The pressure decay rate again increases with forcing amplitude, exhibiting a similar speed-up ratio of approximately 2.5 between $\hat a_v=0.35$ and $\hat a_v=0.49$.
	However, several key differences emerge. Notably, the initial pressure burst is more pronounced at the 70\% fill level. This is attributed to the smaller ullage volume, where a similar evaporation rate produces a larger pressure rise.
	Conversely, the subsequent pressure decay is slower at 70\% fill than at 50\% for the same forcing acceleration. This behavior is attributed to the reduced interface area and the lower level of subcooling and superheating of the liquid and ullage phases at 70\% filling, as reflected by the Jakob number.
	Finally, while the thermal response of the wall remains consistent across all cases (i.e. $\hat \tau_{\text{diff}}\approx5$ and $\hat \tau_w\approx35$), the onset of destratification ($t^\star$) occurs sooner at 70\% fill, indicating that the higher liquid level facilitates the development of the development of the standing wave and sloshing-induced thermal destratification.

	\subsubsection{Heat transfer estimation}
	
	The EKF-based inverse method provides a time-resolved estimation of the heat transfer coefficients for each experimental run. This section presents the results for the cases that exhibited strong destratification: \( S1 \), \( S2 \), \( S4 \), and \( S5 \).
	
	\begin{figure*}[hbtp]              
		\centering
		\includegraphics[width=\textwidth]{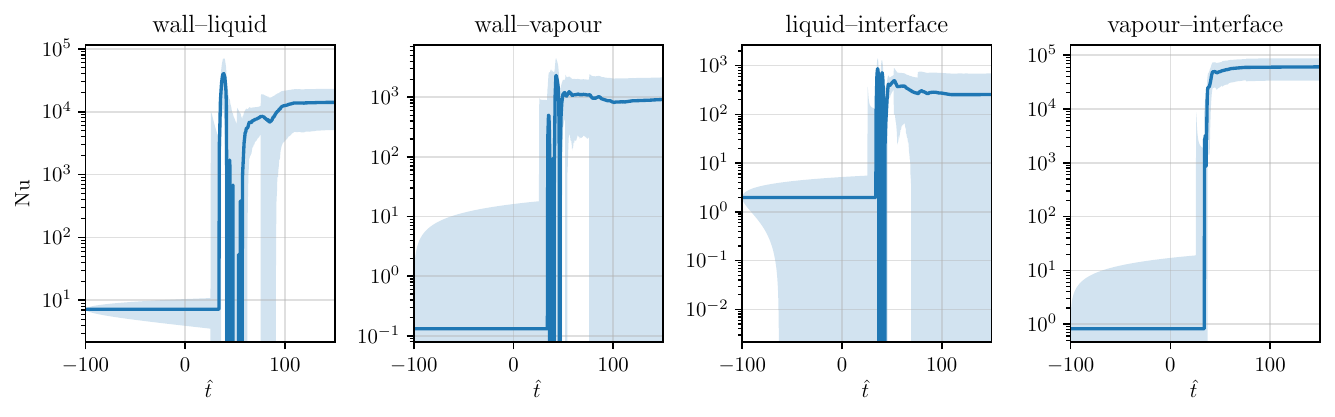}
		\caption{Evolution of the four driving Nusselt numbers for case $S2$ and associated filter uncertainty. From left to right: $\textrm{Nu}_{w,l}$, $\textrm{Nu}_{w,v}$, $\textrm{Nu}_{l,i}$, and $\textrm{Nu}_{v,i}$}
		\label{fig:Nusselt_profiles}
	\end{figure*}

	A representative evolution of the inferred Nusselt numbers is shown for case \( S2 \) in \autoref{fig:Nusselt_profiles}. The results reveal two distinct regimes. During the initial relaxation phase (\( \hat{t} < 0 \)), the estimated heat transfer coefficients remain nearly constant, with uniformly low Nusselt numbers across all interfaces. This behavior persists after the onset of forcing, up to the beginning of the thermal destratification at \( \hat{t} = t^\star \). Once destratification begins, the filter captures a sharp, step-like increase in all Nusselt numbers-by one to four orders of magnitude, depending on the forcing conditions. This pronounced jump marks the transition from a conduction- and natural-convection–dominated regime to one governed by intense forced convection driven by bulk fluid motion.

	The EKF provides time-resolved uncertainty on the inferred coefficients via its covariance. The evolution of these uncertainty bands in \autoref{fig:Nusselt_profiles} reflects a balance between the small “allowance” for the coefficient variation added at each prediction step, i.e. the process noise $Q_{h_{ij}}$, and the information extracted at the update step, which depends on how the sensitivity of the measurements on the coefficients and on the measurement noise $\mathbf R$.
	During relaxation the system is quasi-steady, temperatures and pressure are nearly constant, and the measurements carry little information about $h_{ij}$. The AEKF applies small gains, so the persistent process noise $\mathbf{Q}_{h_{ij}}$ is not fully counteracted and the uncertainty gradually increase despite accurate thermodynamic state predictions.
	Near the pressure-drop onset ($t\!\approx\!t^\star$) we temporarily inflate the coefficient process noise $\mathbf{Q}_{h_{ij}}$ to relax the constant-parameter assumption and let the filter track the rapid regime change. This raises the predicted covariance and the Kalman gain, so $h_{ij}$ can follow the sharp step, and the uncertainty bands widen accordingly. After the short transient $Q_{h_{ij}}$ is reset, but the inflated covariance propagates forward and can contract only through informative updates. As the system approaches equilibrium, the temperature differences between control values are small, leaving the model insensitive to the heat transfer coefficient values. Consequently, the high uncertainty persists, despite accurate state predictions.

	Table~\ref{tab:nusselt} summarizes the time-averaged Nusselt numbers for the pre-sloshing relaxation phase, $\overline{\mathrm{Nu}}^{r}_{ij}$ (i.e. $\hat t<0$), and for the sloshing-induced thermal destratification phase, $\overline{\mathrm{Nu}}^{s}_{ij}$ (i.e. $\hat t\geq t^\star$).

	\begin{table}[hbtp]
		\centering
		\begin{tabular}{lcccc}
			\toprule
			&$S1$ &$S2$ &$S4$ & $S5$ \\
			\midrule
			
			$\overline{\mathrm{Nu}}_{wl}^r$ & $7.7$ & $7.2$ & $8.1\times10^{-1}$ & $1.3$ \\
			$\overline{\mathrm{Nu}}_{wv}^r$ & $1.4\times 10^{-2}$ & $1.3\times 10^{-1}$ &  $5.4\times10^{-1}$ & $7.6\times 10^{-2}$ \\
			$\overline{\mathrm{Nu}}_{li}^r$ & $1.9$   & $2.0$  & $3.4\times10^{-1}$ & $1.1$ \\
			$\overline{\mathrm{Nu}}_{vi}^r$ & $3.8\times10^{-1}$ & $8.3\times10^{-1}$ & $4.9$ & $7.2\times10^{-1}$ \\
			\midrule
			\addlinespace
			$\overline{\mathrm{Nu}}_{wl}^s$ & $9.4\times 10^{3}$ & $1.2\times 10^{4}$ & $6.5\times 10^{3}$ & $7.3\times 10^{3}$ \\
			$\overline{\mathrm{Nu}}_{wv}^s$ & $2.8\times 10^{3}$ & $9.3\times 10^{2}$ & $1.2\times 10^{3}$ & $4.1\times 10^{3}$ \\
			$\overline{\mathrm{Nu}}_{li}^s$ & $7.9\times 10^{1}$ & $2.8\times 10^{2}$ & $9.0$ & $1.4\times 10^{1}$ \\
			$\overline{\mathrm{Nu}}_{vi}^s$ & $8.8\times 10^{2}$ & $5.7\times 10^{4}$ & $3.7\times 10^{3}$ & $3.9\times 10^{3}$ \\
			\bottomrule
		\end{tabular}
		\caption{Averaged Nusselt numbers during relaxation $\overline{\mathrm{Nu}}^r$  and sloshing-induced thermal mixing $\overline{\mathrm{Nu}}^s$ obtained with AEKF.}
		\label{tab:nusselt}
	\end{table}

	During the initial relaxation phase, the mean Nusselt numbers are larger on the liquid side than on the vapor side, both at the wall and across the interface. Because $\mathrm{Nu}_{ij}=h_{ij}R/\kappa_i$ and the ullage has a lower thermal conductivity than the liquid, observing $\mathrm{Nu}_{\text{liquid}}>\mathrm{Nu}_{\text{vapor}}$ implies that the liquid-side heat-transfer coefficients are substantially larger than the vapor-side ones in quiescent conditions. In addition, the wall-vapor and vapor-interface temperature differences are smaller than their liquid-side counterparts, further confirming the predominance of liquid-side exchange at rest. Consistently, the net interfacial heat balance is negative ($Q_{li}+Q_{vi}<0$), yielding a positive condensation rate $\dot m_c=-(Q_{li}+Q_{vi})/\mathcal{L}_v>0$ and an increase in liquid mass over the relaxation period. As the temperature differences and computed Nusselt coefficients are quasi-steady, the condensation rate is nearly constant. 
	It is observed that, across all test cases, $\overline{\mathrm{Nu}}^{r}_{li}\!<\!\overline{\mathrm{Nu}}^{r}_{wl}$ and $\bar T_w > T_i$ for $\hat t<=0$, indicating that the the wall–liquid pathway dominates the convective exchange in the quiescent phase.
	
	A consistent trend during relaxation is \(\overline{\mathrm{Nu}}^{r}_{wl}>\overline{\mathrm{Nu}}^{r}_{li}\) in all cases (Table~\ref{tab:nusselt}). Physically, the near-wall liquid sustains buoyant natural-convection boundary layers driven by the wall–liquid temperature difference, with the no-slip condition enhancing heat transfer along the vertical wall. In contrast, the free surface remains close to saturation and nearly shear-free, so the interfacial liquid-side driving \(\Delta T\) and shear are weaker, and interfacial exchange is further limited by vapor-side resistance. The result is larger wall–liquid coefficients than liquid–interface coefficients in quiescent conditions.
	
	During the destratification event, the time-averaged Nusselt numbers \(\overline{\mathrm{Nu}}^{s}_{ij}\) increase by several orders of magnitude compared to the relaxed state. Rather than simply indicating a shift from natural to forced convection, these values demonstrate that evaporation and the associated pressure decay are strongly *mixing-induced* phenomena. Among all pathways, the wall–liquid interface exhibits one of the largest increases in Nusselt number, reflecting the substantial heat flux required to overcome the high thermal inertia of the liquid phase. The enhanced wall–liquid exchange progressively reduces both wall superheating and liquid subcooling, driving these two control volumes toward equilibrium. Remarkably, the average \(\overline{\mathrm{Nu}}^{s}_{wl}\) remains nearly constant across all forcing conditions, leading to a consistent characteristic decay of wall temperature, as also verified experimentally. This invariance further supports the interpretation that the wall response is governed primarily by its intrinsic diffusive timescale rather than by the details of the external forcing.
	
	
	Recalling that the heat equation \eqref{eq:heat_eq_nd} with Neurman/Robin boundary conditions admits an analytical solution of the form

	\begin{equation}
		\Theta_w(\hat t)=\Theta_w^0e^{-\hat t/\hat \tau_w}
	\end{equation} 
	with $\hat \tau_w=1/(\mathrm{Fo}\lambda^2)$ with $\lambda$ the first eigenvalue such that $\lambda \tan(\lambda)=\overline{\mathrm{Bi}}$, \autoref{tab:biot_ekf} provides the dimensionless time constants and the associated Biot numbers. The values inferred here are in line with the experimental results in \autoref{tab:timescales}.
	\begin{table}[htbp]
		\centering
		\begin{tabular}{ccccc}
			\toprule
			& $S1$ & $S2$ & $S4$ & $S5$ \\
			\midrule
			$\overline{\mathrm{Bi}}$ & 1.13 & 1.39 & 0.99 & 1.15\\
			$\hat \tau_w$ & 37 & 32 & 41 & 37\\
			\bottomrule
		\end{tabular}
		\caption{Averaged values of Biot number and corresponding theoretical non-dimensional characteric time constant of wall temperature $\hat \tau_w$.}
		\label{tab:biot_ekf}
	\end{table}
	
	On the vapor side, $\overline{\mathrm{Nu}}^{s}_{vi}$ also increases significantly, which quickly sets the ullage to near-saturation conditions ($\overline \Theta_v\!\to\!\overline \Theta_i$). Notably, the step increase in the interfacial heat transfer coefficient appears earlier on the vapor side than on the liquid, leading to a short, critical evaporation event, which translates into a pressure burst visible across all experiments. Once $\overline \Theta_v\approx \overline \Theta_i$, the vapor-to-interface heat flux becomes negligible despite the large value of heat transfer coefficient. The subsequent ullage internal energy variations stem mainly in the wall-vapor exchange and condensation.
	Finally, although the liquid-interface Nusselt typically shows the smallest relative increase from quiescent conditions, the associated heat flux remains the dominant driver of the thermodynamic evolution. Indeed, the liquid-side temperature differences are larger, and the sustained interface heat removal from the liquid maintains a positive condensation rate after the ullage has reached saturation. We note that lower fill-levels and stronger accelerations lead to greater value $\overline{\text{Nu}}^s_{li}$, leading to stronger condensation and pressure decay rates consistent with the experimental observations.

\begin{figure*}[hbtp]                
		\centering
		\begin{subfigure}[b]{0.48\textwidth} 
			\includegraphics[width=\textwidth]
			{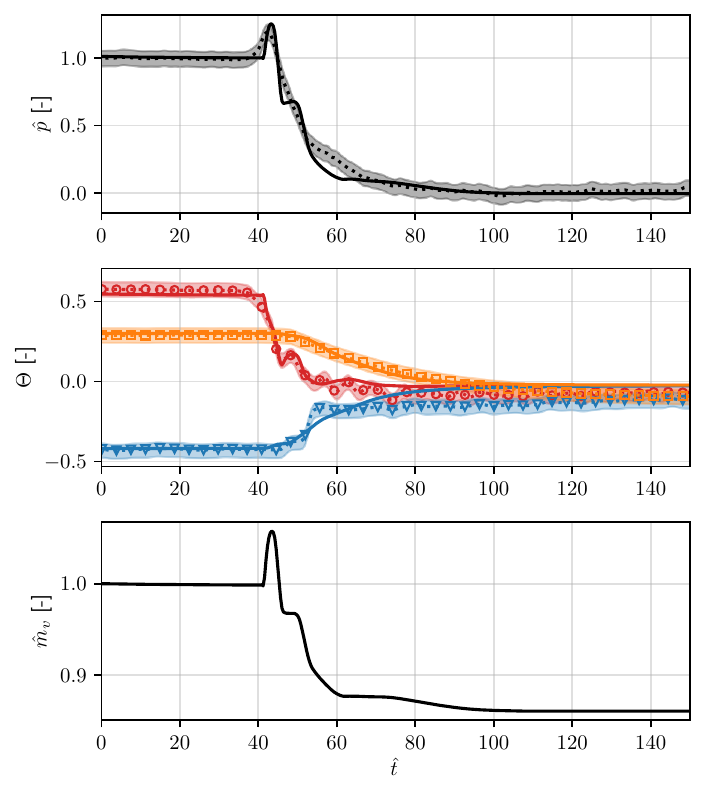}
			\caption{Case $S1$: $\hat a_v=0.35$, $H/2R=0.52$}
		\end{subfigure}  
		\begin{subfigure}[b]{0.48\textwidth} 
			\includegraphics[width=\textwidth]
			{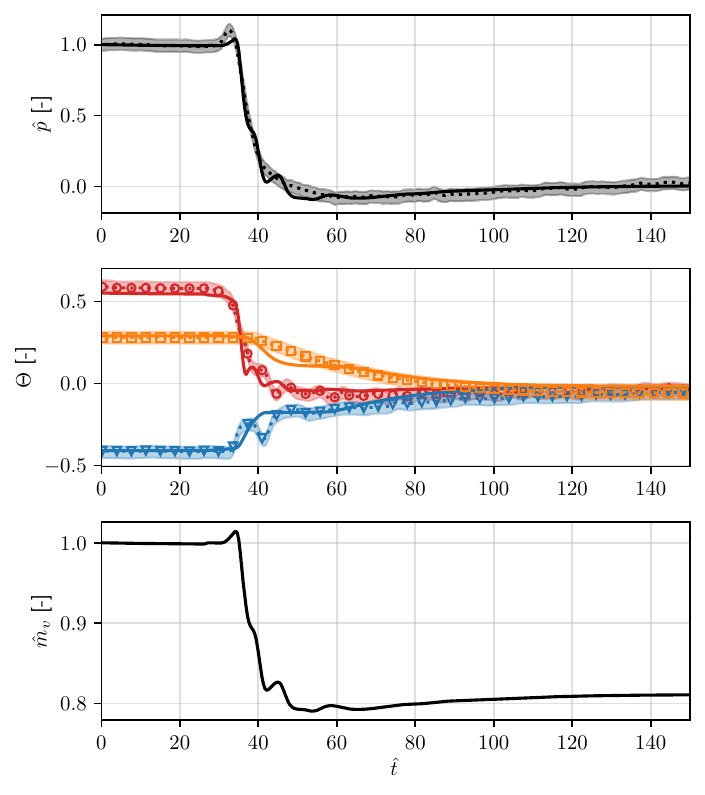}
			\caption{Case $S2$: $\hat a_v=0.49$, $H/2R=0.50$}
		\end{subfigure}  
		\begin{subfigure}[b]{0.48\textwidth} 
			\includegraphics[width=\textwidth]
			{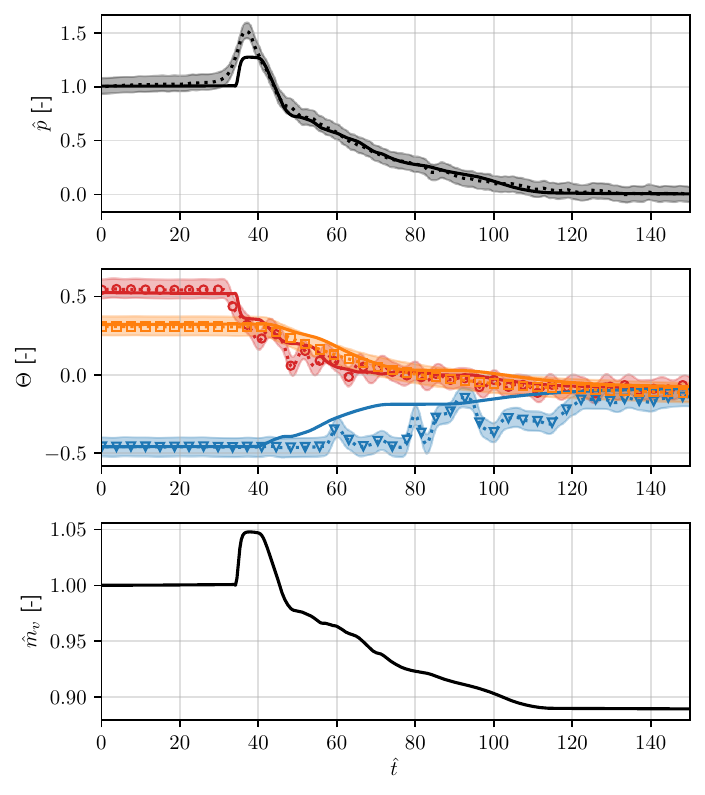}
			\caption{Case $S4$: $\hat a_v=0.35$, $H/2R=0.67$}
		\end{subfigure}  
		\begin{subfigure}[b]{0.48\textwidth} 
			\includegraphics[width=\textwidth]
			{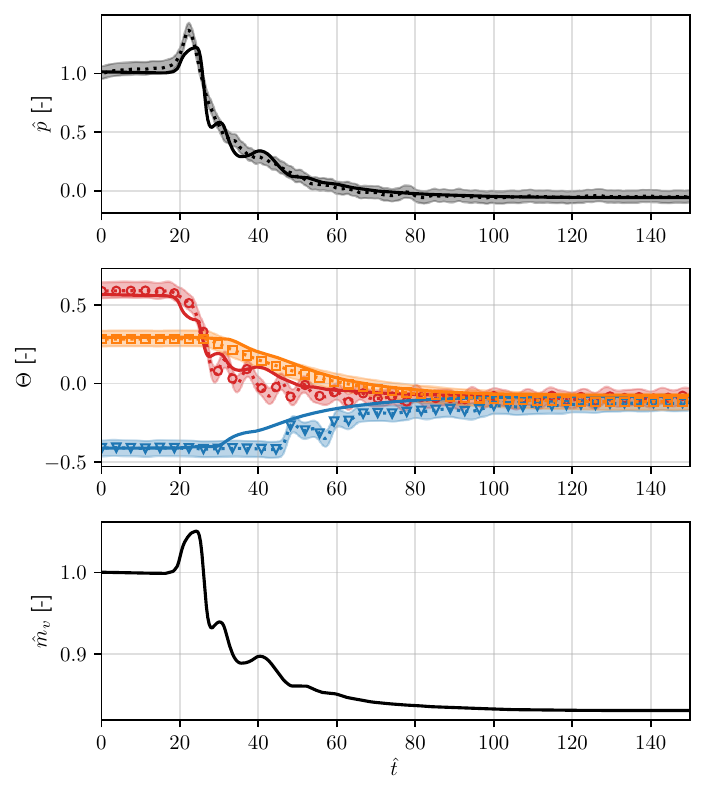}
			\caption{Case $S5$: $\hat a_v=0.49$, $H/2R=0.67$}
		\end{subfigure}  
		\caption{Model–data comparison of the non-dimensional thermodynamic response using the lumped-order model with AEKF-inferred heat-transfer coefficients. Cases: $S1$ (a), $S2$ (b), $S4$ (c), and $S5$ (d). Top: pressure $\hat p$; middle: control-volume mean temperatures $\overline{\Theta}_l$ (liquid, triangles), $\overline{\Theta}_w$ (wall, squares), and $\overline{\Theta}_v$ (ullage, circles); bottom: vapor mass $\hat m_v$. Solid lines denote model predictions; markers denote measurements.}
		\label{fig:EKF_state_evolution}
	\end{figure*}

	Figure~\ref{fig:EKF_state_evolution} compares the predicted non-dimensional pressure and temperatures $\{\hat p,\,\overline{\Theta}_v,\,\overline{\Theta}_l,\,\overline{\Theta}_w\}$ to the experimental records across all cases, and reports the computed vapor mass evolution $\hat m_v$ to illustrate the phase-change mechanism. During the initial quiescent phase, the small Nusselt numbers yield a near-constant pressure and wall and liquid temperatures. In the 70\% fill cases ($S_4$ and $S_5$), the vapor mass remains essentially constant, whereas in the 50\% cases ($S_1$ and $S_2$), the higher liquid-interface Nusselt produces a mild net condensation, which appears as a slight under-prediction of the vapor temperature.
	
	At the onset of thermal destratification, the inferred vapor–interface Nusselt rises first, increasing the heat flux to the free surface. In the model, the pressure rate has two contributions (mass exchange and ullage heating). A burst could in principle come from a rise of $\overline T_v$, but the ullage is already the warmest control volume and no positive $d\overline T_v/dt$ is observed at the burst. The model therefore attributes the transient increase in $p$ to a short surge of evaporation at the interface (positive $dm_v/dt$), which produces the observed inflection in $\overline T_v$ and the brief pressure spike. We note that fast, localized processes during jet–roof impact (e.g. contact of warm liquid with warm wall patches) may also contribute, but they are not resolved in the present framework. 
	
	The pressure burst is followed by a strong pressure collapse as the liquid-interface Nusselt jumps, switching back the main phase change process to condensation. The resulting decrease in vapor mass and internal energy drives the pressure drop. The pressure decay rate is well captured across all cases, suggesting a correct estimate of interfacial heat transfer. During this process the vapor temperature follows closely the saturation temperature. 
	As the collapse proceeds, the strong jump in wall-liquid heat exchanges occurs, initiating the convergence of the liquid and wall temperature to the equilibrium point. The thermal conduction along the wall thickness delays the cooling of the external side of the wall, showing an excellent match with the experimental data. 
	
	After roughly 80 forcing periods from the onset of destratification, the model and the experiment reach the same equilibrium state in pressure and temperature.

	Qualitatively, $\hat p(t)$ closely follows $\hat m_v(t)$, suggesting that interfacial mass exchange is the primary driver of the pressure dynamics. To analyze the link between these two quantities, one can note  that differentiating \eqref{eq:p_state} and neglecting ullage-volume changes (i.e.\ $dV_v/dt\simeq 0$), the pressure rate splits into a phase-change (mass) term and a thermal (ullage-temperature) term:
	\begin{equation}\label{eq:dp_dt}
		\frac{dp}{dt}
		= \underbrace{-\frac{1}{V_v}\left(\frac{\partial p}{\partial \rho}\right)_{T}\,\dot m_{\mathrm c}}_{(dp/dt)_m}
		\;+\;
		\underbrace{\left(\frac{\partial p}{\partial T}\right)_{\rho}\,\frac{d\overline{T}_v}{dt}}_{(dp/dt)_T}.
	\end{equation}
	
	The relative roles of the two mechanisms over a time window that spans both the burst and the decay (i.e., $\hat t \in \big[t^\star-2,\;t^\star+3\hat \tau_p\big]$, can be quantified in terms of absolute-mean shares:
	\begin{equation}\label{eq:dpdt_share}
		\delta p_m=\frac{\int_{t^\star-2}^{t^\star+3\hat \tau_p} |(dp/dt)_m| dt}{\int_{t^\star-2}^{t^\star+3\hat \tau_p}\left[|(dp/dt)_m|  +  |(dp/dt)_T| \right]dt}
	\end{equation}
	
	This contribution is reported in \autoref{tab:dpdt_abs_shares}. It appears that the phase change consistently dominates the pressure dynamics, representing about 80\% of the mean absolute share of the variation across all test cases.

	\begin{table}[hbtp]
		\centering
		\small
		\begin{tabular}{ccccc}
			\toprule
			& $S1$ & $S2$ & $S4$ & $S5$ \\
			\midrule
			$\delta p_m$ & 0.821 & 0.791 & 0.863 & 0.868 \\
			\bottomrule
		\end{tabular}
		\caption{Absolute-mean share of \(|dp/dt|\) over \([t^\star-2,\,t^\star+3\hat \tau_p]\), attributed to phase change effects.}
		\label{tab:dpdt_abs_shares}
	\end{table}
	
\begin{figure}[hbtp]
		\centering
		\includegraphics[width=0.7\linewidth]{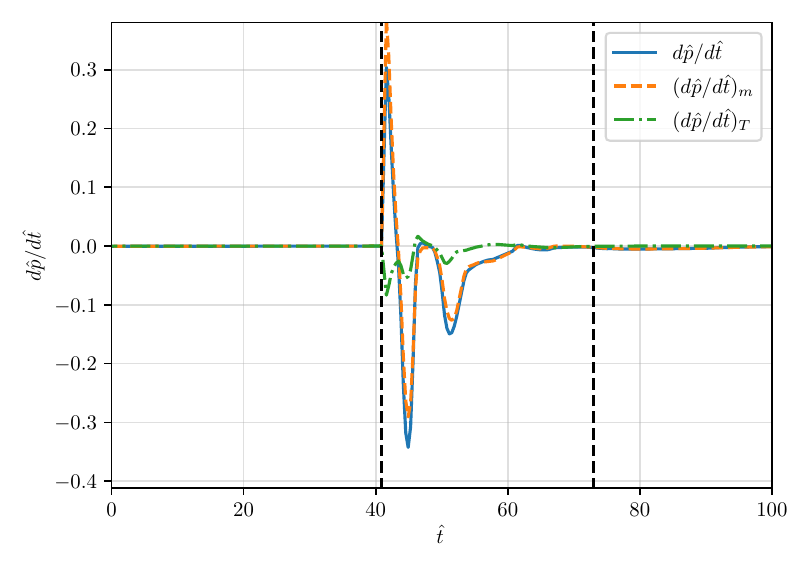}
		\caption{Instantaneous pressure-rate decomposition for $S1$: phase-change contribution (dashed), ullage-temperature contribution (dash–dotted), and sum (solid). Vertical dashed lines mark the integration bounds $t^\star-2$ and $t^\star+3\hat \tau_p$ used in \eqref{eq:dpdt_share}.}
		\label{fig:dp_dt_S1}
	\end{figure}

	Finally, Figure~\ref{fig:dp_dt_S1} illustrates the decomposition of the instantaneous pressure rate into the phase-change and ullage-temperature terms of \eqref{eq:dp_dt}. The mass-exchange contribution tracks the total signal closely and sets both the pressure burst at $\hat t \approx t^\star$ with a sudden strong evaporation and the subsequent condensation-driven pressure drop. In contrast, the thermal term remains smaller and mostly negative. During the burst, it opposes the rise in pressure and contributes slightly to the sloshing-induced drop.

	Table~\ref{tab:ekf-RMSE} quantifies discrepancies between the model prediction and the measurements using phase-specific RMSE on non-dimensional variables. During the initial relaxation ($\hat t\in[-100,0]$), errors are small: pressure remains at $O(10^{-2})$, wall and liquid temperatures are essentially flat and well captured (RMSE $\leq 0.008$), and the vapor temperature shows the largest but still modest discrepancies (0.011–0.025), consistent with its higher sensitivity to interfacial exchange. During thermal mixing ($\hat t\in[t^\star,150]$), the largest RMSE appears in the liquid temperature. This finding is attributed to the low spatial sampling in the liquid, as only one temperature sensor is located in the bulk liquid. Consequently, the liquid temperature readings are affected by sloshing-induced wetting intermittency during the transient to equilibrium. This effect inflates the mismatch on the transient liquid temperature prediction despite the overall trend being captured.
	
	\begin{table}[hbtp]
		\centering
		\small
		\begin{tabular}{lcccc}
			\toprule
			\multicolumn{5}{c}{Relaxation $\hat t\in[-100,0]$}  \\
			\midrule
			& $S_1$ & $S_2$ & $S_4$ & $S_5$ \\
			\midrule
			$\hat p$           & 0.005 & 0.002 &  0.004 & 0.013  \\
			$\overline{\Theta}_l$   & 0.002 & 0.001 & 0.008 & 0.000 \\
			$\overline{\Theta}_v$   & 0.020 & 0.025 & 0.011 & 0.013  \\
			$\overline{\Theta}_w$   & 0.007 & 0.006 & 0.008 & 0.007 \\
			\midrule
			\multicolumn{5}{c}{Thermal mixing $\hat t\in[t^\star,150]$} \\
			\midrule
			& $S_1$ & $S_2$ & $S_4$ & $S_5$ \\
			\midrule
			$\hat p$            & 0.049 & 0.026 & 0.035 & 0.036 \\
			$\overline{\Theta}_l$    & 0.088 & 0.048 & 0.135 & 0.074 \\
			$\overline{\Theta}_v$    & 0.048 & 0.035 & 0.042 & 0.042 \\
			$\overline{\Theta}_w$    & 0.039 & 0.039 & 0.045 & 0.037 \\
			\bottomrule
		\end{tabular}
		\caption{Phase-specific RMSE (dimensionless) between model and experiment for observable thermodynamic quantities in cases $S1$, $S2$, $S4$, and $S5$.}
		\label{tab:ekf-RMSE}
	\end{table}

	\subsubsection{Sensitivity analysis}
	We quantify the sensitivity of the predicted thermodynamic state, $\mathbf{s}(t)$, to the heat-transfer closures $\{h_{ij}\}$ using a Monte Carlo sensitivity analysis. Each coefficient is modeled as an independent Gaussian random variable, with mean centered on the expected value (computed from the AEKF results) and standard deviation equal to 25\% of the mean.
	For each $h_{ij}$, we randomly sample $N=200$ heat transfer coefficient trajectories while keeping the other parameters fixed to the baseline. These parameters are propagated in the lumped-parameter to obtain the associated thermodynamic evolution.

	\begin{figure*}[hbtp]                 
		\centering
		\begin{subfigure}[b]{0.47\textwidth} 
			\includegraphics[width=\textwidth]{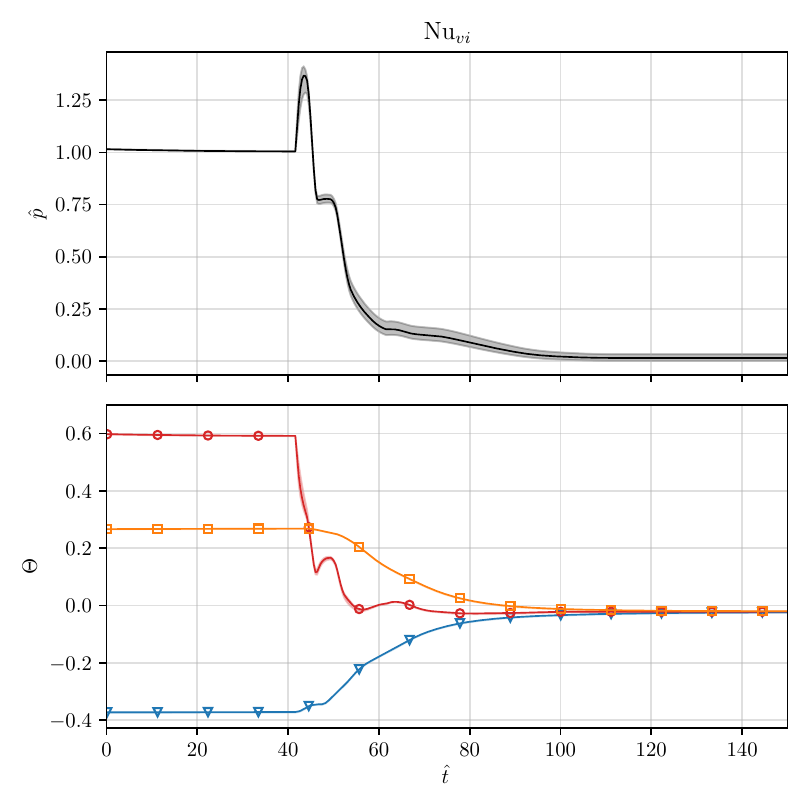}
			\caption{}
		\end{subfigure}  
		\begin{subfigure}[b]{0.47\textwidth} 
			\includegraphics[width=\textwidth]{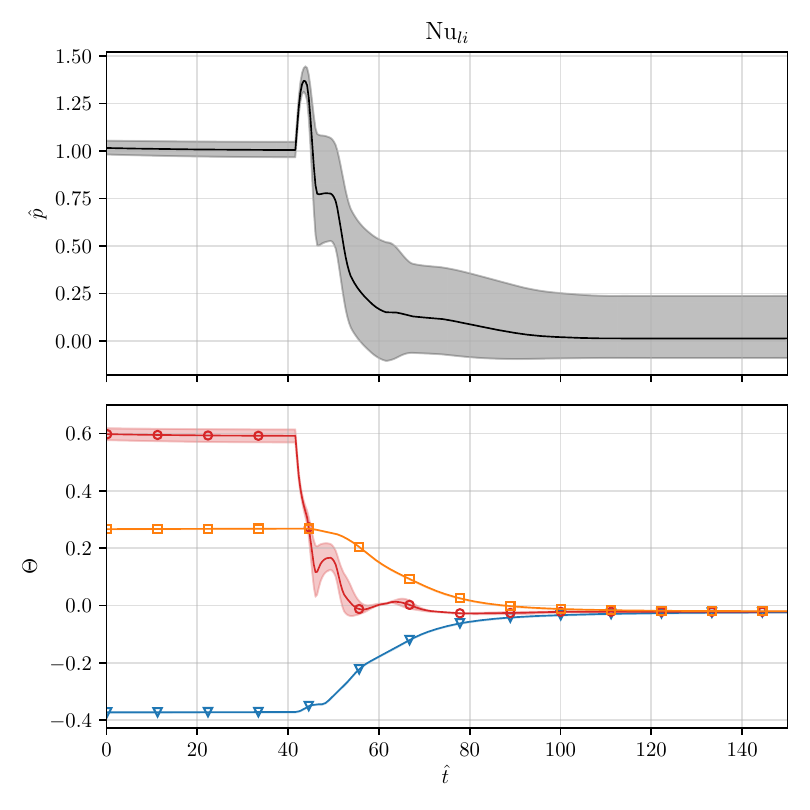}
			\caption{}
		\end{subfigure}  
		\begin{subfigure}[b]{0.47\textwidth} 
			\includegraphics[width=\textwidth]{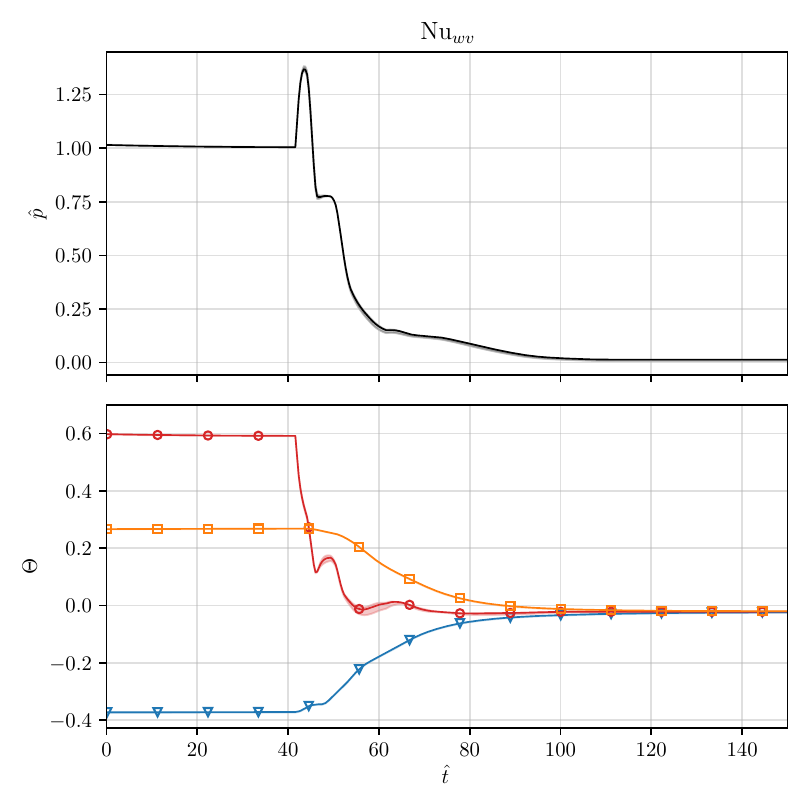}
			\caption{}
		\end{subfigure}  
		\begin{subfigure}[b]{0.47\textwidth} 
			\includegraphics[width=\textwidth]{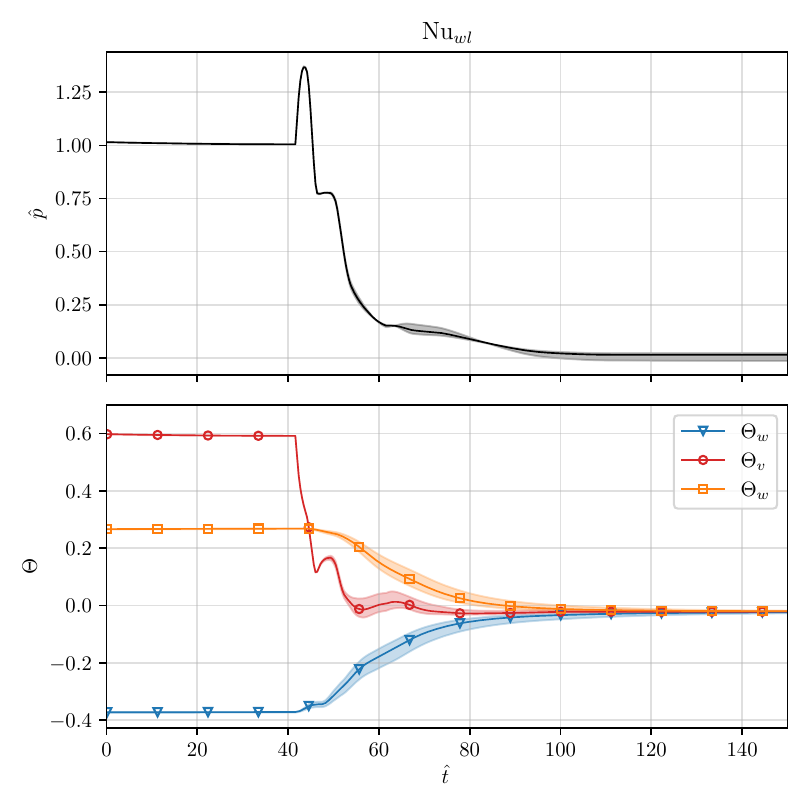}
			\caption{}
		\end{subfigure}  
		\caption{Sensitivity of the thermodynamic response in case $S1$ to perturbations of the inferred Nusselt numbers. Panels: $\mathrm{Nu}_{vi}$ (a), $\mathrm{Nu}_{li}$ (b), $\mathrm{Nu}_{wv}$ (c), and $\mathrm{Nu}_{wl}$ (d).  The shaded band shows the response envelope due to the $\pm25\%$ variation; the solid line is the baseline prediction. Top subplots show the non-dimensional pressure $\hat p(\hat t)$; bottom subplots show control-volume mean temperatures (circles: ullage $\overline{\Theta}_v$, squares: wall $\overline{\Theta}_w$, triangles: liquid $\overline{\Theta}_l$).}
		\label{fig:MC-sensitivity}
	\end{figure*}
	
	\autoref{fig:MC-sensitivity} reports the predicted pressure and temperatures for case $S1$. The solid lines correspond to the baseline solution, and the shaded envelopes show the effect of perturbing the selected coefficient.
	
	The results show that the predicted pressure is predominantly affected by the liquid-interface heat transfer coefficient. This is explained by the fact that the liquid holds most of the thermal energy of the fluid, and \(\textrm{Nu}_{li}\) determines how quickly this energy is converted into a phase change at the interface. The resulting phase-change rate directly controls the evolution of vapor mass and internal energy, which, through the equation of state, manifests as changes in pressure and vapor temperature. In contrast, the impact of this parameter on the bulk liquid temperature is negligible, because the interfacial heat exchange remains too small to alter the large thermal inertia of the liquid. During the mixing event, the sensitivity to this parameter grows as the mean value of the associated Nusselt rise from $o(1)$ to $o(10^2)$, so the perturbation amplitude increases, but the same mechanism controls the impact of the thermodynamic state. 
	
	Other heat-transfer pathways play secondary roles on the pressure evolution. During the initial relaxation phase, this behavior is explained by the weak mean values of other Nusselt numbers compared to the liquid-interface one (see \autoref{tab:nusselt}). As the jump in heat exchange occurs during thermal mixing, the vapor-interface heat exchange pathway has an impact on the pressure burst amplitude as it controls the short evaporation event. The influence of this parameter becomes negligible again as the jump in liquid-interface Nusselt occurs and the vapor starts condensing again. Contrarily, the wall-vapor parameter has barely any impact on the system dynamics due to the small thermal inertia enclosed in the vapor. Finally, the wall-liquid heat transfer route has a small impact on the transient pressure and temperatures evolution during the liquid warming phase, but does not affect the equilibrium point.

	\section{Conclusions}\label{sec.conclusions}
	
	This work presented an experimental characterization of the thermodynamic response of a partially filled horizontal cylinder to vertical parametric excitation. The investigation combined a decoupled experimental approach to first investigate the isothermal kinematic sloshing response before quantifying the resulting non-isothermal effects. Experiments focused on excitation frequencies near twice the natural frequency of the first longitudinal symmetric mode $(2,0)$. A lumped-parameter model, coupled with an Augmented Extended Kalman Filter, was then employed to infer the time-resolved heat transfer coefficients governing the thermodynamic evolution of the system.
	
	Data-driven modal decomposition of the images obtained during the isothermal campaign allowed for the computation of the natural frequencies, and it revealed that the tank's domes have a significant impact on the values predicted by linear potential flow theory for a geometry with straight ends. This preliminary investigation also highlighted the existence of a critical forcing amplitude consistent with linear stability theory, beyond which strong sloshing motion develops. At 50\% filling, the resonant sloshing mode $(2,0)$ was observed to alternate periodically with the first longitudinal antisymmetric mode $(1,0)$. This observed mode competition suggests strong non-linear modal interactions.
	
	The non-isothermal experiments demonstrated that the low-amplitude surface wave ripples obtained below the instability threshold did not affect the thermal stratification of the system, nor the heat and mass processes. Conversely, above the threshold, the onset of the middle-jet triggered a strong thermal mixing, characterized by a pressure burst, followed by a collapse onto the saturated equilibrium point of the system and complete thermal destratification. The characteristic time for the onset of the thermal mixing was shown to decrease with the fill-level. Finally, the pressure decay rate was shown to increase with the forcing amplitude, with the characteristic pressure decay time decreasing by more than half as acceleration was raised of 40\%.
	
	The AEKF-based inference of the heat transfer coefficients provided critical insight into the underlying physics. The transition to strong sloshing was marked by a sudden, step-like increase in all heat transfer coefficients, with Nusselt numbers jumping by two to four orders of magnitude. A direct pressure-rate decomposition showed that phase change is the primary driver of the pressure dynamics, representing about 80\% of the pressure change, while ullage temperature variations provide a secondary, mostly damping contribution. A sensitivity analysis highlighted the primary role of the liquid-interface heat transfer on the sloshing-induced pressure drop as it drives the rate of vapor condensation. The successful application of the AEKF establishes a robust methodology for analyzing transient heat transfer in complex multiphase systems, providing interpretable parameters that generalize across fill levels and forcing.

	This study provides the first experimental benchmark on the thermodynamic impact of vertical sloshing in a horizontal tank. Future experimental work should focus on establishing a comprehensive regime map linking the sloshing response to the excitation parameters, and further investigating the non-linear mode interactions observed at 50\% filling. The present thermodynamic benchmark should then be expanded to include other resonant and off-resonant conditions, different excitation directions, and a wider range of fluids and tank geometries. Such a extensive dataset is required for developing and validating sloshing heat transfer models, which are essential for the safe design and operation of full-scale cryogenic systems.

	\section*{Acknowledgment}
	
	The experiments were designed and carried out as part of the MACHEN project funded by Airbus. S.A.A. is supported by the Flemish Agentschap Innoveren \& Ondernemen under the framework ``Fundamenteel onderzoek 2024-2025 in Klimaatverandering en Energietransitie'', under Grant No HBC.2023.0897. 
	M.A.M is supported by the European Research Council (ERC, grant agreement No 101165479 RE-TWIST StG). Views and opinions expressed are however those of the authors only and do not necessarily reflect those of the European Union or the European Research Council. Neither the European Union nor the granting authority can be held responsible for them.

	\appendix

	\section{Derivation of the Mathieu equation}\label{appA}
	
	We briefly outline the derivation of \eqref{eq:modal_ODE} for vertical sloshing in a container of arbitrary geometry. The general setting is the linearized potential flow such that the velocity field satisfies $\bm{u}=\nabla \phi$, with $\phi$ the velocity potential, and incompressibility requires the potential to satisfy Laplace's equation $\nabla \cdot \bm{u}=\nabla ^2 \phi=0$. At the interface, the liquid elevation elevation \(\eta(\bm{r},t)\) and the flow potential $\phi(\mathbf{x}=\eta(t),t)$ are linked by the linearized kinematic and dynamic conditions 
	
	\begin{align}
		\partial_t \eta &= \partial_z \phi, \label{eq:kinematic_lin}\\[6pt]
		\partial_t \phi + g_{\mathrm{eff}}(t)\,\eta &= 0, \label{eq:dynamic_lin}
	\end{align} with \(g_{\mathrm{eff}}(t) = g - z_e\omega^2_e\cos(\omega_e t)\) the effective gravity.
	
	In the linearized regime, the interface elevation \(\eta(\bm{r},t)\) is represented as a superposition of spatial eigenfunctions \(\xi_{n,m}(\mathbf{r})\) weighted by temporal amplitudes \(b_{n,m}(t)\),
	\begin{equation}
		\label{exp_1}
		\eta(\bm{r},t) = \sum_{n,m} \xi_{n,m}(\bm{r})\, b_{n,m}(t).
	\end{equation}
	
	The velocity potential is expanded analogously as
	
	\begin{equation}
		\label{exp_2}
		\phi(\mathbf{x},t) = \sum_{n,m} \psi_{n,m}(\mathbf{x})\, a_{n,m}(t),
	\end{equation}
	where \(\psi_{n,m}\) are the potential eigenfunctions in the liquid domain. Substituting these representations into \eqref{eq:kinematic_lin} and using standard Galerkin procedure for projecting onto \(\xi_{n,m}\) yields
	\begin{equation}
		\label{proj_1}
		\dot b_{n,m}(t) 
		= a_{n,m}(t)\,\frac{\langle \xi_{n,m}, \partial_z \psi_{n,m}\rangle}
		{\langle \xi_{n,m}, \xi_{n,m}\rangle},
	\end{equation}
	with \(\langle \cdot,\cdot\rangle\) denoting the appropriate inner product on the horizontal domain. The explicit form of the inner product is immaterial for the present derivation; only the orthogonality of the eigenfunctions with respect to this product is required.
	
	Similarly, introducing the expansion \eqref{exp_2} into \eqref{eq:dynamic_lin} and projecting onto \(\psi_{n,m}\) yields
	\begin{equation}
		\label{proj_2}
		\dot a_{n,m}(t) 
		= -\,g_{\mathrm{eff}}(t)\,b_{n,m}(t)\,
		\frac{\langle \psi_{n,m}, \xi_{n,m}\rangle}
		{\langle \psi_{n,m}, \psi_{n,m}\rangle}.
	\end{equation}
	
	Eqs \eqref{proj_1} and \eqref{proj_2} can be combined into a single equation by first differentiating \eqref{proj_1} in time and eliminating the time derivatives $\dot a_{n,m}(t) $ to obtain: 
	
	\begin{equation}
		\label{ODE_in_b}
		\ddot b_{n,m}(t) 
		+ g_{\mathrm{eff}}(t)\, b_{n,m}(t)\,\mathcal{K}_{n,m}= 0.
	\end{equation} with 
	
	\begin{equation}
		\mathcal{K}_{m,n}=\frac{\langle \psi_{n,m}, \xi_{n,m}\rangle}
		{\langle \psi_{n,m}, \psi_{n,m}\rangle}\,
		\frac{\langle \xi_{n,m}, \partial_z \psi_{n,m}\rangle}
		{\langle \xi_{n,m}, \xi_{n,m}\rangle}\,\in\mathbb{R}^{+}.
	\end{equation}
	
	It is worth stressing this quantity solely depends on the eigenfunctions and not the forcing. To identify it, we now impose that \eqref{ODE_in_b} converges to the unforced conditions at the limit $g_{\mathrm{eff}}(t)\rightarrow g$. This means that  
	
	\begin{equation}
		\label{Kappas}
		\mathcal{K}_{m,n}=\frac{\omega^2_{n,m}}{g}\,.
	\end{equation}
	
	Introducing \eqref{Kappas} in \eqref{ODE_in_b} gives \eqref{eq:modal_ODE}.

	

	
	
	
	
	\bibliography{cas-refs}

	
	

\end{document}